\documentclass[apj]{emulateapj}
\pdfoutput=1
\usepackage{graphicx}
\usepackage[space]{grffile}
\usepackage{latexsym}
\usepackage{amsfonts,amsmath,amssymb}
\usepackage{url}
\usepackage[utf8]{inputenc}
\usepackage{hyperref}
\hypersetup{colorlinks=false,pdfborder={0 0 0}}
\usepackage{textcomp}
\usepackage{longtable}
\usepackage{multirow,booktabs}
\usepackage{natbib}
\usepackage{float}
\usepackage{color,soul}
\usepackage{verbatim}

\begin{document}

\slugcomment{To Appear in the Astrophysical Journal}

\title{The Skeleton of the Milky Way\\}


\author{Catherine Zucker}
\affil{Astronomy Department, University of Virginia, Charlottesville, VA, 22904}
\affil{Harvard-Smithsonian Center for Astrophysics, Cambridge, MA, 02138}

\author{Cara Battersby and Alyssa Goodman}
\affil{Harvard-Smithsonian Center for Astrophysics, Cambridge, MA, 02138}

\email{catherine.zucker@cfa.harvard.edu}






\begin{abstract}
Recently, \citet{Goodman_2014} argued that the very long, very thin
infrared dark cloud ``Nessie'' lies directly in the Galactic mid-plane
and runs along the Scutum-Centaurus arm in position-position-velocity
({\emph{p-p-v}}) space as traced by lower density $\textrm{CO}$ and
higher density $\mathrm{NH_3}$ gas. Nessie was presented as the first
``bone'' of the Milky Way, an extraordinarily long, thin, high-contrast
filament that can be used to map our Galaxy's ``skeleton." Here, we
present evidence for additional bones in the Milky Way Galaxy, arguing
that Nessie is not a curiosity but one of several filaments that could
potentially trace Galactic structure. Our ten bone candidates are all
long, filamentary, mid-infrared extinction features which lie parallel
to, and no more than 20 pc from, the physical Galactic
mid-plane. We use $\textrm{CO}$, $\mathrm{N_2H^+}$, $\textrm{HCO}^+$
and $\mathrm{NH_3}$ radial velocity data to establish the
three-dimensional location of the candidates in {\emph{p-p-v}} space. Of
the ten candidates, six also: have a projected aspect ratio of
$\ge$ 50:1; run along, or extremely close to, the Scutum-Centaurus
arm in \emph{p-p-v} space; {\emph{and}} exhibit no abrupt shifts in
velocity. The evidence presented here suggests that these candidates mark the
locations of significant spiral features, with the bone called filament
5 (``BC\_18.88-0.09") being a close analog to Nessie in the Northern
Sky. As molecular spectral-line and extinction maps cover more of the
sky at increasing resolution and sensitivity, it should be possible to find more bones
in future studies. \end{abstract}



\section{Introduction}
Many surprisingly fundamental questions remain about the structure of the Milky Way. For instance, does the Milky Way have two \citep{Jackson_2008,Francis_2009,Dobbs_2012} or four \citep{Reid_2009,Bobylev_2014,Urquhart_2013} major spiral arms? What is the precise location of these arms in position-position-velocity (\textit{p-p-v}) space? What is the nature of inter-arm structure---is it made of well-defined spurs or more web-like structures? Does it even make sense to describe the Milky Way as a grand design spiral, and is it fruitful to count the number of spiral arms or prescribe terms such as ``log-spiral", ``spur", ``feather" or ``pitch angle" to structure whose detailed nature is not yet known? Simply put, the spiral structure of the Milky Way is far from solved, and an understanding of its true structure continues to elude us, largely due to the fact that we are embedded in the galaxy we are attempting to delineate. 

Much of our current understanding of the Milky Way's three-dimensional structure stems from radial velocity measurements of gas. Making use of the Milky Way's rotation curve \citep{McClure_Griffiths_2007,Reid_2014}, we can translate line-of-sight velocities into distances and construct a gross three-dimensional model of our Galaxy, though it is often difficult to disentangle features when they accrue along any line-of-sight. Thanks to a wealth of spectroscopic surveys, velocity-resolved observations are readily available for much of the Galaxy's molecular, atomic, and ionized gas. Extended tracers, like CO \citep{Dame_2001} or HI \citep{Shane_1972} provide the best constraints on the Galaxy's overall anatomy. To probe finer structure, observations of high-mass star forming regions can also provide kinematic and distance information for high-density gas. For instance, measurements of trigonometric parallaxes and proper motions of masers from the BeSSeL survey produce accurate locations for several spiral arm segments, along with their associated pitch angles \citep{Reid_2014}. Likewise, the Bolocam Galactic Plane Survey \citep[BGPS,][]{Schlingman_2011,Shirley_2013,Ellsworth_Bowers_2013,Ellsworth_Bowers_2015}, the Millimetre Astronomy Legacy Team 90 GHz Survey \citep[MALT90,][]{Foster_2011,Jackson_2013}, the $\textrm{H}_2\textrm{O}$ Southern Galactic Plane Survey \citep[HOPS,][]{Purcell_2012,Walsh_2011}, and ATLASGAL follow-up spectral line surveys \citep{Beuther_2012, Wienen_2012} have produced hundreds of high-spectral resolution velocity measurements of the dense gas in molecular clouds. Analyses of extinction data from surveys like Pan-STARRS1 complement this emission line data and can also be used to create three-dimensional models of the Galaxy's structure \citep{Green_2014,Schlafly_2014}, with high resolution on the plane of the Sky, but much coarser resolution along the line of sight.

To supplement existing extinction maps and observations of cold gas tracers and high-mass star forming regions, observations of star-forming complexes, young open clusters, embedded clusters, and Classical Cepheids can constrain the substructure of the Milky Way. For instance, \citet{Russeil_2003} create a new catalog of star forming complexes---identified as a combination of HII regions, diffuse ionized gas, molecular clouds, and OB stars---and derive a kinematic distance to each complex. When taking into account error bars, they do not find any visual evidence of large-scale spiral structure. However, they do identify the Carina, Perseus and Local arms as well-defined spiral segments, suggesting that the Milky Way is a grand design spiral composed of several prominent sub-features. 

Similarly, \citet{Vazquez_2008} use optical observations of young open clusters, in combination with CO radio observations of molecular clouds, to trace spiral structure in the third Galactic quadrant. In particular, well-traced by both stars and CO, they classify the Outer (Cygnus) arm as a grand design spiral arm extending from $l=190^\circ - 255^\circ$. In parallel, \citet{Majaess_2009} find that young open clusters (ages $<$ 10 Myear) and short-period Classical Cepheids (with typical ages of 40-80 Myears) are young enough to have drifted minimally from their birthplaces within spiral arms, making them both viable tracers of spiral features. Using a combination of young open clusters and short-period classical Cepheids, \citet{Majaess_2009} confirm Sagittarius-Carina to be a major spiral arm and note the existence of another spiral feature concentrated in the Cygnus-Vulpecula region. Finally, \citet{Carraro_2011} and \citet{Camargo_2015} use UBVI photometry of young stars and WISE images of embedded clusters, respectively, to trace major spiral segments. In the former, distance measurements place two of the three major reddening groups of young stars within major spiral arms. In the latter, analysis of the distribution of embedded clusters find them preferentially located within the thin disk and along spiral arms, with the current catalog of embedded clusters tracing the Sagittarius-Carina, Perseus, and Outer arms.

While the tools available for probing the Milky Way's internal structure are diverse, none has especially high three-dimensional resolution over wide areas, nor are they capable of probing the densest gas on Galactic scales. To address these issues, \citet{Goodman_2014} recently suggested that extraordinarily elongated filamentary infrared dark clouds (IRDCs), termed ``bones" could potentially be used to constrain the structure of the Milky Way. \citet{Goodman_2014} presented an extended version of the IRDC called ``Nessie" \citep[discovered by][]{Jackson_2010} as the first bone of the Milky Way.  They found that Nessie was at least $3^\circ$ ($\sim 160$ pc), and possibly as long as $8 ^\circ$ ($\sim 430$ pc) in length, while being less than 0.1$^\circ$ ($\sim 0.6$ pc) wide. They also conclude that Nessie lies very near the geometric mid-plane of the Milky Way Galaxy, at the 3.1 kpc distance to the Scutum-Centaurus arm.  Analysis of the radial velocities of ${\rm NH}_3$ emission and CO emission shows that Nessie runs along the Scutum-Centaurus arm in {\it p-p-v} space, suggesting that it forms a dense spine of that arm in physical space as well \citep{Goodman_2014}

Until very recently, no simulations had the spatial resolution to predict filaments as narrow as the bones, if they were to exist.  In 2014, numerical simulations from \citet{Smith_2014}, using the AREPO moving mesh code, revealed dense filaments, with aspect ratios and column densities similar to Nessie, forming within and parallel to the mean plane of a simulated spiral galaxy.  A detailed analysis of Nessie's properties, along with these new simulation results, suggests Nessie may be the first in a class of objects that could trace our Galaxy's densest spiral features \citep{Goodman_2014}. It is reassuring to recognize that Nessie should be the easiest object of its kind to find. Nessie is located in the closest major spiral arm to the the Sun, perpendicular to our line of sight, slightly offset from the Galactic center. This placement makes Nessie clearly visible against the bright background of the Galactic center, and more elongated than objects more distant or more inclined to our line-of-sight.

In this paper, we use large-scale mid-infrared imaging of the Galactic plane to search for bone candidates near locations where currently-claimed spiral arms {\it should} lie on the Sky (\S2.1). It is critical to appreciate, as explained in detail in \citet{Goodman_2014}, that the Sun's 25 pc elevation above the Galactic mid-plane gives viewers on Earth a (very-foreshortened) top-down perspective view of the Galaxy's structure, so that arms lie at predictable offsets (e.g. $b=-0.4 ^\circ $ for Nessie) from the IAU zero of Galactic Latitude. We predict and exploit these offsets in our search for bone candidates. We then follow-up on promising candidates using radial velocity measurements of high and low density gas tracers, to establish velocity contiguity and any potential association with pre-existing spiral arm traces in \textit{p-p-v} space (\S2.2). Next, we develop criteria for objects to be termed Galactic ``bones" (\S2.3), sort the filaments by these criteria (\S2.4) and analyze the strongest candidates, noting comparisons to related searches for spiral-tracing filaments (\S3). When used in conjunction with other methods outlined above (i.e. CO and HI observations, extinction mapping, high-mass star forming regions, masers, embedded clusters, short-period Classical Cepheids, young open clusters, star-forming complexes), these new bones have the potential to pin down the Milky Way's Galactic structure on the Sky to pc-scale resolution in regions in the vicinity of the nearest bones (\S4).

\section{Methodology}
\subsection{Visual Search}
To search for more bones, we looked for them around where they are expected to lie in $l,b,v$ space, according to our current understanding of the Milky Way's structure, which, as we stress, is far from complete. We began by calculating the expected $l,b$ paths of Galactic arms using a log-spiral approximation as described in recent literature \citep{Dame_2011,Vallee_2008} and assuming a 25 pc height above the mid-plane for the Sun \citep[see][and references therein]{Goodman_2014}. The predicted positions of the Galactic arms (Scutum-Centaurus, Carina-Sagittarius, Norma, and Perseus) were then overlaid on \textit{Spitzer} GLIMPSE/MIPSGAL \citep{Benjamin_2003,Churchwell_2009} images in World Wide Telescope (WWT)\footnote{http://www.worldwidetelescope.org/}---a tool that facilitates easy visualization of several layers of data at scales from the full sky down to the highest-resolution details. We restricted our initial search to the MIPSGAL footprint ($|l|<62^\circ, |b|<1^\circ$), with particular attention given to the region $-51^\circ<l<31^\circ$, as the Scutum-Centaurus arm (the closest major spiral arm from our vantage point) has tangent points limited by these longitudes. Panning along the full \textit{Spitzer}/MIPSGAL Survey in WWT, we searched for largely continuous, filamentary, extinction features near and roughly parallel to the Galactic mid-plane, where all of the overlain arm traces lie. This visual inspection yielded about fifteen initial bone candidates. A video showing how this search worked in WWT is available on YouTube\footnote{http://tinyurl.com/morenessies}, and the original WWT Tour, of which the video shows a capture, is available at the Bones of the Milky Way Dataverse\footnote{http://dx.doi.org/10.7910/DVN/29934}.

\subsection{Probing Velocity Structure}
For features that appear associated with spiral arms on the 2-D plane of the sky, radial velocity data are needed to establish whether 3-D association with a spiral feature is likely.  Any good bone candidate must have similar line-of-sight velocities along its full length (i.e. no abrupt shifts in velocity of more than 3 km s$^{-1}$ per 10 pc along the bone), and more importantly, the measured radial velocities should be very close to those predicted by the Milky Way's rotation curve for arms at a known distance. To probe the velocity structure of the initial bone candidates identified in WWT, we employed radial velocity data from five separate radio surveys:  HOPS \citep{Purcell_2012,Walsh_2011}, MALT90 \citep{Foster_2011,Jackson_2013}, BGPS spectral-line follow-up \citep{Schlingman_2011,Shirley_2013,Ellsworth_Bowers_2013}, GRS \citep{Jackson_2006} and ThrUMMS \citep{Barnes_2011}. The HOPS, MALT90, and BGPS surveys are all geared towards probing dense regions hosting the early stages of high-mass star formation. We utilize $\textrm{NH}_3$ emission from HOPS, $\mathrm{N_2H^+}$ from MALT90, and $\textrm{HCO}^+$ from BGPS. All three of these  lines trace dense molecular gas ($\sim 10^{4}\textrm{ cm}^{-3}$), and are often found in dense, cool clouds with temperatures less than 100 K \citep{Purcell_2012,Shirley_2013}. As IRDCs tend to harbor cool, high density clumps of gas which fuel the formation of massive stars, all three of these data sets contain spectra for hundreds of regions within the longitude range of the potential Galactic bones. To complement these high density gas tracers, we probe the puffier envelopes ($\sim 10^{2}\textrm{ cm}^{-3}$) surrounding these bones using high resolution $^{13}\rm{CO}$ data from the GRS and ThrUMMS surveys. 

\begin{figure*}
\begin{center}
\plotone{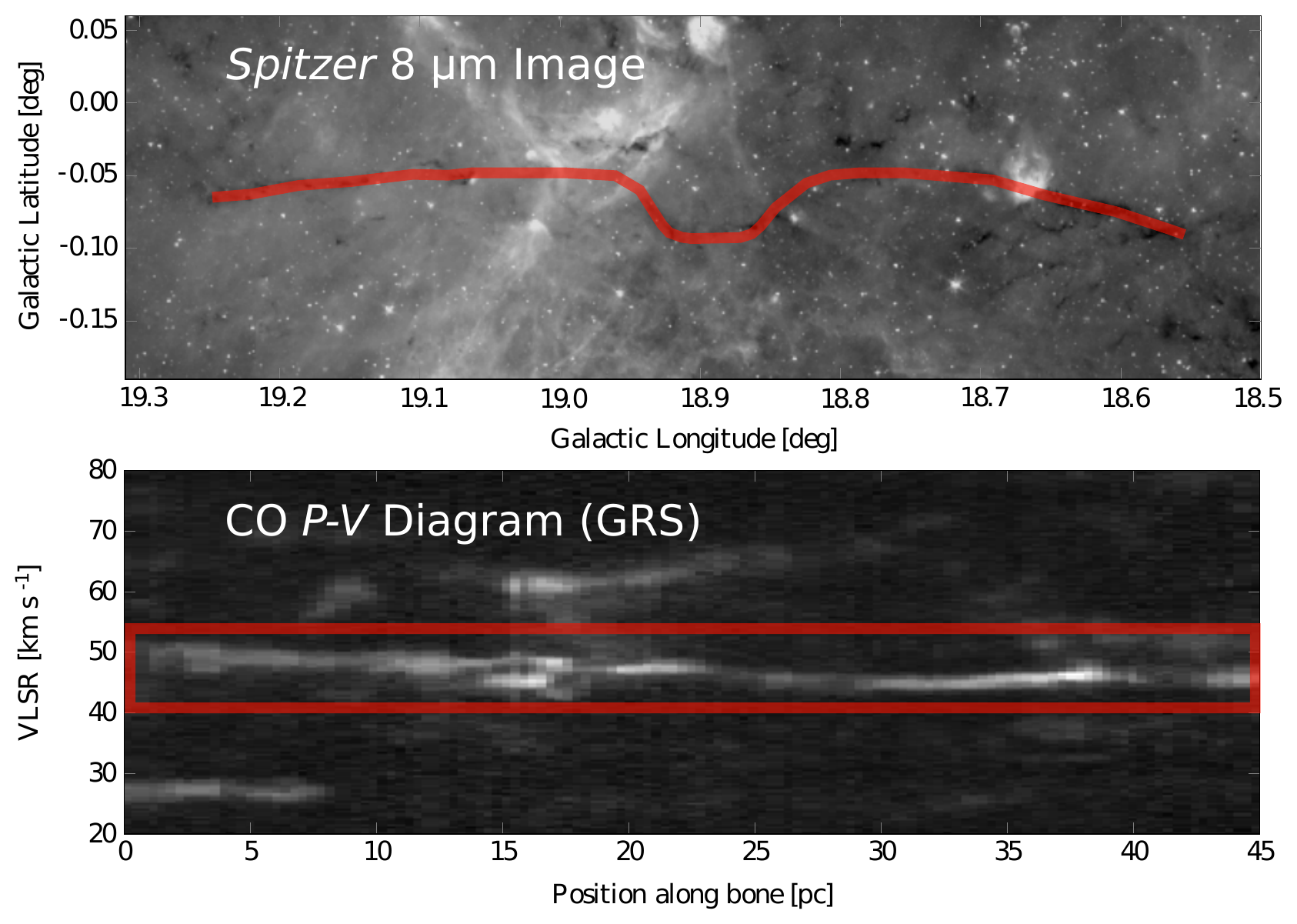} 
\caption{\label{fig:filament5_slice} Results of performing a slice extraction along the filamentary extinction feature of our strongest bone candidate, filament 5. The top panel shows a \textit{\textit{Spitzer}}-GLIMPSE $8\micron$ image of filament 5, and the red trace indicates the curve (coincident with the extinction feature) along which a \textit{p-v} slice was extracted. The bottom panel shows the \textit{p-v} slice, with the red-boxed region indicating the emission corresponding to filament 5. 
}
\end{center}
\end{figure*}

We investigate the velocity structure of our filaments in two ways: first, whenever possible, we establish the velocity contiguity of our candidates, as traced by lower density gas, by performing a slice extraction along each filamentary extinction feature in Glue\footnote{www.glueviz.org/en/stable/index.html}, a visualization tool that facilitates the linking of data sets. We link spectral \textit{p-p-v} cubes from the GRS and ThrUMMS survey with GLIMPSE-\textit{Spitzer} mid-infrared images and obtain velocity as a function of position along a path that traces the entire extinction feature; for a demonstration of how this was done, see Figure \ref{fig:demoI} in the appendix. The results of the slice extraction along the path of one of our strongest bone candidates is shown in Figure \ref{fig:filament5_slice}. We are able to establish velocity contiguity, as traced by lower density $^{13}\rm{CO}$ gas, for all bone candidates lying within the coverage range of the GRS survey ($ 18^\circ < l < 56^\circ$) and the ThrUMMS survey ($300^\circ < l < 358^\circ$). 

Next, we ensure that the candidates are contiguous in velocity space as traced \textit{mainly} by high-density emission from the HOPS, MALT90, and BGPS surveys. In cases where HOPS, MALT90, and BGPS catalog data are not available along the extinction feature, we also extract spectra from GRS and MALT90 \textit{p-p-v} cubes using the spectrum extractor tool in Glue. We once again link spectral \textit{p-p-v} cubes from the GRS or MALT90 surveys with GLIMPSE-\textit{Spitzer} mid-infrared images and use the spectrum-extractor tool to obtain velocities along different regions of the extinction feature; a demonstration of the procedure used to extract velocities at different coordinates along the filament in Glue is shown in Figure \ref{fig:demoII} of the appendix. Since CO traces lower density gas ($\sim 10^2 \textrm{ cm}^{-3}$) and $\mathrm{N_2H^+}$, $\textrm{HCO}^+$, and $\textrm{NH}_3$ trace higher density gas ($\sim 10^4 \textrm{ cm}^{-3}$), the dense gas sources provide more relevant estimates of the velocity of cold, dense, filamentary IRDCs. However, where dense gas sources are not available, the complete and unbiased high resolution GRS survey, although less desirable, allows us to roughly gauge the velocity along entire lengths of filaments. 

\begin{figure*}
\begin{center}
\epsscale{1.0}
\plotone{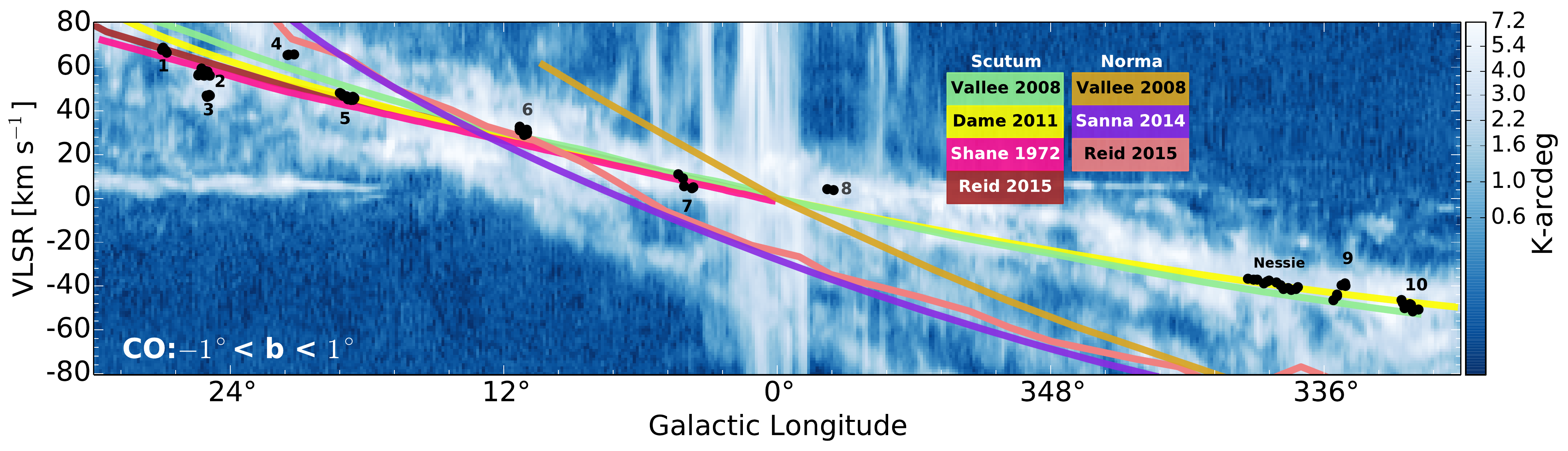} 
\caption{Position-velocity summary of bone candidates and spiral arm models. Blue background shows $^{12}\rm{CO}$ emission from \citet{Dame_2001}, integrated between $-1^\circ < b < 1^\circ$. Black dots show measurements of BGPS-, HOPS-, MALT90-, and GRS-determined velocities, with particular candidate filaments identified by number (see Table \ref{tab:candidates} for further identification), or, in the case of Nessie, by name. Lines of varying color show predicted \textit{p-v} spiral arm traces from the literature (see text for references).}
\label{fig:skeleton}
\end{center}
\end{figure*}

By overlaying the HOPS-, MALT90-, BGPS-, and GRS-determined velocities on a \textit{p-v} diagram of CO emission, we establish whether these filaments are associated with an existing spiral arm trace. For this study, we first use the whole-galaxy \citet{Dame_2001} CO survey to roughly locate each of the arms in \textit{p-p-v} space. We then overplot spiral fits to CO and HI for various spiral arm models, to determine whether candidates are consistent with previously claimed spiral arm traces (Figure \ref{fig:skeleton}). Of the approximately fifteen candidates identified visually, ten of these candidates are within $\approx$ 10 km s$^{-1}$ of the Scutum-Centaurus and Norma arms. We show these ten candidates on a \textit{p-v} diagram in Figure \ref{fig:skeleton}. In addition to showing our bone candidates, we show several different predictions of the positions of the Scutum-Centaurus and Norma arms toward the inner Galaxy in \textit{p-v} space, from \citet{Dame_2011}, \citet{Sanna_2014}, \citet{Shane_1972}, and \citet{Vallee_2008}. The \citet{Shane_1972} fit should be taken with reservation at low longitudes, as the HI observations terminated at $l=22^\circ$; the fit has been extrapolated to $l=0^\circ$ by \citet{Sato_2014}, under the assumption that it must pass through the origin in the absence of non-circular motion. We also include Scutum-Centaurus and Norma-4kpc fits from M. Reid \& T. Dame (2015, in preparation), derived from trigonometric parallax measurements of high-mass star forming regions taken as part of the BeSSeL survey \citep{Reid_2014}. M. Reid \& T. Dame (2015, in preparation) produce fits with ($l,b,v$) loci that follow Giant Molecular Clouds that trace the arms, producing a rough log-spiral approximation determined by trigonometric parallax rather than an assumed Galactic rotation curve. 

We emphasize that there is a significant amount of discrepancy between the various log-spiral models, particularly between the \citet{Sanna_2014} or M. Reid \& T. Dame (2015, in preparation) fit to the Norma arm and the \citet{Vallee_2008} fit to the same arm. Compared to the former models, the \citet{Vallee_2008} fit is inconsistent with CO observations and does not account for expanding motion fixed at  $l=0^\circ=-29.3\; \rm km \;s^{-1}$, obtained from CO absorption spectra towards the Galactic center\citep{Sanna_2014}. In general, we also caution that the log-spiral fits should be used as rough guides to delineate major spiral features and that there is little evidence that the Milky Way actually follows such a clean mathematical model. Scientists have difficulty fitting log-spirals to nearby face-on spiral galaxies such as M31, a fact that underlines the monumental challenge of inferring a similar model for the Milky Way while embedded in the Galactic disk \citep[cf.][]{Carraro_2015}. As a result, the unreliabiliy of the various log-spiral models was taken into account when establishing the bone criteria, outlined in \S2.3 below; in particular, we note the leniency of criterion 4, which permits the filaments to lie as much as 10 km s$^{-1}$ from the global-fit to \textit{any} Milky Way arm and still qualify as a bone.

\subsection{Establishing ``Bone" Criteria}
After narrowing down our list to ten filaments with kinematic structure consistent with existing spiral arm models, we develop a set of criteria for an object to be called a ``bone":

\begin{enumerate}
\item{Largely continuous mid-infrared extinction feature}
\item{Parallel to the Galactic plane, to within $30^\circ$}
\item{Within 20 pc of the physical Galactic mid-plane, assuming a flat galaxy}
\item{Within 10 km s$^{-1}$ of the global-log spiral fit to any Milky Way arm}
\item{No abrupt shifts in velocity (of more than ~3 km s$^{-1}$ per 10 pc) within extinction feature}
\item{Projected aspect ratio $\ge$ 50:1} 
\end{enumerate}

The names and coordinates for the ten bone-candidate filaments, along with their average LSR velocities, the number of bone criteria they satisfy, and a ``quality rating" are listed in Table \ref{tab:candidates}. In Table \ref{tab:properties}, we summarize physical parameters for all ten bone candidates, including estimates of distance, length, radius, volume, mass, and aspect ratio. We calculate mass by \textit{estimating} an average H$_2$ column density of $2 \times 10^{22}$ cm$^{-2}$, consistent with the minimum IRDC peak column density to be included in the \citet{Peretto_2009} catalog of 11,303 IRDCs. We calculate distances assuming all of our bone candidates (see Figure \ref{fig:skeleton}) are associated with the \citet{Dame_2011} Scutum-Centaurus arm. When available, we also provide distance measurements (independent of any presumed association with a spiral arm model) from the \citet{Ellsworth_Bowers_2015} catalog, which cites distances to 1710 molecular clouds from the BGPS survey, derived using a Bayesian distance probability density function. 

\begin{deluxetable*}{ccccc}
\setlength{\tabcolsep}{0pt}
\tabletypesize{\scriptsize}
\tablecolumns{5}
\tablewidth{0pt}
\tablehead{\colhead{1} & \colhead{2} & \colhead{3} & \colhead{4} & \colhead{5}\\\colhead{Official Name}& \colhead{Referenced Name} & \colhead{Average VLSR (km s$^{-1}$)} & \colhead{Criteria Satisfied} & \colhead{Quality Rating}}
\startdata
BC\_026.94-0.30 & Filament 1 & 68 & All &  A\\ 
BC\_025.24-0.45 & Filament 2 & 57 & All & A\\
BC\_024.95-0.17 & Filament 3 & 47 & 1,2,3,5 & C\\
BC\_021.25-0.15 & Filament 4 & 66 & 1,2,3,4,5 &C\\
BC\_018.88-0.09 & Filament 5 & 46 &All & A\\
BC\_011.13-0.12 & Filament 6 & 31 & 1,2,3,4,5 & B\\
BC\_004.14-0.02 & Filament 7 & 8 & All & B\\
BC\_357.62-0.33 & Filament 8 & 4 & 1,2,3,4,5 & B\\
BC\_335.31-0.29 & Filament 9 & -42 & All & B\\
BC\_332.21-0.04 & Filament 10 & -49 & All & B\\
\enddata
\tablecomments{(1) Central Galactic coordinates for our filaments, prefixed with ``BC" (bone candidate) and ordered by Galactic longitude. (2) Name by which each bone candidate is referred to throughout this paper. (3) Average VLSR of the bone candidate, computed by averaging the velocities of the sources for each filament seen in Figure \ref{fig:skeleton}. (4)  Number of bone criteria satisfied (see Section 2.3). (5) We assign a quality rating to each bone candidate dictated by how strongly they satisfy the bone criteria; a score of ``A" is given if the candidate strongly or moderately satisfies all bone criteria. A score of ``B" is given if the candidate strongly or moderately satisfies five criteria, but weakly satisfies (or fails to satisfy) one criterion. A score of ``C" is given if the candidate strongly or moderately satisfies some criteria, but weakly satisfies (or fails to satisfy) two or more criteria.}
\label{tab:candidates}
\end{deluxetable*}

\begin{deluxetable*}{ccccccccccc}
\setlength{\tabcolsep}{5pt} 
\tabletypesize{\scriptsize}
\tablecolumns{13}
\tablewidth{0pt}
\tablehead{\colhead{1} & \colhead{2} & \colhead{3} & \colhead{4} & \colhead{5} & \colhead{6} & \colhead{7} & \colhead{8} & \colhead{9} & \colhead{10}\\ \colhead{Official} & \colhead{Referenced} & \colhead{Scutum/Bayesian} & \colhead{Length} & \colhead{Radius} & \colhead{Length} & \colhead{Radius} & \colhead{Volume} & \colhead{Mass} & \colhead{Aspect} \\ \colhead{Name} & \colhead{Name} & \colhead{Distance} & \colhead{} & \colhead{} & \colhead{} & \colhead{} & \colhead{} & \colhead{} & \colhead{Ratio} \\ \colhead{ } & \colhead{} & \colhead{kpc} & \colhead{$^\circ$} & \colhead{$^\circ$} & \colhead{pc} & \colhead{pc} & \colhead{cc} & \colhead{$\rm{M}_{\rm{suns}}$} & \colhead{ }}
\startdata
BC\_026.94-0.30 & Filament 1 & 4.6/$4.00\substack{+0.48 \\ -0.52}$ & 0.16 & 0.0015 & 13/11 & 0.12/0.10 & 1.6E+55/1.0E+55 & 1.7E+03/1.3E+3 & 53 \\  
BC\_025.24-0.45 & Filament 2 & 4.3/$3.54\substack{+0.38 \\ -0.42}$& 0.63 & 0.0025 & 47/39 & 0.19/0.15 & 1.4E+56/7.9E+55  & 9.8E+03/6.6E+3 & 126 \\  
BC\_024.95-0.17 & Filament 3 & 4.3/$3.00\substack{+0.40 \\ -0.44}$& 0.18 & 0.0025 & 14/9 & 0.19/0.13 & 4.0E+55/1.4E+55 & 2.8E+03/1.4E+3 & 36 \\ 
BC\_021.25-0.15 & Filament 4 & 3.9/$4.02\substack{+0.30 \\ -0.34}$ & 0.20 & 0.0025 & 14/14 & 0.17/0.18 & 3.3E+55/3.7E+55 & 2.5E+03/2.7E+3 & 40 \\ 
BC\_018.88-0.09 & Filament 5 & 3.7/$3.36\substack{+0.50 \\ -0.58}$ &  0.70 & 0.0025 & 45/41 & 0.16/0.15 & 1.0E+56/7.5E+55 & 8.0E+03/6.6E+3 & 140 \\ 
BC\_011.13-0.12 & Filament 6 & 3.3/$3.24\substack{+0.60 \\ -0.76}$ &  0.38 & 0.0075 & 22/21 & 0.43/0.42 & 3.5E+56/3.3E+56  & 1.0E+04/1.0E+4 & 25 \\ 
BC\_004.14-0.02 & Filament 7 & 3.1/-- & 0.69 & 0.005 & 37/-- & 0.27/-- & 2.3E+56/-- &  1.1E+04/-- & 69 \\ 
BC\_357.62-0.33 & Filament 8 & 3.0/-- & 0.40 & 0.005 & 21/-- & 0.26/-- & 1.2E+56/-- &  6.0E+03/-- & 40 \\ 
BC\_335.31-0.29 & Filament 9 & 3.2/-- &  0.60 & 0.005 & 34/-- & 0.28/-- & 2.2E+56/-- &  1.0E+04/-- & 60 \\ 
BC\_332.21-0.04 & Filament 10 & 3.3/-- & 0.90 & 0.005 & 52/-- & 0.29/-- & 3.6E+56/-- & 1.6E+04/-- & 90 \\ 
\enddata
\tablecomments{Comparison of the physical properties of the bone candidates, based on a similar table from \citet{Goodman_2014}. We \textit{estimate} an average H$_2$ column density of $2 \times 10^{22} \rm{cm}^{-2}$, consistent with the minimum peak column density to be included in the \citet{Peretto_2009} catalog of 11,303 IRDCs; this corresponds to an assumed equivalent Av. of 20 magnitudes. (1) Central Galactic coordinates for our filaments, prefixed with ``BC" (bone candidate) and ordered by Galactic longitude. (2) Name by which each bone candidate is referred to throughout this paper. (3) We cite two distance calculations; the first assumes association with the Scutum-Centaurus arm as fitted by \citet{Dame_2011}; the second, when available, lists distances from \citet{Ellsworth_Bowers_2015}, which were derived by applying a Bayesian distance probability density function to 1710 molecular clouds from the BGPS survey. (6) Length in pc is calculated first using the Scutum-Centaurus arm distance and then using the Bayesian distance. (7) Radius in pc is calculated first using the Scutum-Centaurus arm distance and then using the Bayesian distance. (8) We assume the filaments are cylindrically shaped and calculate the volume based on measured radius and length, first using the Scutum-Centaurus arm distance and then the Bayesian distance. (10) Aspect ratio does not account for projection effects. }
\label{tab:properties}
\end{deluxetable*}

\begin{figure*}
\begin{center}
\epsscale{1.0}
\plotone{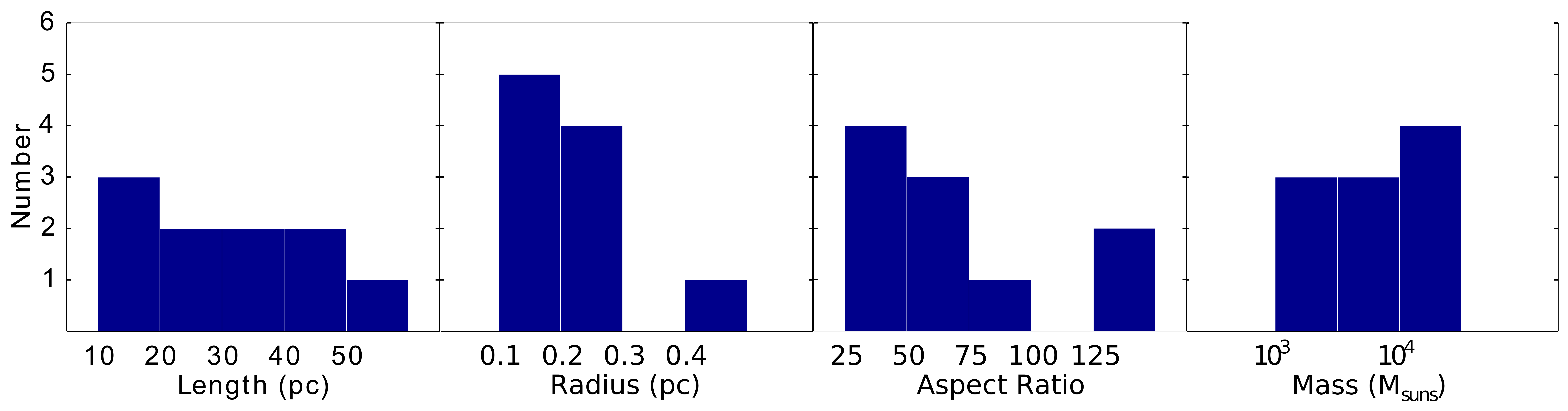} 
\caption{Distributions of length, radius, aspect ratio, and mass for the ten bone candidates, based on data from Table \ref{tab:properties}.}
\label{fig:histograms}
\end{center}
\end{figure*}

In Figure \ref{fig:histograms} we show histograms of the distributions of length, radius, aspect ratio, and mass (derived assuming association with the Scutum-Centaurus arm) for the ten bone candidates. The lengths of the candidates range from 13-52 pc, with three candidates having a length less than 20 pc. The other seven candidates are fairly uniformly distributed beyond a length of 20 pc. The radii of the candidates range from 0.12-0.43 pc, with a majority (nine candidates) having a radius below 0.3 pc; the notable exception is filament 6 (``the snake"), which has a radius of 0.43 pc. The distribution of aspect ratios is skewed towards lower values, with four candidates possessing an aspect ratio less than 50:1 and another three between 50:1-75:1. Finally, six of the candidates have a mass between $10^3$ $\rm{M}_{\rm{suns}}-10^4$ $\rm{M}_{\rm{suns}} $, with four candidates having a mass at or slightly above $10^4$ $\rm{M}_{\rm{suns}}$.

\subsection{Description of Bone Candidates}

Of the ten filaments with velocities consistent with Galactic rotation, \textbf{six} of these meet all six bone criteria: \textbf{filament 1 (``BC\_026.94-0.30"), filament 2 (``BC\_025.24-0.45"), filament 5 (``BC\_018.88-0.09"), filament 7 (``BC\_004.14-0.02"), filament 9 (``BC\_335.31-0.29"), and filament 10 (``BC\_332.21-0.04")}, to varying degrees of excellence. We note that filament 10 has likely been disrupted by stellar feedback, making its aspect ratio and velocity structure more difficult to define. Since we predict that all Galactic bones will likely be destroyed by stellar feedback and/or Galactic shear, we include it here as part of a larger attempt to build a catalog of bones at all stages of their evolution. We also include filament 9, even though it has a \textit{p-v} orientation perpendicular to predicted fits of the Scutum-Centaurus arm (see Figure \ref{fig:skeleton}). As spurs and inter-arm structures are likely to lie close to the physical Galactic mid-plane, but with velocity gradients angled with respect to predicted arm fits, we do {\it not} require that a bone be parallel to arm \textit{p-v} traces, so as not to exclude potential spurs, feathers, or other inter-arm features.

Of the four remaining filaments that do not meet all six criteria---filament 8 (``BC\_357.62-0.33"), filament 6 (``BC\_011.13-0.12"), filament 3 (``BC\_024.96-0.17"), and filament 4 (``BC\_021.25-0.15")---all of them fail criterion 6 (aspect ratio $\ge$ 50:1).  As our criterion 6 does not allow for projection effects in imposing an aspect ratio limit, we emphasize that those filaments lying more tangential to our line-of-sight will appear foreshortened, and could very well meet the 50:1 minimum limit if projection effects were removed. The first of these candidates, filament 8, shows particular promise, lying within 2-3 pc of the physical Galactic mid-plane and tracing a prominent peak of CO emission in both \textit{p-v} and \textit{p-p} space (see Figures \ref{fig:candid8pv} and \ref{fig:candid8pp} in the appendix).  The second filament, filament 6 (``the snake"), has already been well-studied from a star formation perspective, hosting over a dozen protostellar cores likely to produce regions of high-mass star formation \citep{Wang_2014,Henning_2010}. From a Galactic bone perspective, the snake strongly satisfies all criteria except number 6---it lies within 15 pc of the physical Galactic mid-plane and 5 km s$^{-1}$ from the \citet{Dame_2011} Scutum-Centaurus global-log fit to CO, also tracing a prominent peak of CO emission in \textit{p-v} space (see Figure \ref{fig:snake_pv} in the appendix). The remaining two filaments, filaments 3 and 4, are both awarded a quality rating of ``C." Filament 3 lies slightly more than 10 km s$^{-1}$ from the \citet{Shane_1972} fit to HI for the Scutum-Centaurus arm (just above the upper limit of criterion 4) while filament 4 appears to be a potential interarm filament, lying between the Scutum-Centaurus and Norma-4kpc arms in \textit{p-v} space. We also note a small break in the extinction feature of filament 4, though the filament has been confirmed to be contiguous in velocity space as traced by $^{13}\rm{CO}$ gas from the GRS Survey.  

In summary, it is important to emphasize that some of the above criteria will likely be modified in the long run, as we learn more about spiral structure. Given our limited \textit{a priori} knowledge of the Galaxy's structure, it is presently easier to confirm bones that are spine-like, lying along arms with velocities predicted by extant modeling (e.g. filament 5), and harder to find spurs off those arms or inter-arm features (e.g. filament 9), the velocities of which are hard to predict well. Similarly, since criterion 6 does not allow for projection effects in imposing an aspect ratio limit, bones which otherwise meet all criteria could fail the aspect ratio test if they lie close to the tangents of spiral arms. This task is, of course, hindered by the fact that there is little consensus regarding the number and placement of major spiral arms in our Galaxy, let alone more nuanced structure like interarm spurs or feathers, so any declaration regarding these structures would be premature. As we learn more about spiral structure from simulations, modeling, and new observational data, these criteria will be adjusted to allow for the detection of finer bone-like features that could potentially represent spurs, inter-arm structures, and/or foreshortened structures lying close to our line of sight.

\section{Analysis of New Bones}
\begin{figure}
\begin{center}
\epsscale{1.0}
\plotone{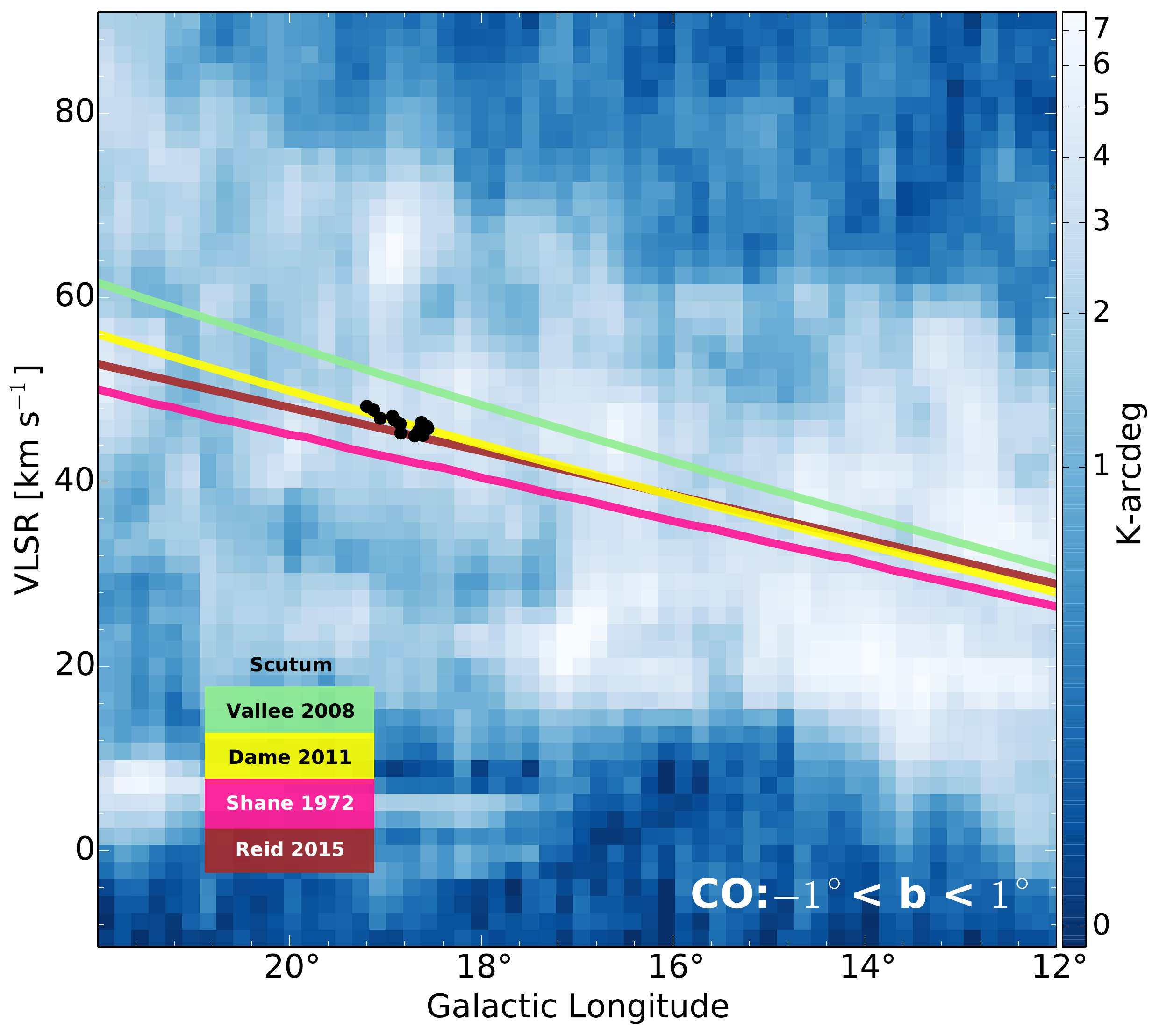} 
\caption{\label{fig:Candid5_pos_vel} Position-velocity diagram of $\textrm{CO}$,  $\textrm{NH}_3$, and $\textrm{HCO}^+$ emission for filament 5. Blue background shows $^{12}$CO (1-0) emission integrated between $-1^\circ < \textrm{b} < 1 ^\circ$ \citep{Dame_2001}. Black dots show HOPS (NH$_3$ emission), BGPS (HCO$^+$ emission) and GRS (high resolution $^{13}\rm{CO}$ emission) sources associated with filament 5. The colored lines show fits to the Scutum-Centaurus arm (see text for references).%
}
\end{center}
\end{figure}

\begin{figure*}
\begin{center}
\plotone{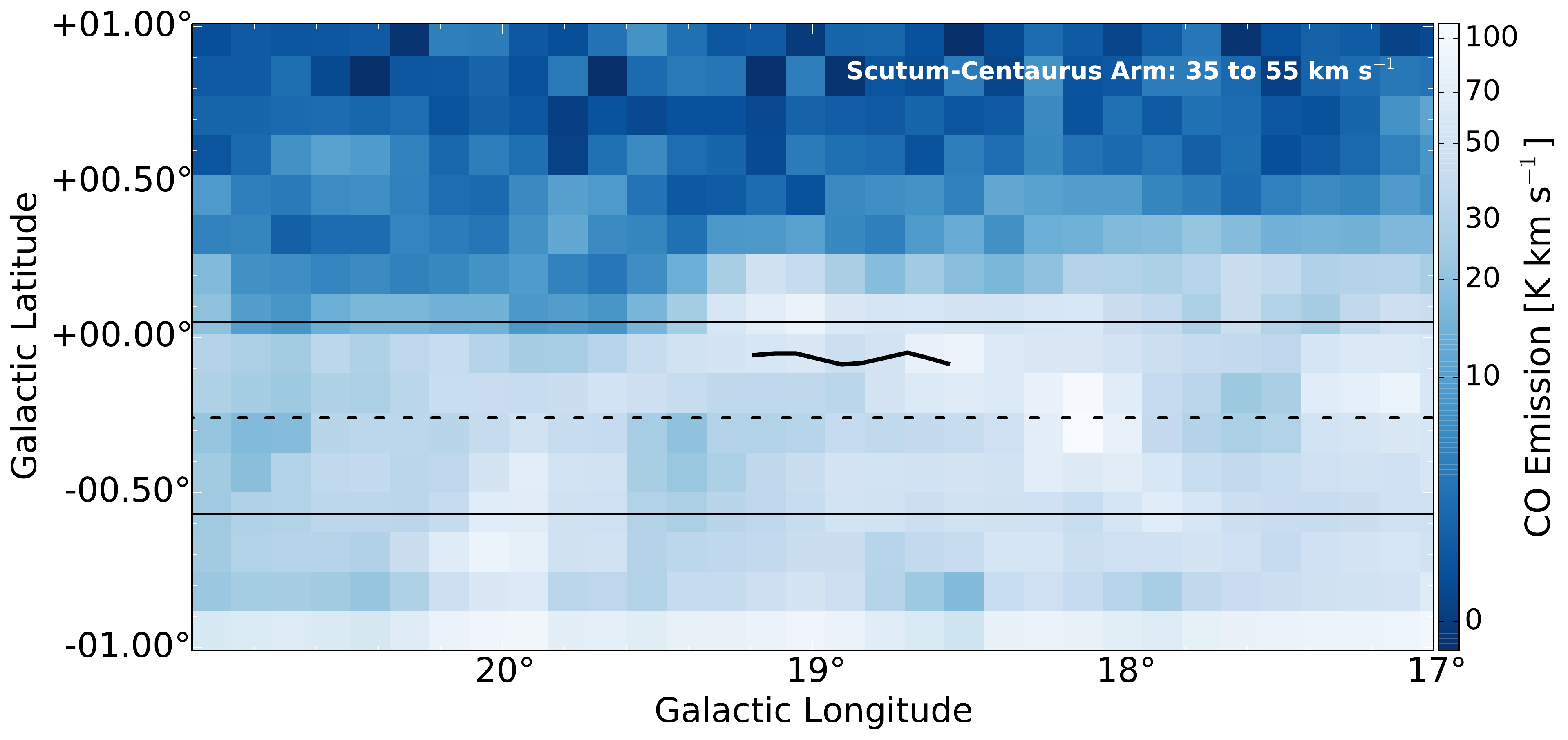} 
\caption{\label{fig:Candid5_pos_pos} Plane of the sky map integrated between 35 and 55 km s$^{-1}$, the approximate velocity range of the Scutum-Centaurus arm in the region around filament 5. A trace of filament 5, as it would appear as a mid-IR extinction feature, is superimposed on the $^{12}\rm{CO}$ emission map \citep{Dame_2001}. The black dashed line indicates the location of the physical Galactic mid-plane, while the solid black lines indicate $\pm$ 20 pc from the Galactic mid-plane at the 3.7 kpc distance to filament 5, assuming the candidate is associated with the \citet{Dame_2011} Scutum-Centaurus model.%
}
\end{center}
\end{figure*}

\begin{figure*}
\begin{center}
\plotone{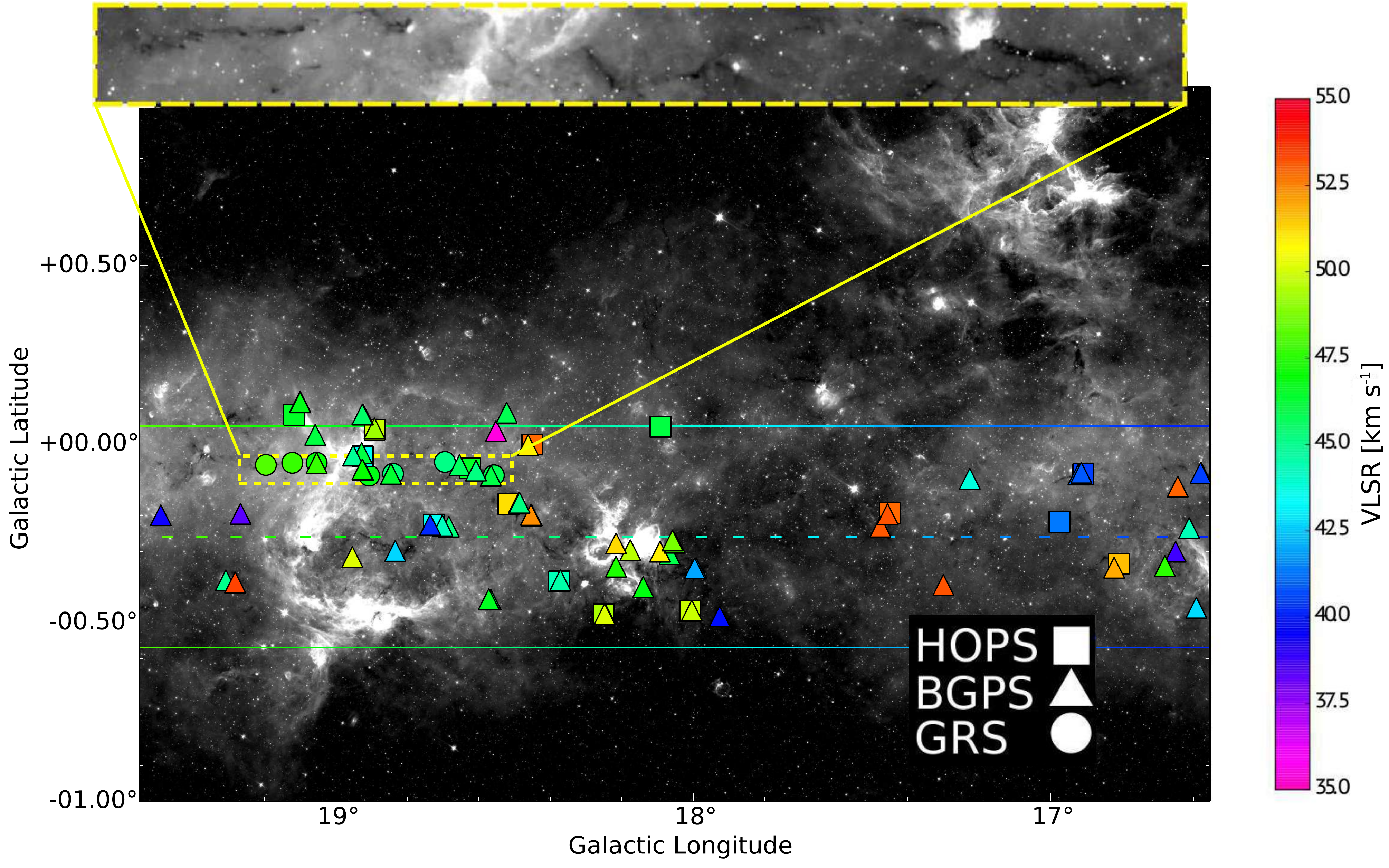} 
\caption{\label{fig:Candid5_with_tilt} Filament 5 lies within $\approx$ 15 pc of the physical Galactic mid-plane. The background is a GLIMPSE-\textit{Spitzer} 8 $\mu\textrm{m}$ image. The dashed line is color-coded by \citet{Dame_2011} LSR velocity and indicates the location of the physical Galactic mid-plane. The solid colored lines indicate $\pm$ 20 pc from the Galactic mid-plane at the 3.7 kpc distance to filament 5, assuming the candidate is associated with the \citet{Dame_2011} Scutum-Centaurus model. The squares, triangles, and circles correspond to HOPS, BGPS, and GRS sources, respectively.  A closer look at filament 5 can be seen in the inset.%
}
\end{center}
\end{figure*}

Filament 5 is our strongest bone candidate, in that it is highly elongated ($0.7^\circ$  or 45 pc, with an aspect ratio of 140:1) and \textit{exactly} (within 1-2 km s$^{-1}$) along a previously-claimed spiral arm trace in \textit{p-p-v} space, although its orientation makes it somewhat less elongated than Nessie on the sky. In Figure \ref{fig:Candid5_pos_vel} we show a \textit{p-v} diagram in the longitude range of filament 5 and overlay fits to the Scutum-Centaurus arm from \citet{Shane_1972}, \citet{Vallee_2008}, \citet{Dame_2011}, and M. Reid \& T. Dame (2015, in preparation). We see that the HOPS-, BGPS-, and GRS-determined velocities associated with filament 5 are highly correlated with the \citet{Dame_2011} and the M. Reid \& T. Dame (2015, in preparation) global-log fits to CO and HI, suggesting that filament 5 is marking a ``spine" of the Scutum-Centaurus arm in this longitude range. Moreover, filament 5 also lies along a CO peak in longitude-latitude space, as evident in Figure \ref{fig:Candid5_pos_pos}. By overlaying a trace of the mid-IR extinction feature of filament 5 on a plane of the sky map (integrated in Scutum-Centaurus's velocity range in the region around filament 5) we see that filament 5 lies in the center of the most intense CO emission, which suggests that it may be a spine of Scutum-Centaurus as traced by lower density CO gas. Finally, Figure \ref{fig:Candid5_with_tilt} shows that filament 5 lies within $\approx$ 15 pc of the true physical mid-plane, with both the Sun's 25 pc elevation above the IAU mid-plane and the small tilt ($+0.12^\circ$) of the plane caused by the offset of the Galactic Center from the IAU (0,0) having been accounted for in this view. All these lines of evidence taken together indicate that filament 5 (``BC\_018.88-0.09") is Nessie's counterpart in the first quadrant, suggesting that Nessie is not a curiosity, but one of several bones that trace significant spiral features. 

Our study is not the first follow-up to the Nessie work \citep{Goodman_2014} to look for more long filaments associated with spiral structure. \citet{Ragan_2014} and \citet{Wang_2015} have undertaken similar studies. However, this is the first study to specifically look for bones in regions we are most likely to find them, that is, elongated along the Galactic plane. Moreover, our study offers a set of criteria capable of defining this new class of objects (i.e. Galactic ``bones"). 

\begin{figure*}
\begin{center}
\plotone{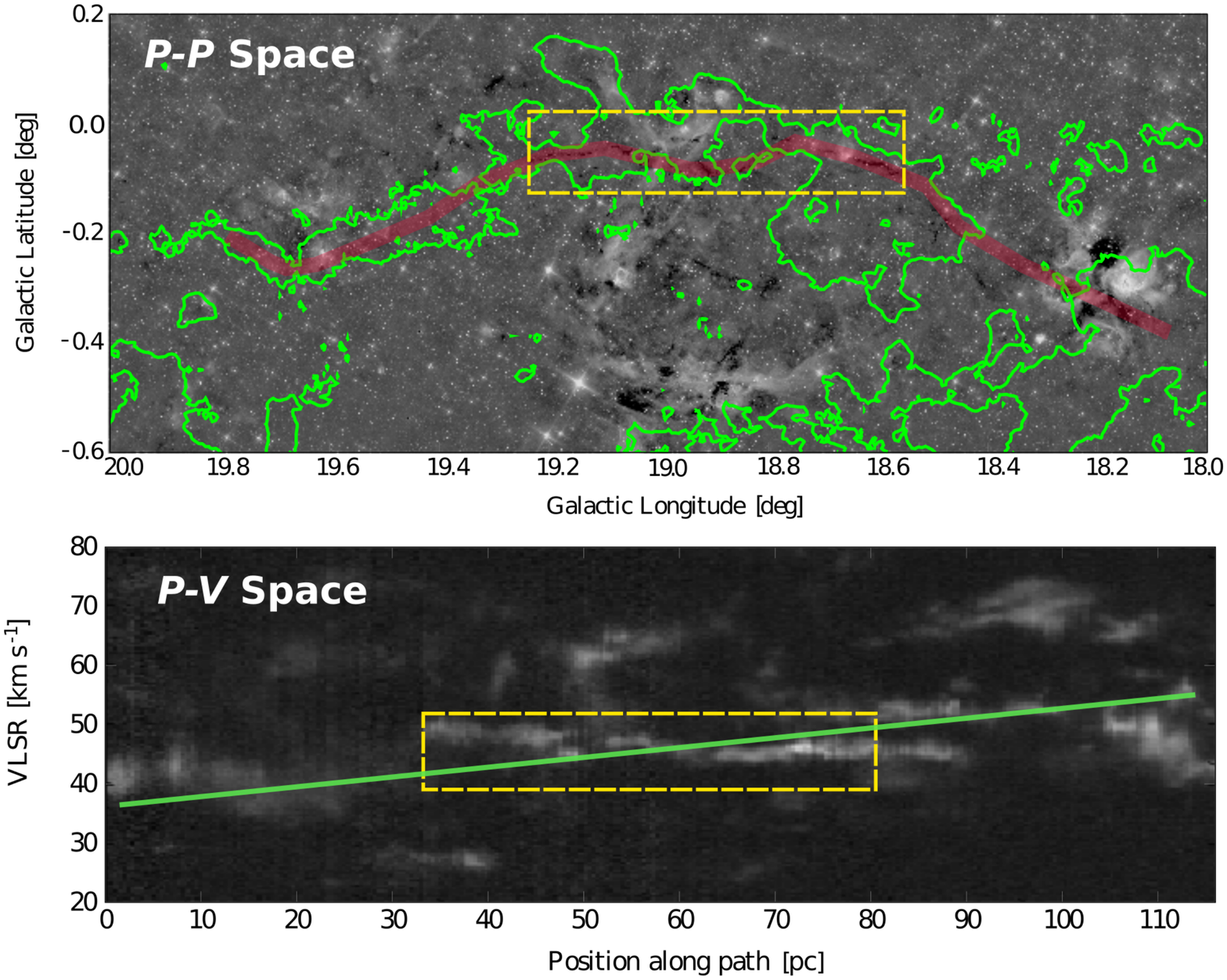} 
\caption{\label{fig:ragan_comp} \textit{Top:} Position-position analysis of filament 5 as it compares to the larger feature, GMF20.0-17.9 \citep{Ragan_2014}, of which filament 5 is a subset. In green, we overlay the GRS $^{13}\rm{CO}$ integrated intensity contours which define GMF20.0-17.9, and we box the region corresponding to filament 5 in yellow. In red, we show a path that connects the \citet{Ragan_2014} IRDCs and traces filament 5.   \textit{Bottom:} Position-velocity analysis of filament 5, as it compares to GMF20.0-17.9. We show the results of extracting a slice (from a $^{13}\rm{CO}$ GRS FITS cube) along the red path in the upper panel, which runs through the \citet{Ragan_2014} IRDCs and our filament 5. As seen inside the yellow boxed region in the lower panel, the section of the path that corresponds to filament 5 is remarkably kinematically contiguous, with velocities ranging between 45 and 49 km s$^{-1}$. In contrast, \citet{Ragan_2014} group the 37 km s$^{-1}$ emission at x=0 pc with the 50 km s$^{-1}$ emission at x=115 pc and connect these two points with a straight line on a longitude-velocity diagram (green line in lower \textit{p-v} panel).}
\end{center}
\end{figure*}

\citet{Ragan_2014} undertook a blind search ({\it not} restricted to latitudes where the mid-plane should lie) for long thin filaments ($>1^\circ$) in the first quadrant of the Milky Way, using near and mid-infrared images. In addition to confirming that Nessie lies along the Scutum-Centaurus arm, \citet{Ragan_2014} find seven Giant Molecular Filaments (GMFs) of which only one, GMF 20.0-17.9, is said to be associated with Galactic structure (declared a spur of the Scutum-Centaurus arm). Our strongest bone candidate, filament 5, is a subsection of GMF 20.0-17.9, but, unlike \citet{Ragan_2014}, we argue that filament 5 runs right down the spine of the Scutum-Centaurus arm in \textit{p-v} space. We believe the discrepancy arises due to a difference in methodology. \citet{Ragan_2014} group neighboring IRDCs into a single filament, despite breaks in the extinction feature and kinks in velocity structure. Since grouping several IRDCs to make a longer structure violates our criteria 1 and 5, we only consider the continuous and kinematically coherent part of the filament, which is remarkably parallel to the Scutum-Centaurus arm in \textit{p-v} space. Likewise, in Figure 4 from \citet{Ragan_2014} (analogous to our Figure \ref{fig:skeleton}), they represent filaments as straight lines connecting velocities measured at the tips of the filaments while we represent filaments as sets of points whose velocities are determined by the BGPS, HOPS, MALT90, and GRS surveys. We compare our \textit{p-p} and \textit{p-v} analysis of filament 5 with the analysis from \citet{Ragan_2014} in Figure \ref{fig:ragan_comp}.

\citet{Ragan_2014} find little or no association of their GMFs with Galactic structure, and they suggest that observations are perhaps not as sensitive to spiral arm filaments in the first quadrant, or that the observed frequency and orientation of spiral arm filaments in the first quadrant is different than the fourth. Our three bone candidates with an ``A" quality rating (filaments 1,2, and 5) all lie in the first quadrant, so we speculate that Galactic bones are not subject to the same fourth quadrant bias that GMFs are potentially prone to. We also emphasize that GMFs are part of a broader class of objects and that bones are expected to be high-contrast subsections of particular GMFs that align with spiral structure. The lengths of the GMFs range from 60-230 pc, while our longest bone candidates is only 52 pc. By definition, GMFs are meant to be larger structures composed of several smaller, high-contrast elements, so no bone in itself will realistically be classified as a GMF. As is the case with filament 5, we expect that there will be significant overlap between the GMF and bone catalogs in the future, as our Galactic bones should be a subset of any spiral tracing GMFs yet to be discovered. 

Like \citet{Ragan_2014}, \citet{Wang_2015} search for large-scale filaments and establish their relationship to Galactic structure after the fact. Rather than searching for filaments elongated along the Galactic plane, \citet{Wang_2015} search for the longest, coldest, and densest filaments (aspect ratio $>>10$) in Hi-GAL images, within the longitude range of $15^\circ < l < 56^\circ$. Filaments were initially identified using Hi-GAL 350 and 500 $\mu\rm{m}$ emission. Temperature and column density maps were created for each candidate, and those which exhibited systematically lower temperatures with respect to the background were selected. As in our study, \citet{Wang_2015} confirm velocity contiguity by extracting a \textit{p-v} slice along the curvature of each filament. 

\citet{Wang_2015} highlight nine filaments as their most prominent, with one of the nine being Nessie.  Only one of their filaments (their ``G11", our filament 6) overlaps with our sample, and is not classified as a bone due to its short aspect ratio ($\approx$ 25:1).  Seven other \citet{Wang_2015} filaments fail one or more of our bone criteria. G24, G26, and G47 lie between 39-62 pc above the physical Galactic mid-plane (outside our $\pm 20$ pc criterion), while G28 and G64 have aspect ratios of around 38:1 and 19:1 (less than our 50:1 minimum aspect ratio criterion). Additionally, G29 and G49 are not largely continuous mid-infrared extinction features, violating our criterion 1. This is not surprising, as the \citet{Wang_2015} study was designed to identify filaments emitting at longer Hi-GAL wavelengths, which are not necessarily seen continuously in absorption at mid-infrared wavelengths. In future studies, \citet{Wang_2015} plan to extend their search to the entire Galactic plane. Despite differences in methodology, there should also be some degree of overlap between the \citet{Wang_2015} catalog and our bones catalog.

An in-depth analysis of the other nine bone candidates can be found in the appendix.

\section{Discussion}
Our initial search was intentionally limited: we specifically searched the Galaxy for the most prominent bones, near previously-modeled major spiral features, in a narrow longitude range ($|l| < 62^{\circ}$). As we expand our longitude range, bones' visual association with spiral structure could become more evident as the spiral models themselves span a greater range in velocity. Through this more comprehensive search, there are potentially hundreds of bone-like filaments discoverable in the Milky Way. If we can find enough bones, it should be possible to piece them together to delineate the major structural features of our Galaxy, using other relevant information as well (from maser measurements, 3D extinction mapping, and kinematic constraints on both gas and stars). Astronomers have been trying to accurately model the location of the spiral arms in \textit{p-p-v} space for decades. A bones-based approach might be able to resolve some of the discrepancies amongst the many arm models shown in Figure \ref{fig:skeleton}. The level of disagreement on arm locations is large enough that finding even just a handful of Galactic bones marking sections of the spines of spiral arms could tie arm fits down with high fidelity at particular positions in \textit{p-p-v} space. These ``spinal" anchors could have especially large weights in statistical fits that seek to combine many measures of the Milky Way's skeletal structure. Moreover, even filaments that are not currently confirmed bone candidates might provide insight to Galactic structure in the future. Filament 3, which has been awarded a grade of ``C", is just over 10 km s$^{-1}$ from any \textit{p-v} fit to the Scutum-Centaurus arm, failing criterion 4. Despite this, it traces the CO emission in its longitude range better than existing fits from \citet{Vallee_2008} and \citet{Dame_2011}, which both intersect the bottom of a CO hole at $l\approx25^\circ$ (see Figure \ref{fig:skeleton} and the appendix section on filament 3). Thus, one should expect this process to be iterative, and the criteria should be continually refined as more bones are found and new lines of evidence become available. 

Along with increasing the known bone population, it should be possible to improve simulations in hopes of answering key questions about bones' origin and evolution.  For instance, synthetic observations of simulations could shed light on what fraction of highly-elongated dense clouds appear to be: a) aligned with arms; b) spur-like; c) inter-arm; or d) random long thin clouds unaligned with Galactic structure. And, the simulations could pin down the likely origins of these types of objects, in part by predicting different velocity, density, or mass profiles for objects with different origins.  

Along the same vein, It is important to note that not all long skinny filaments are expected to be associated with Galactic structure. Studies prior to \citet{Ragan_2014} offer at least two examples of long molecular clouds that are not obviously bone-like. The ``Massive Molecular Filament" G32.02+0.06, studied by \citet{Battersby_2014}, does not appear to be tracing an arm structure. Likewise, the 500-pc long molecular ``wisp" discussed by \citet{Li_2013} also does not presently appear directly related to Galactic structure. Neither of these two clouds currently lies in any special position in \textit{p-p-v} space. It is possible that these are bone remnants, disrupted by feedback or Galactic shear, but, without better Galaxy modeling, it is very hard to speculate on what fractions of long thin clouds were formerly bones, are currently bones, or were never bones. 

While the \citet{Smith_2014} Galaxy models are the first that provide high enough resolution to simulate incredibly long and thin bones, they do not include stellar feedback or magnetic fields---either of which could cause disruptions in the appearance of the currently-simulated bone-like features. In the future, it should be possible to utilize more comprehensive, targeted high-resolution synthetic observations (e.g. of dust absorption and emission and of CO spectra), based on high-resolution simulations like the ones in \citet{Smith_2014}. Finally, it would be prudent to use these simulations to estimate the biases inherent in our selection criteria (how many spurious ``bones" should one expect to find randomly, by the chance alignment of discontinuous IRDC peaks?). 

Though promising, we caution that our results are preliminary and that significant advancements must be made within the larger field of Galactic structure before we can definitively confirm association (or lack thereof) with spiral features. As a brief example, we cite the current debate over the nature and placement of the Local arm, as discussed in \citet{Carraro_2015}. For instance, \citet{Xu_2013} suggest that the Local arm might be consistent with a grand design spiral feature: they use trigonometric parallax measurements of water masers to determine the pitch angle of the feature. They find that the Local arm's small pitch angle is inconsistent with being a short spur; this, combined with its active star formation and relatively long length ($>$ 5 kpc) provides support that it is similar to the nearby Perseus or Sagittarius arms and could possibly be an independent spiral section. In contrast, \citet{Vazquez_2008} provide evidence that the Local arm is a spur or bridge and not a major spiral feature. \citet{Vazquez_2008} combine optical and radio observations of young open clusters and molecular clouds to trace spiral structure in the third Galactic quadrant, with observations strongly indicating that the Perseus arm is being bifurcated by the Local arm in this quadrant. The fact that no agreement can be reached even within our solar neighborhood underlies the importance of continued mapping of both bones and spiral features before any definitive claims can be made.

While challenging, in the long run, it could be feasible to combine future high resolution synthetic observations with a wealth of existing data sets to build a \textit{skeletal model of the Milky Way}. When used in conjunction with BeSSeL maser-based rotation curves \citep{Reid_2014}, CO \citep{Dame_2001} and HI \citep{Shane_1972} \textit{p-v} fitting, 3D extinction mapping \citep{Schlafly_2014b}, HII region arm mapping \citep{Anderson_2012}, observations of embedded clusters, young open clusters, short-period Classical Cepheids, and star-complexes \citep{Majaess_2009, Russeil_2003, Vazquez_2008, Carraro_2011, Camargo_2015}, and GAIA results, bones have the potential to not only clarify Galactic structure at unprecedented resolution, but also to resolve dated questions related to Galactic structure that have been plaguing astronomers for decades. 

\section{Conclusion}
A search for Galactic bones has been undertaken in the region $|l|<62^\circ, |b|<1^\circ$, using large-scale \textit{\textit{Spitzer}} GLIMPSE/MIPSGAL images of the Galactic plane in combination with radial velocity measurements of high and low density gas. The search has produced six filaments which meet all established bone criteria: these filaments are all 1) largely continuous mid-infrared extinction features that lie 2) parallel to and 3) within 20 pc of the physical Galactic mid-plane (assuming a flat galaxy); they also lie within 4) 10 km s$^{-1}$ of a pre-existing spiral arm trace in \textit{p-v} space, in addition to being 5) contiguous in velocity space and 6) possessing an aspect ratio of at least 50:1. Several other candidates fail the minimum aspect ratio bone criterion, and could be reclassified if the aspect ratio test is redefined to allow for projection effects. 

The strongest candidate, filament 5 (``BC\_18.88-0.09") runs remarkably parallel to the physical Galactic mid-plane and lies just 15 pc above that plane.  It also exhibits remarkable velocity contiguity and runs \textit{exactly} (within 1-2 km s$^{-1}$) along the \citet{Dame_2011} fit to the Scutum-Centaurus arm in \textit{p-v} space. Filament 5 also possesses an aspect ratio of at least 140:1---many times greater than that of a typical Giant Molecular Cloud. The evidence we present in this paper suggests that filament 5 and our other classified Galactic bones might mark the location of significant spiral features and could be used to pin down the accuracy of spiral arm models to within one pc in regions near bones. Follow-up work (i.e. extinction mapping; obtaining higher resolution spectra with a suite of dense gas tracers) is needed in order to characterize the properties of the bone sample with greater precision. Ultimately, if we can reliably identify hundreds of Milky Way bones, it should be possible to combine the ``Skeleton" suggested by bones with other tracers of Galactic structure, in order to piece together a much better view of the Milky Way's structure than we have now. 

\section{Acknowledments}
This paper greatly benefited from the knowledge and support of several collaborators. Tom Dame and Mark Reid provided invaluable expertise on Galactic structure and spiral arm traces in \textit{p-p} and \textit{p-v} space. Chris Beaumont and Tom Robitaille provided much helpful advice on Glue and their co-development of the Astropy Spectral-Cube package saved the author countless hours reducing and analyzing \textit{p-p-v} cubes. Sarah Ragan generously provided her Python scripts for extracting and plotting the Vall{\'{e}}e spiral arm models in \textit{p-v} space.  Finally, Alberto Pepe, Nathan Jenkins, and Deyan Ginev provided guidance on the use of Authorea. We are also grateful to ApJ Scientific Editor, Butler Burton, and to an anonymous referee, for providing helpful insights concerning measurements of Galactic structure.

This work is supported in part by the National Science Foundation REU and Department of Defense ASSURE programs under NSF Grant no. 1262851 and by the Smithsonian Institution.

This publication makes use of molecular line data from the Boston University-FCRAO Galactic Ring Survey (GRS). The GRS is a joint project of Boston University and Five College Radio Astronomy Observatory, funded by the National Science Foundation under grants AST-9800334, AST-0098562, and AST-0100793. This publication also makes use of data from the H2O Southern Galactic Plane Survey and the The Millimetre Astronomy Legacy Team 90 GHz Survey.

This work is based [in part] on observations made with the \textit{Spitzer} Space Telescope, which is operated by the Jet Propulsion Laboratory, California Institute of Technology under a contract with NASA.

The Glue software is developed by the Seamless Astronomy group at the Center for Astrophysics, under contract to NASA's James Webb Space Telescope program. WorldWide Telescope has been developed and supported by Microsoft Research, in part through gifts to the Seamless Astronomy program, and it is now an open-source resource available on GitHub\footnote{https://github.com/WorldWideTelescope}.

The first draft of this paper was written in Authorea\footnote{https://www.authorea.com/}, a collaborative, online, editor for research. 
\clearpage
\bibliographystyle{apj}
\bibliography{full_article_emulate}

\appendix
\subsection{Glue Demonstrations}
\begin{figure}[h!]
\begin{center}
\epsscale{.9}
\plotone{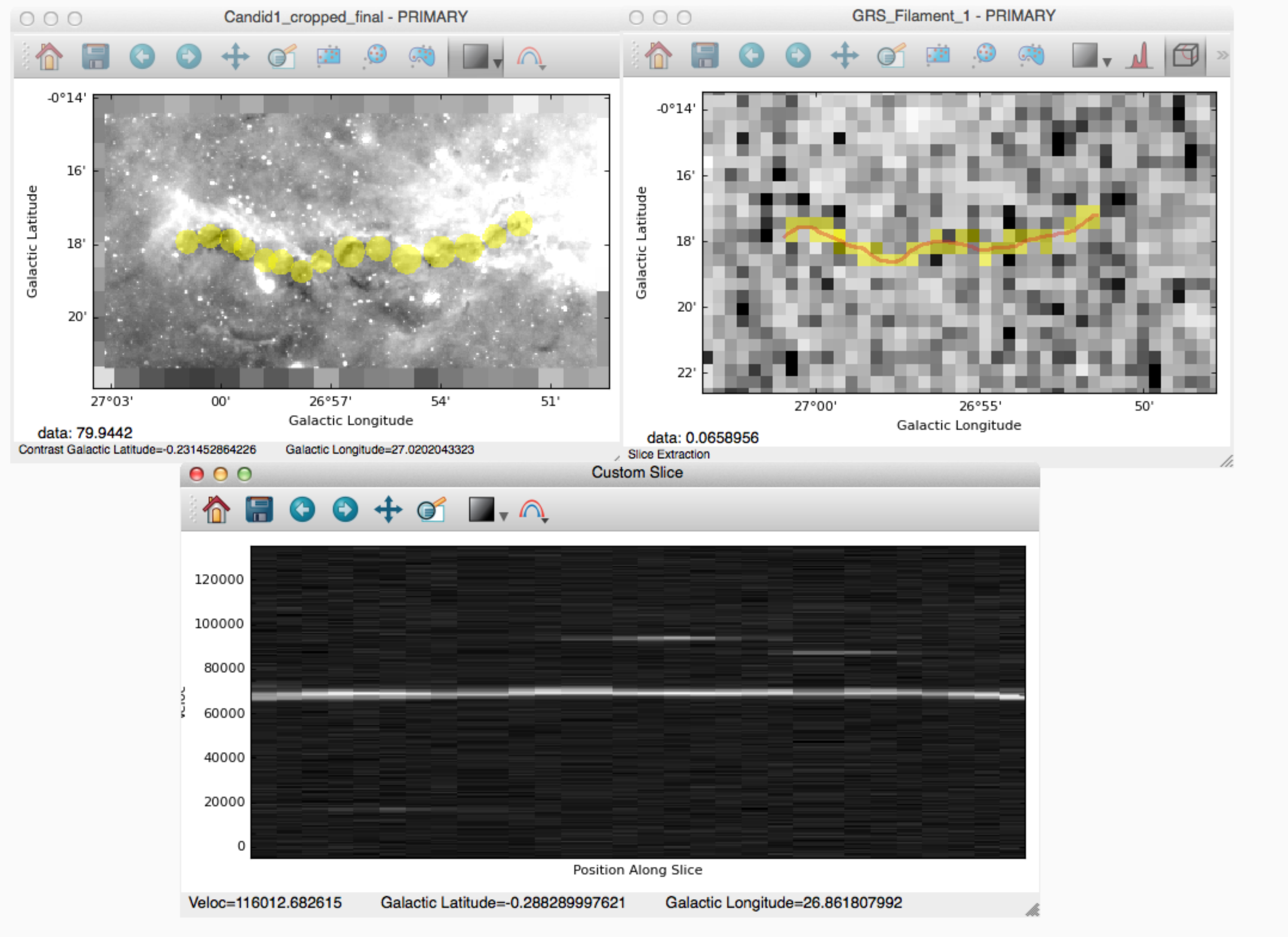} 
\caption{Extracting a \textit{p-v} slice along an extinction feature in Glue. The upper left image shows yellow circular regions overlaid on a \textit{\textit{Spitzer}}-GLIMPSE $8\micron$ image, coincident with the extinction feature. The upper right image shows a GRS $^{13}\rm{CO}$ FITS cube, with the same regions overlaid. The red path marks the curve along which the \textit{p-v} slice is extracted. The bottom image shows the results of the \textit{p-v} extraction along the red curve. 
}
\label{fig:demoI}
\end{center}
\end{figure}

In Figure \ref{fig:demoI} we demonstrate how to extract a \textit{p-v} slice along a path tracing the extinction feature, using the software visualization package Glue.  First, using the ``link data'' function, we link the Galactic latitude and longitude from the \textit{\textit{Spitzer} image} (upper left image of Figure \ref{fig:demoI}) with the Galactic latitude and longitude from the GRS or ThrUMMS FITS cubes (upper right image of Figure \ref{fig:demoI} ). Next, we create yellow circular regions along the extinction feature in the \textit{\textit{Spitzer}} image, marking the path along which the slice will be extracted. Since the data products are ``linked", Glue automatically overlays the same yellow regions on the GRS or ThrUMMS data cube (upper right image of Figure \ref{fig:demoI}). Then, using the ``slice extractor" tool, we trace a path through yellow circular regions on the CO FITS cube, effectively creating a customized \textit{p-v} slice along the extinction feature (bottom image of Figure \ref{fig:demoI}).

\begin{figure}[h!]
\begin{center}
\epsscale{.9}
\plotone{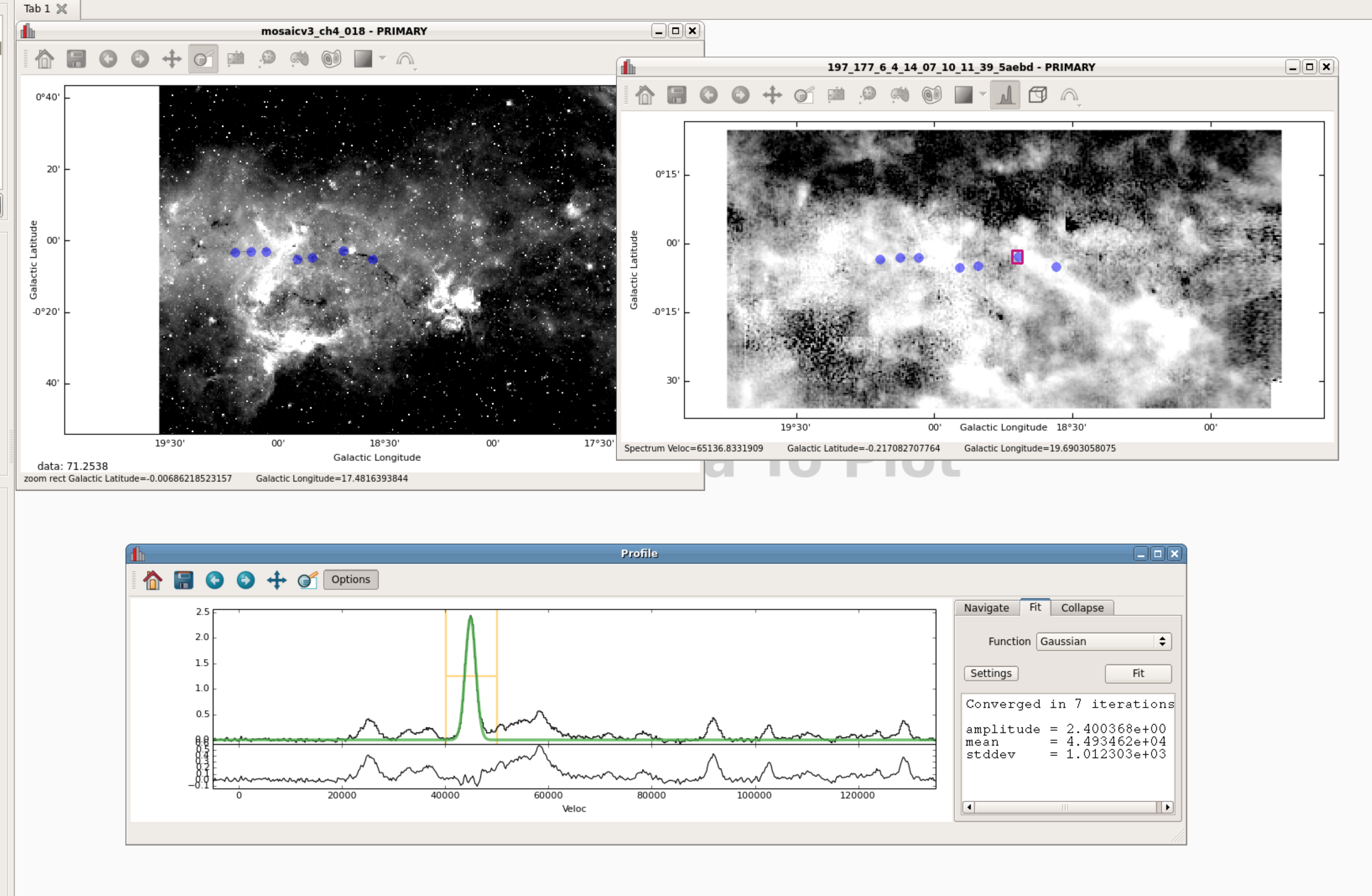} 
\caption{Extracting radial velocities at different regions along the extinction feature, using Glue. The upper left image shows blue circular regions selected along the extinction feature, overlaid on a \textit{\textit{Spitzer}}-GLIMPSE $8\micron$ image. The upper right image shows a $^{13}\rm{CO}$ FITS cube, with the same regions overlaid. The red boxed region in the upper right image indicates the area from which we extract a spectrum. The bottom image shows the spectrum extracted, with a Gaussian fitted around the highest peak in intensity}
\label{fig:demoII}
\end{center}
\end{figure}

In Figure \ref{fig:demoII} we demonstrate how to extract radial velocities at specific Galactic coordinates along the filament, using the software visualization package Glue. First, using the``link data" function, we once again link the Galactic latitude and longitude from the \textit{\textit{Spitzer}} image (upper left image of Figure \ref{fig:demoII}) with the Galactic latitude and longitude from the GRS or MALT90 FITS cubes (upper right image of Figure \ref{fig:demoII}). Then, we select points along the extinction feature (blues circles in upper left image of Figure \ref{fig:demoII}) which are then automatically overlaid on the $^{13}\rm{CO}$ (GRS) or $\textrm{N}_2\textrm{H}^+$ (MALT90) fits cube (blue points in upper right image of Figure \ref{fig:demoII}). We extract a spectrum (bottom image of Figure \ref{fig:demoII}) from the the small red boxed region around the the blue point (top right image of Figure \ref{fig:demoII}), using the ``spectrum extractor" function in Glue. Then, we fit a Gaussian around the highest peak in intensity, to determine a representative gas velocity.

\clearpage

\subsection{\large Filament 1 (``BC\_26.94-0.30"): Grade ``A"}
Filament 1 is a confirmed bone, meeting all six criteria with a quality grade of ``A." With a length of 13 pc and a radius of 0.12 pc, it is the shortest, thinnest, and least massive of our ten candidates, with an aspect ratio of $\approx$ 53:1. It runs \textit{exactly} (within 1-2 km s$^{-1}$) along the M. Reid \& T. Dame (2015, in preparation) \textit{p-v} fit to the Scutum-Centaurus arm and lies within 10 pc of the physical Galactic mid-plane. 

\begin{figure}[h!]
\textbf{Filament 1 (``BC\_26.94-0.30"): Grade ``A"}
\begin{center}
\plotone{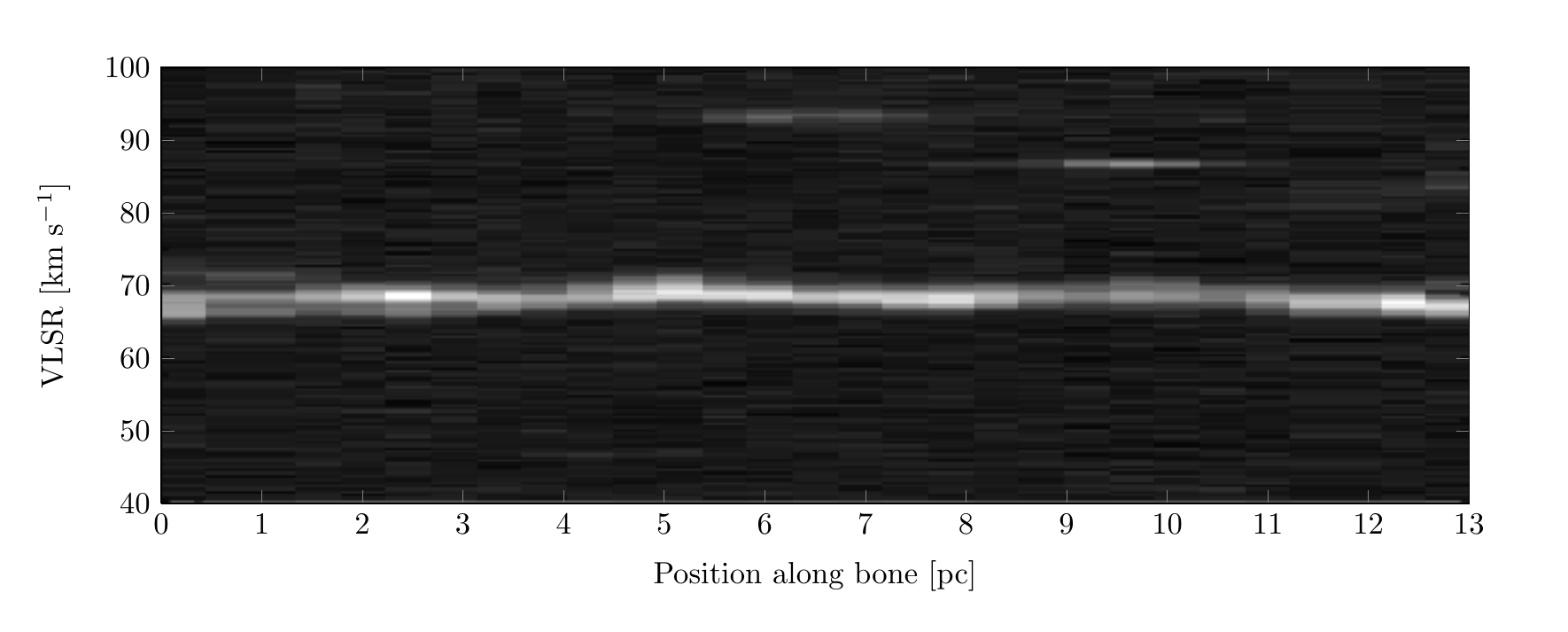}
\caption{Results of performing a slice extraction along the filamentary extinction feature of filament 1, using $^{13}\rm{CO}$ data from the GRS survey.}
\end{center}
\end{figure}

\begin{figure}[h!]
\textbf{Filament 1 (``BC\_26.94-0.30"): Grade ``A"}
\begin{center}
\epsscale{0.6}
\plotone{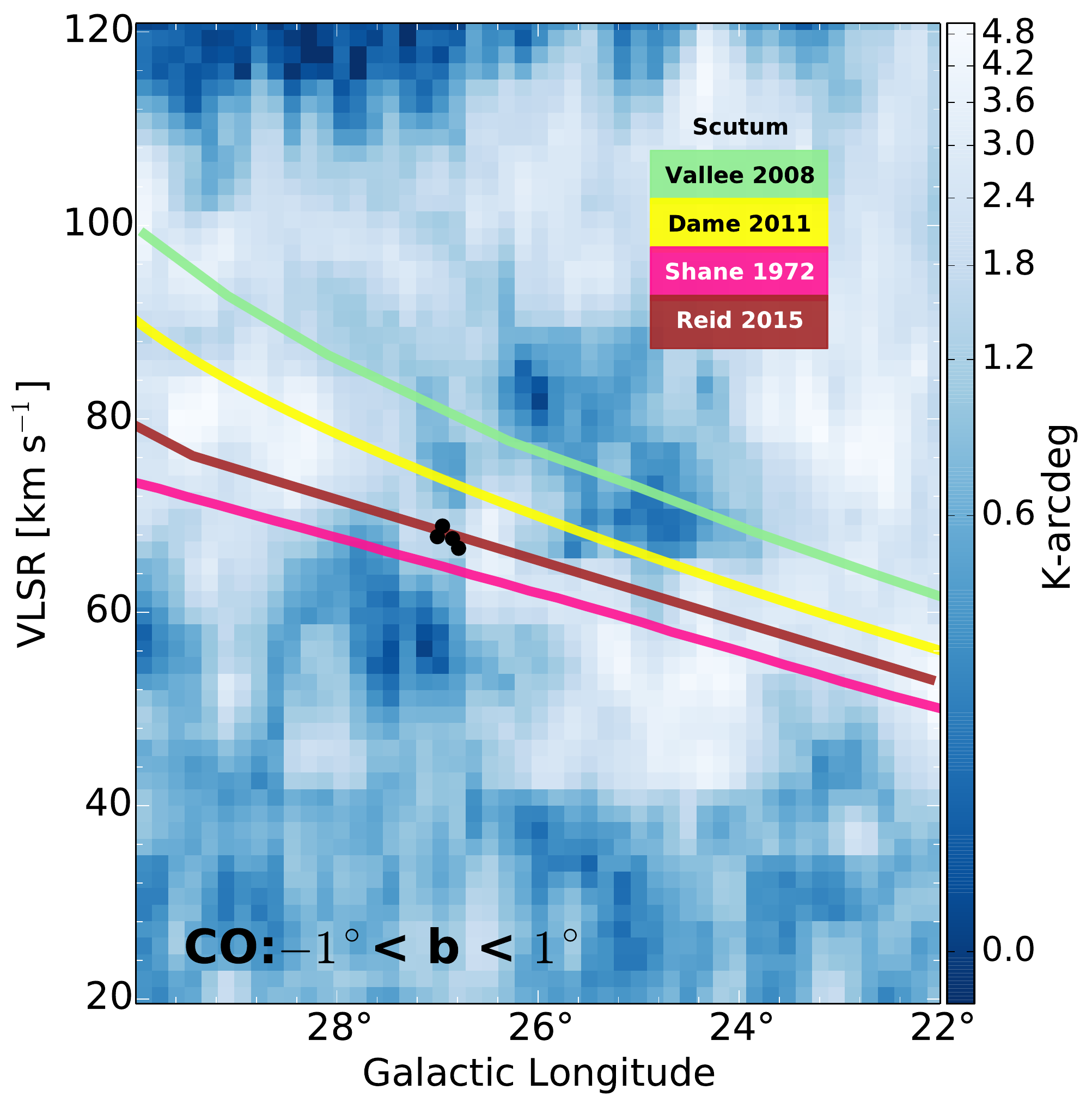}
\caption{Position-velocity diagram of CO and HCO$^+$ emission for filament 1. Blue background shows $^{12}$CO (1-0) emission integrated between $-1^\circ < \textrm{b} < 1^\circ$ \citep{Dame_2001}. Black dots show GRS and BGPS sources associated with filament 1. Colored lines show spiral fits from the literature for the Scutum-Centaurus arm (see text for references).}%
\end{center}
\end{figure}

\begin{figure}[h!]
\textbf{Filament 1 (``BC\_26.94-0.30"): Grade ``A"}
\begin{center}
\plotone{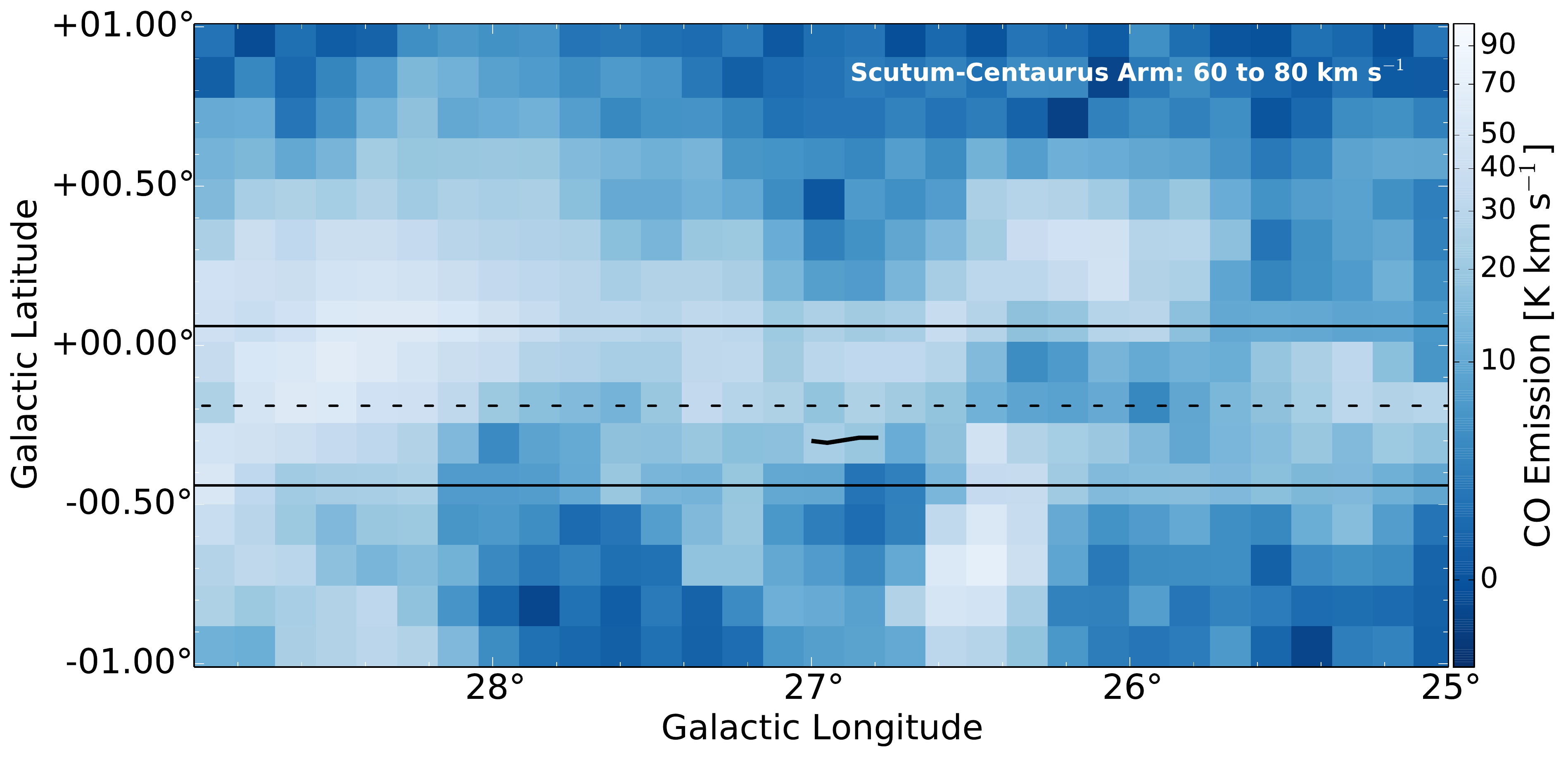}
\caption{Plane of the sky map integrated between 60 and 80 km s$^{-1}$, the approximate velocity range of the Scutum-Centaurus arm in the region around filament 1. A trace of filament 1, as it would appear as a mid-IR extinction feature, is superimposed on the $^{12}\rm{CO}$ emission map \citep{Dame_2001}. The black dashed line indicates the location of the physical Galactic mid-plane, while the solid black lines indicate $\pm$ 20 pc from the Galactic mid-plane at the 4.6 kpc distance to filament 1, assuming the candidate is associated with the \citet{Dame_2011} Scutum-Centaurus model.}
\end{center}
\end{figure}

\begin{figure}[h!]
\textbf{Filament 1 (``BC\_26.94-0.30"): Grade ``A"}
\begin{center}
\plotone{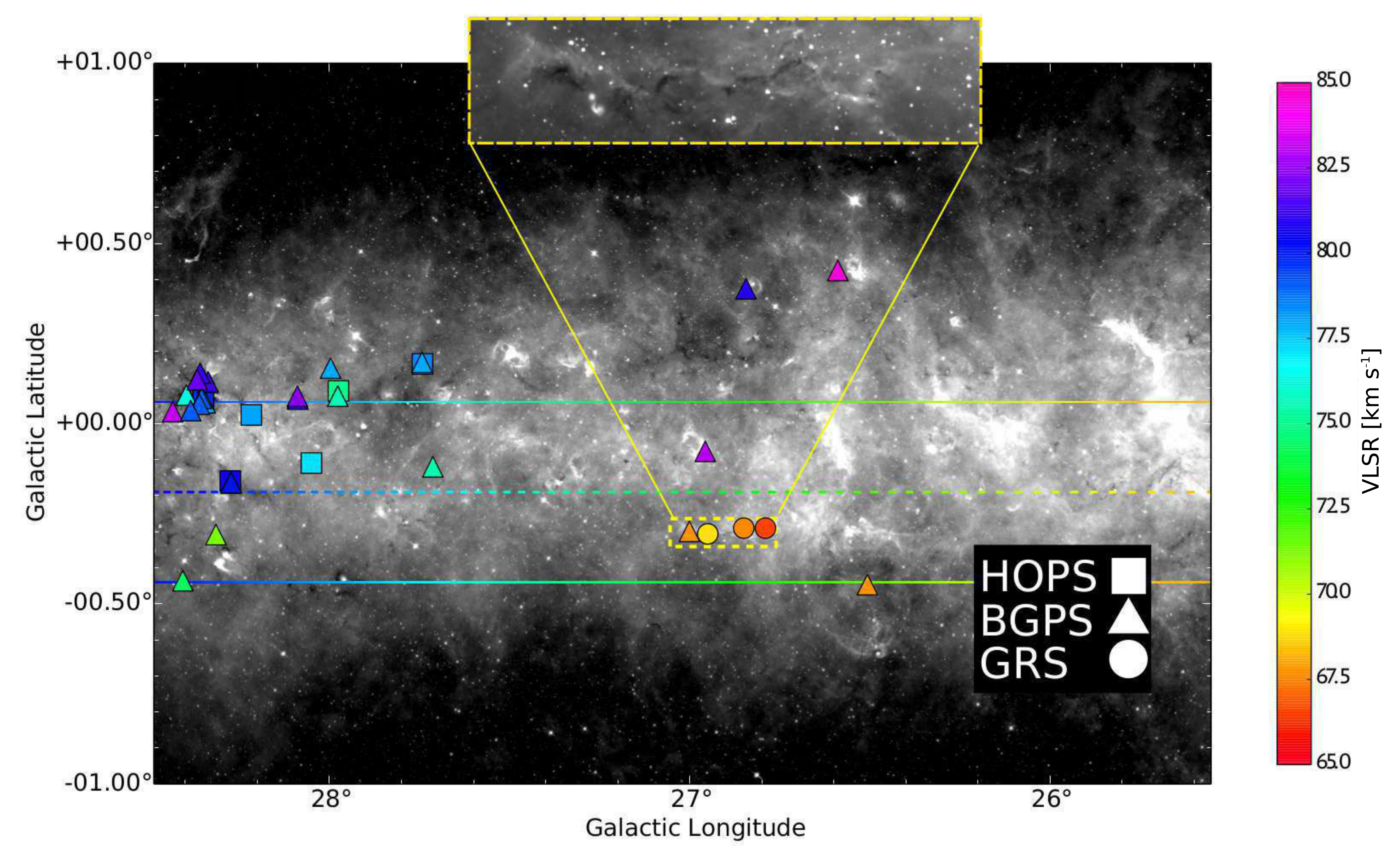}
\caption{Filament 1 lies within $\approx$ 10 pc of the physical Galactic mid-plane. The background is a GLIMPSE-\textit{Spitzer} 8 $\mu\textrm{m}$ image. The dashed line is color-coded by \citet{Dame_2011} LSR velocity and indicates the location of the physical Galactic mid-plane. The solid colored lines indicate $\pm$ 20 pc from the Galactic mid-plane at the 4.6 kpc distance to filament 1, assuming the candidate is associated with the \citet{Dame_2011} Scutum-Centaurus model. The squares, triangles, and circles correspond to HOPS, BGPS, and GRS sources, respectively.  A closer look at filament 1 can be seen in the inset.%
}
\end{center}
\end{figure}

\clearpage

\subsection{\large Filament 2 (``BC\_025.24-0.45"): Grade ``A"}
Filament 2 is a confirmed bone, meeting all six criteria with a quality grade of ``A." Behind filament 5 (the strongest bone), it has the second-longest aspect ratio of our ten candidates, at 126:1. It lies about 2-3 km s$^{-1}$ from the \citet{Shane_1972} \textit{p-v} fit to the Scutum-Centaurus arm, as well as 20 pc from the physical Galactic mid-plane. 

\begin{figure}[h!]
\textbf{Filament 2 (``BC\_025.24-0.45"): Grade ``A"}
\begin{center}
\plotone{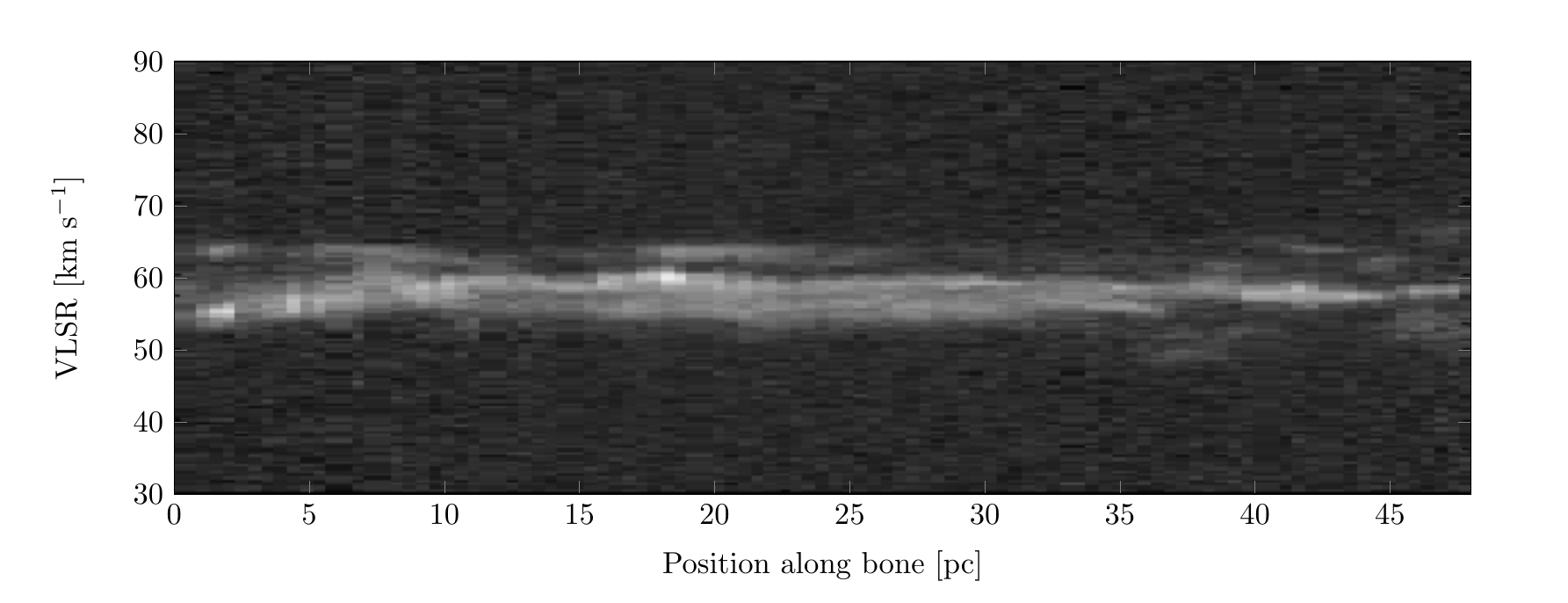}
\caption{The results of performing a slice extraction along the filamentary extinction feature of filament 2, using $^{13}\rm{CO}$ data from the GRS survey.}
\end{center}
\end{figure}

\begin{figure}[h!]
\textbf{Filament 2 (``BC\_025.24-0.45"): Grade ``A"}
\begin{center}
\epsscale{.6}
\plotone{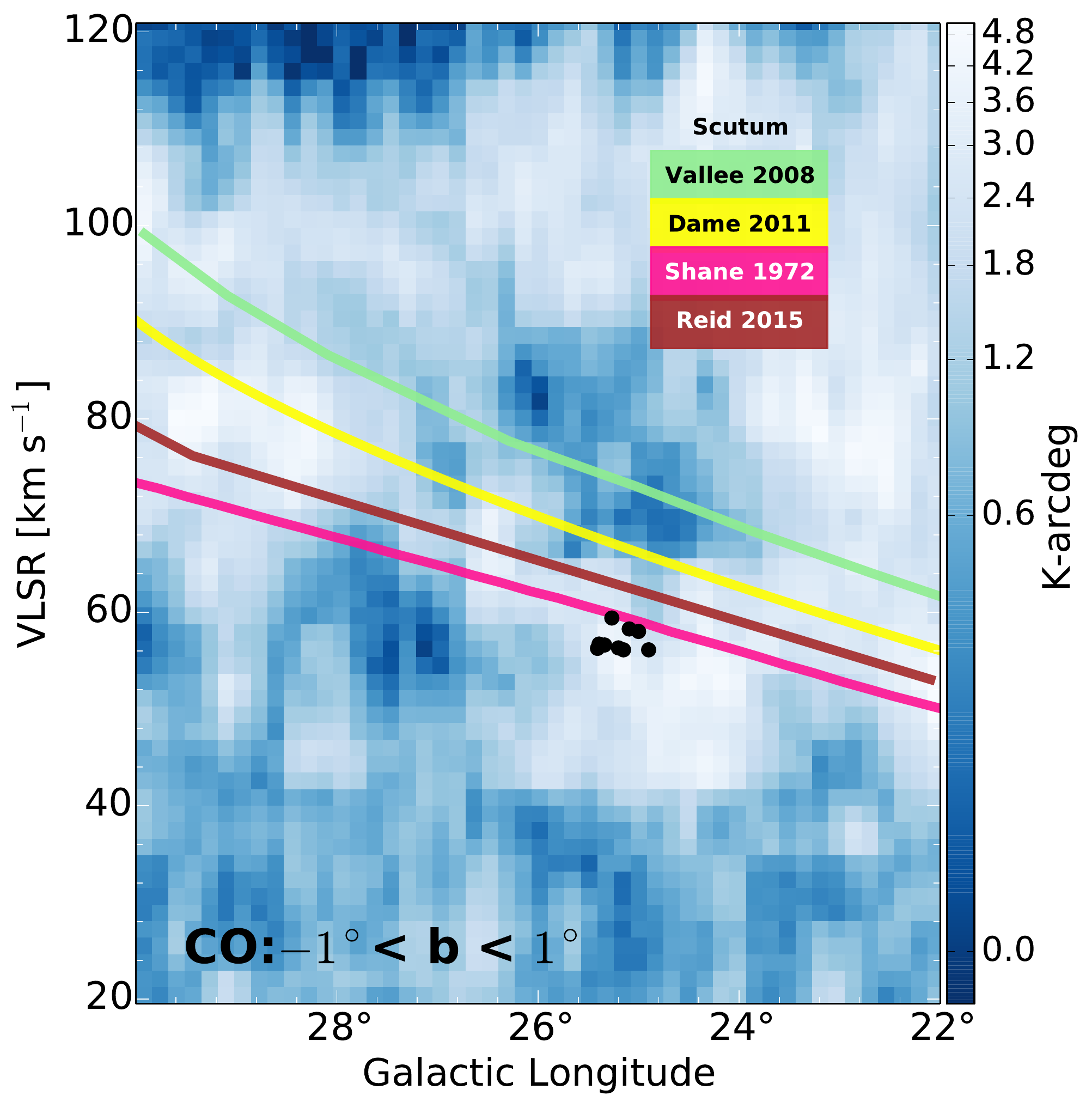}
\caption{Position-velocity diagram of CO, NH$_{3}$, and HCO$^+$ emission for filament 2. Blue background shows $^{12}$CO (1-0) emission integrated between $-1^\circ < \textrm{b} < 1^\circ$ \citep{Dame_2001}. Black dots show HOPS, BGPS, and GRS sources associated with filament 2. Colored lines show spiral fits from the literature for the Scutum-Centaurus arm (see text for references)%
}
\end{center}
\end{figure}

\begin{figure}[h!]
\textbf{Filament 2 (``BC\_025.24-0.45"): Grade ``A"}
\begin{center}
\plotone{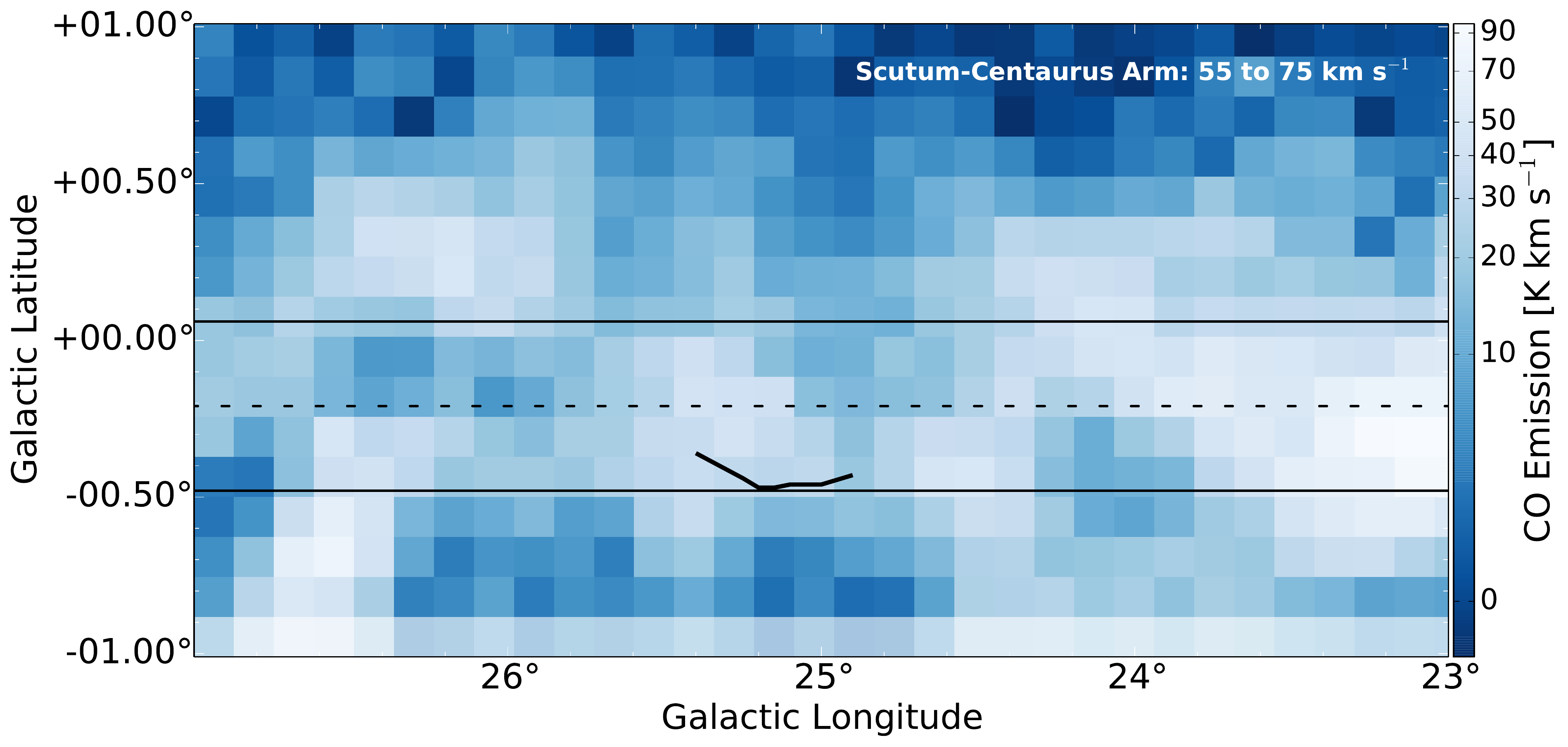}
\caption{Plane of the sky map integrated between 55 and 75 km s$^{-1}$, the approximate velocity range of the Scutum-Centaurus arm in the region around filament 2. A trace of filament 2, as it would appear as a mid-IR extinction feature, is superimposed on the $^{12}\rm{CO}$ emission map \citep{Dame_2001}. The black dashed line indicates the location of the physical Galactic mid-plane, while the solid black lines indicate $\pm$ 20 pc from the Galactic mid-plane at the 4.3 kpc distance to filament 2, assuming the candidate is associated with the \citet{Dame_2011} Scutum-Centaurus model.}
\end{center}
\end{figure}

\begin{figure}[h!]
\textbf{Filament 2 (``BC\_025.24-0.45"): Grade ``A"}
\begin{center}
\plotone{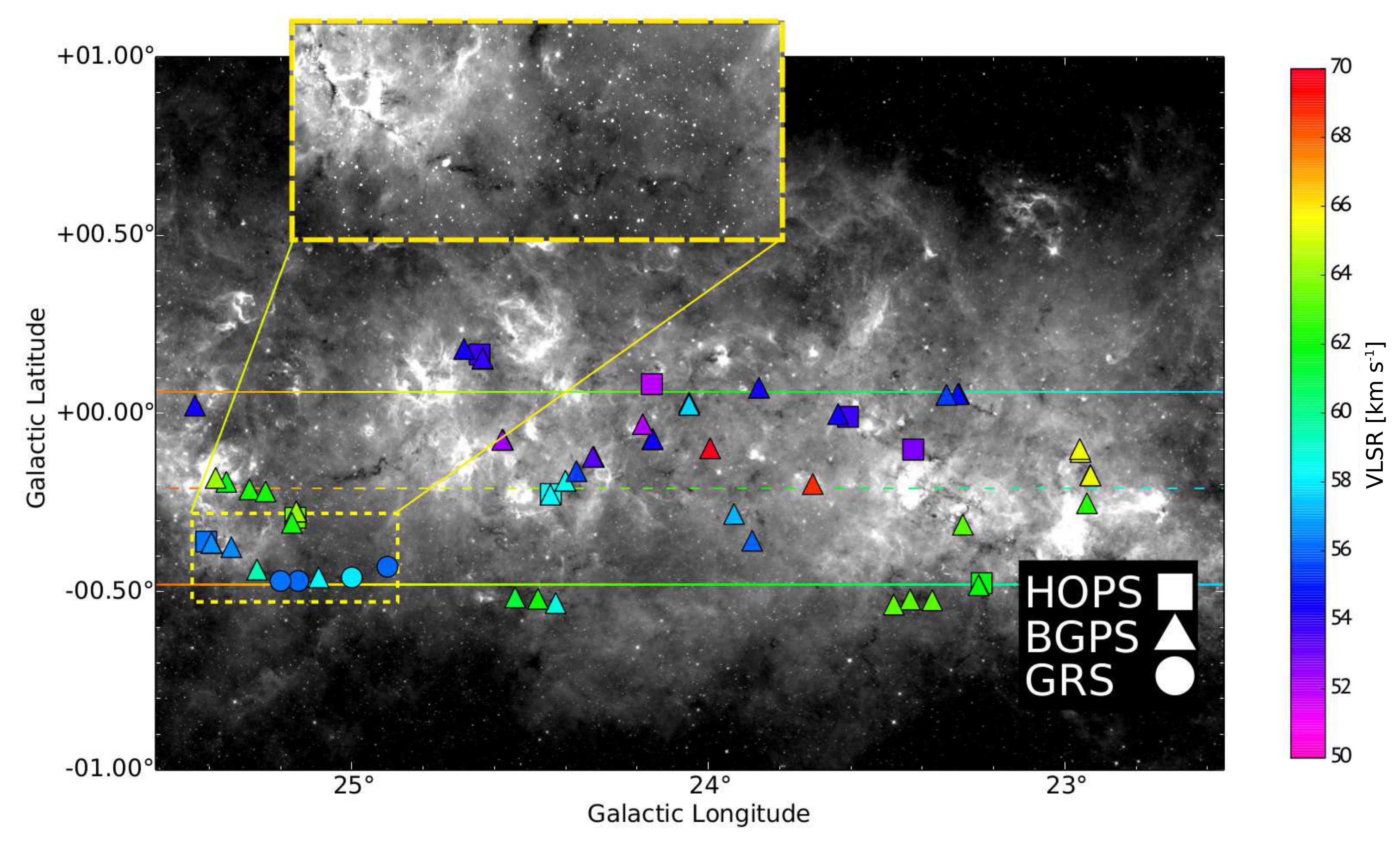}
\caption{Filament 2 lies within $\approx$ 20 pc of the physical Galactic mid-plane. The background is a GLIMPSE-\textit{Spitzer} 8 $\mu\textrm{m}$ image. The dashed line is color-coded by \citet{Dame_2011} LSR velocity and indicates the location of the physical Galactic mid-plane. The solid colored lines indicate $\pm$ 20 pc from the Galactic mid-plane at the 4.3 kpc distance to filament 2, assuming the candidate is associated with the \citet{Dame_2011} Scutum-Centaurus model. The squares, triangles, and circles correspond to HOPS, BGPS, and GRS sources, respectively.  A closer look at filament 2 can be seen in the inset.%
}
\end{center}
\end{figure}

\clearpage

\subsection{\large Filament 3 (``BC\_24.95-0.17"): Grade ``C"}
Filament 3 is not a confirmed bone, receiving a quality grade of ``C." With an aspect ratio of 36:1, it fails criterion 6 (minimum aspect ratio of 50:1). It also fails criterion 4 (within 10 km s$^{-1}$ of the \textit{p-v} fit to any Milky Way arm), lying slightly more than 10 km s$^{-1}$ below the \citet{Shane_1972} fit to HI for the Scutum-Centaurus arm. Despite lying beyond the upper limit of criterion 4, it strongly satisfies criterion 3, falling about 3 pc from the physical Galactic mid-plane. It also traces the CO emission well in \textit{p-v} space, better than the \citet{Vallee_2008} and \citet{Dame_2011} fits to the Scutum-Centaurus arm, which both intersect the bottom of a CO hole at $l\approx25^\circ$ (see Figure \ref{fig:BC24}).

\begin{figure}[h!]
\textbf{Filament 3 (``BC\_24.95-0.17"): Grade ``C"}
\begin{center}
\plotone{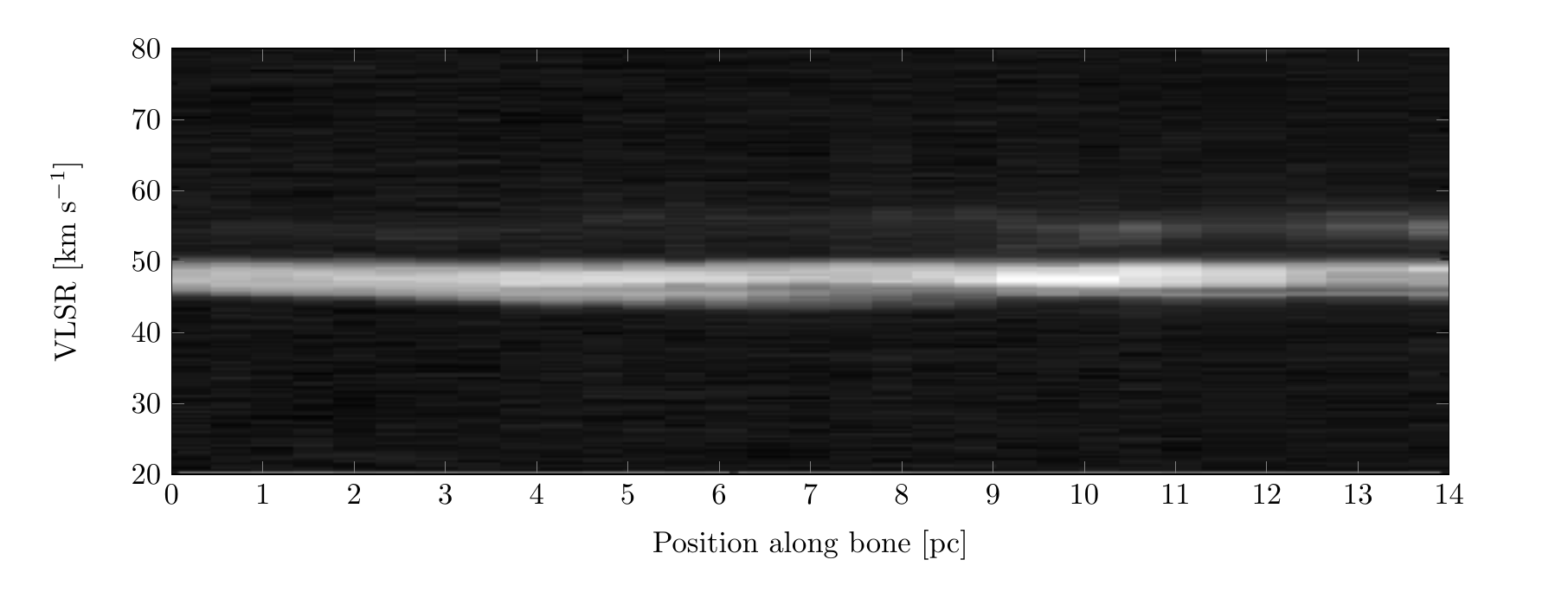}
\caption{Results of performing a slice extraction along the filamentary extinction feature of filament 3, using $^{13}\rm{CO}$ data from the GRS survey.
}
\end{center}
\end{figure}

\begin{figure}[h!]
\textbf{Filament 3 (``BC\_24.95-0.17"): Grade ``C"}
\begin{center}
\epsscale{.6}
\plotone{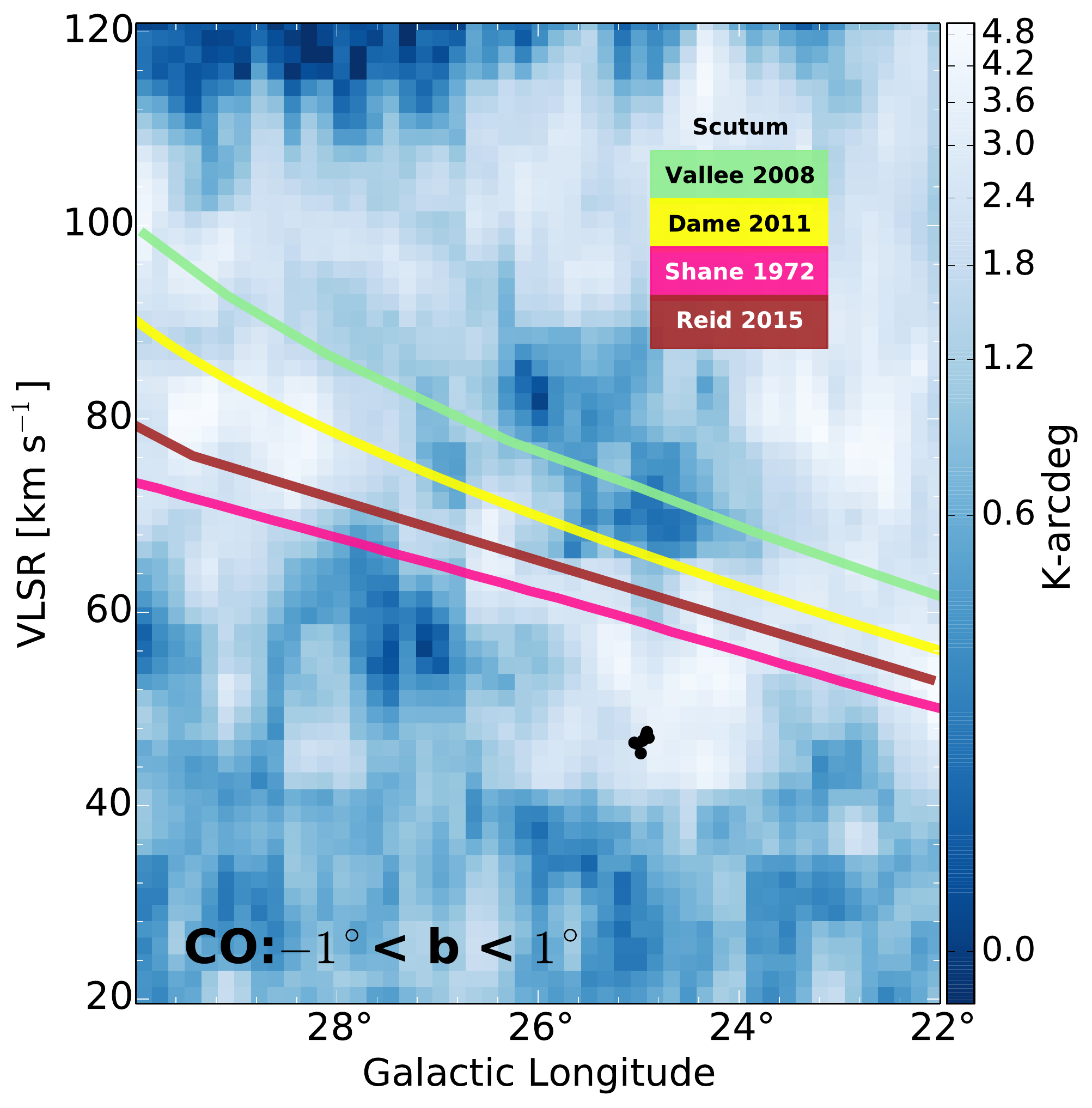}
\caption{ \textit{Top}: Position-velocity diagram of CO, NH$_{3}$, and HCO$^+$ emission for filament 3. Blue background shows $^{12}$CO (1-0) emission integrated between $-1^\circ < \textrm{b} < 1^\circ$ \citep{Dame_2001}. Black dots show HOPS, BGPS, and GRS sources associated with filament 3. Colored lines show fits for the Scutum-Centaurus arm (see text for references). %
}
\label{fig:BC24}
\end{center}
\end{figure}

\begin{figure}[h!]
\textbf{Filament 3 (``BC\_24.95-0.17"): Grade ``C"}
\begin{center}
\plotone{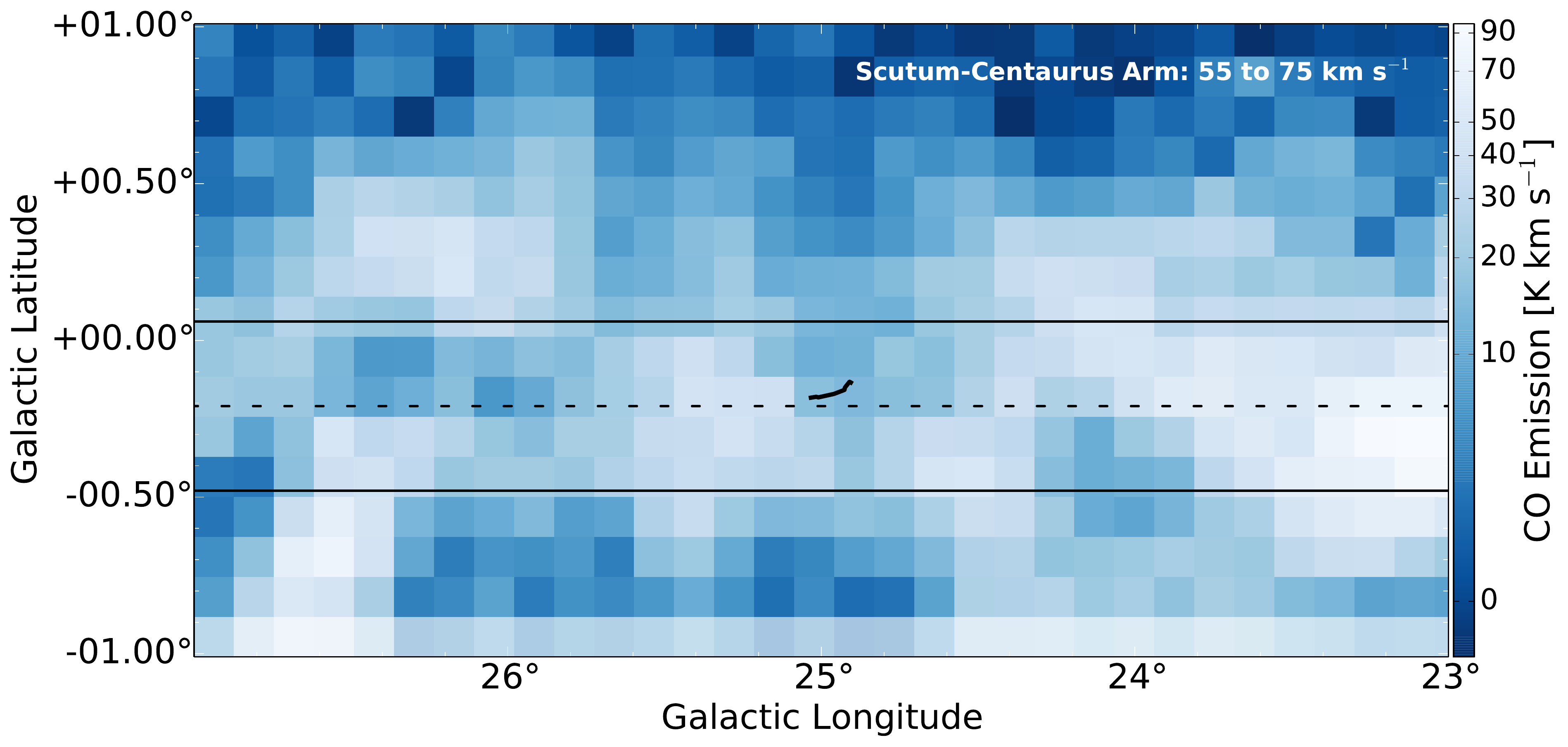}
\caption{Plane of the sky map integrated between 55 and 75 km s$^{-1}$, the approximate velocity range of the Scutum-Centaurus arm in the region around filament 3. A trace of filament 3, as it would appear as a mid-IR extinction feature, is superimposed on the $^{12}\rm{CO}$ emission map \citep{Dame_2001}. The black dashed line indicates the location of the physical Galactic mid-plane, while the solid black lines indicate $\pm$ 20 pc from the Galactic mid-plane at the 4.3 kpc distance to filament 3, assuming the candidate is associated with the \citet{Dame_2011} Scutum-Centaurus model.}
\end{center}
\end{figure}

\begin{figure}[h!]
\textbf{Filament 3 (``BC\_24.95-0.17"): Grade ``C"}
\begin{center}
\plotone{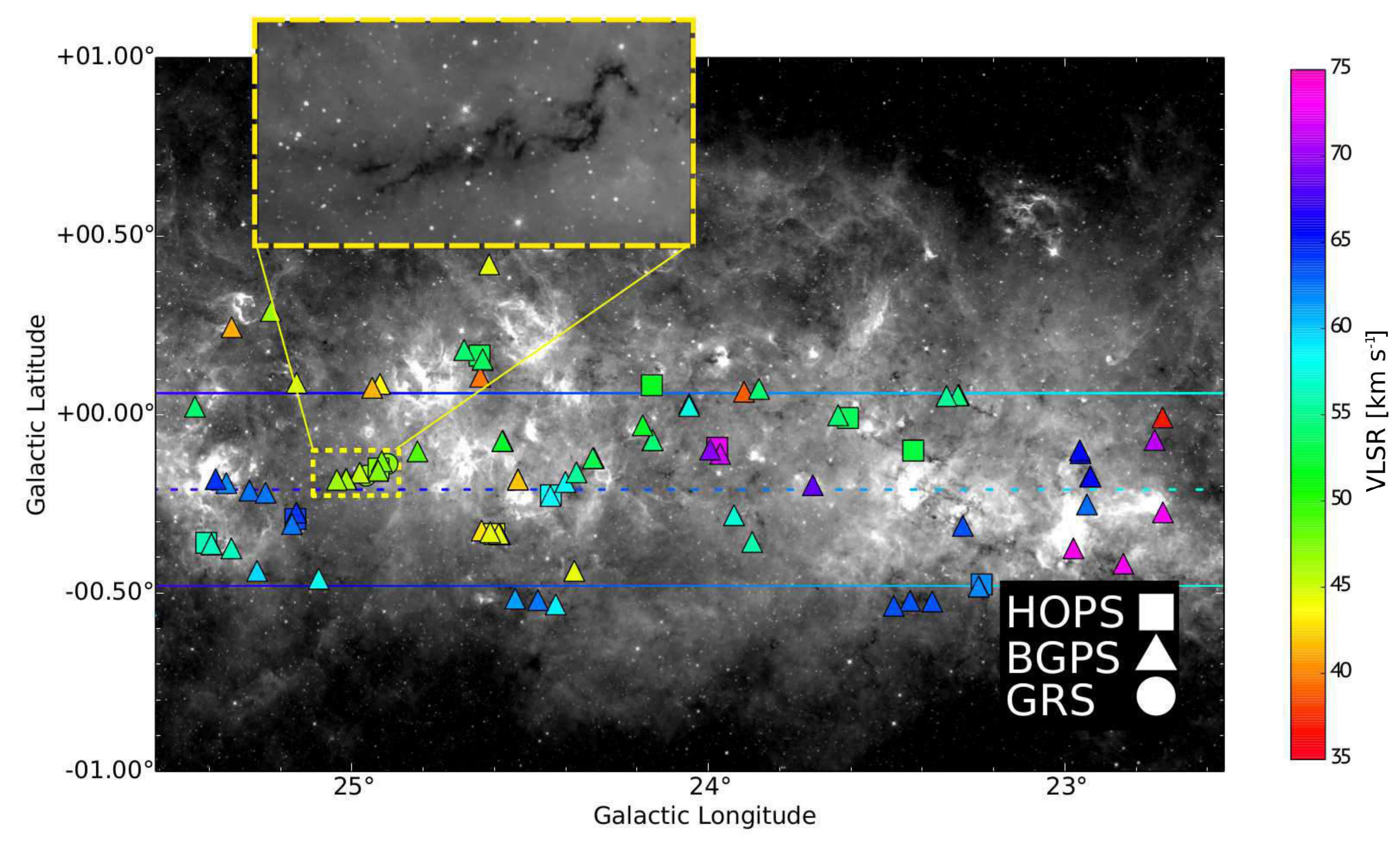}
\caption{Filament 3 lies within $\approx$ 3 pc of the physical Galactic mid-plane. The background is a GLIMPSE-\textit{Spitzer} 8 $\mu\textrm{m}$ image. The dashed line is color-coded by \citet{Dame_2011} LSR velocity and indicates the location of the physical Galactic mid-plane. The solid colored lines indicate $\pm$ 20 pc from the Galactic mid-plane at the 4.3 kpc distance to filament 3, assuming the candidate is associated with the \citet{Dame_2011} Scutum-Centaurus model. The squares, triangles, and circles correspond to HOPS, BGPS, and GRS sources, respectively.  A closer look at filament 3 can be seen in the inset.%
}
\end{center}
\end{figure}

\clearpage

\subsection{\large Filament 4 (``BC\_021.25-0.15"): Grade ``C"}
Filament 4 is not a confirmed bone, receiving a quality grade of ``C." It weakly satisfies criterion 4 and could potentially be an interarm filament, lying between the Scutum-Centaurus and Norma-4kpc arms in \textit{p-v} space. With an aspect ratio of 40:1, it also fails criterion 6. Finally, it only weakly satisfies criterion 1 (largely continuous mid-infrared extinction feature), as there is a small break in the feature around $l=21.25^\circ$.

\begin{figure}[h!]
\textbf{Filament 4 (``BC\_021.25-0.15"): Grade ``C"}
\begin{center}
\plotone{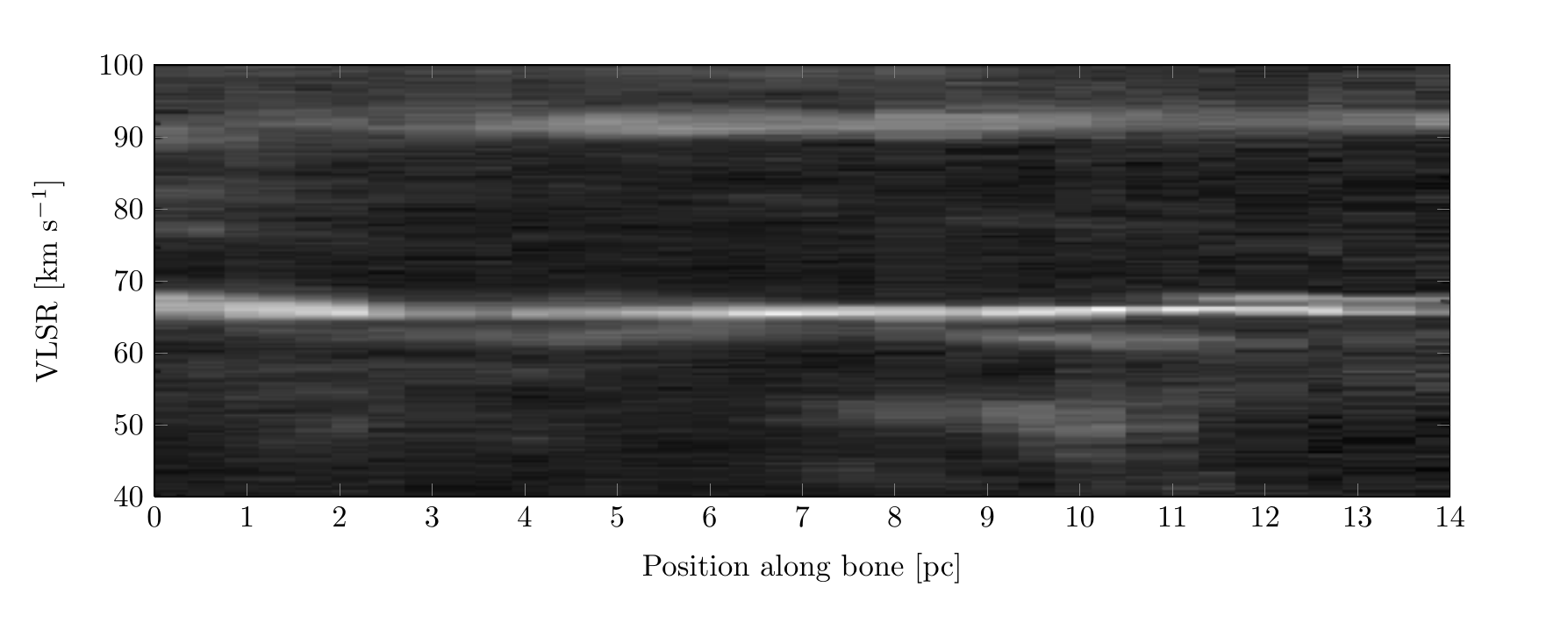}
\caption{Results of performing a slice extraction along the filamentary extinction feature of filament 4, using $^{13}\rm{CO}$ data from the GRS survey.}
\end{center}
\end{figure}

\begin{figure}[h!]
\textbf{Filament 4 (``BC\_021.25-0.15"): Grade ``C"}
\begin{center}
\epsscale{.6}
\plotone{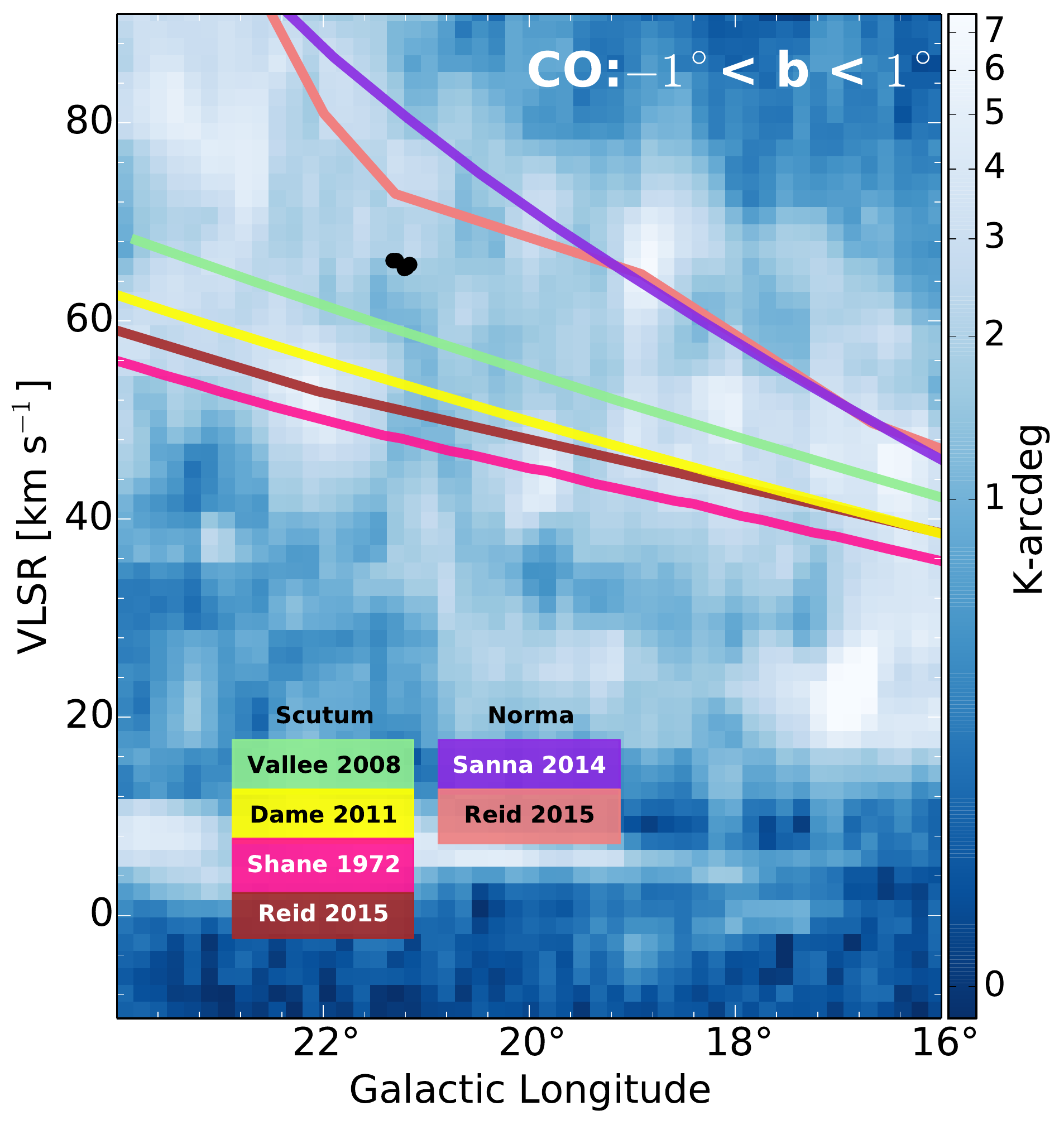}
\caption{Position-velocity diagram of CO and $\rm{HCO}^+$ emission for filament 4. Blue background shows $^{12}$CO (1-0) emission integrated between $-1^\circ < \textrm{b} < 1^\circ$ \citep{Dame_2001}. Black dots show GRS and BGPS sources associated with filament 4. Colored lines show fits for the Scutum-Centaurus and Norma-4kpc arms (see text for references).
}
\end{center}
\end{figure}

\begin{figure}[h!]
\textbf{Filament 4 (``BC\_021.25-0.15"): Grade ``C"}
\begin{center}
\plotone{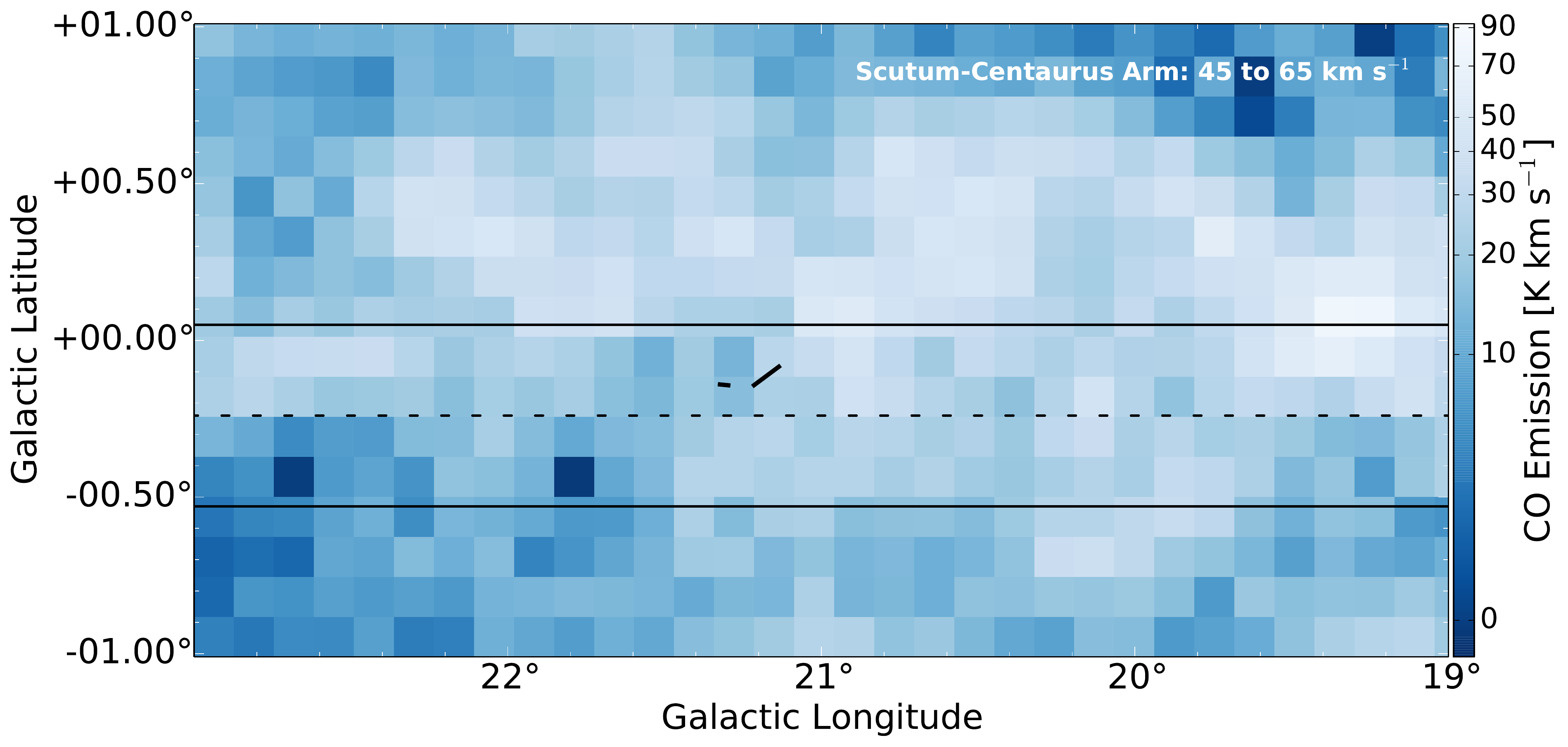}
\caption{Plane of the sky map integrated between 45 and 65 km s$^{-1}$, the approximate velocity range of the Scutum-Centaurus arm in the region around filament 4. A trace of filament 4, as it would appear as a mid-IR extinction feature, is superimposed on the $^{12}\rm{CO}$ emission map \citep{Dame_2001}; there is a visible break in the extinction feature at $l \approx 21.25^\circ$. The black dashed line indicates the location of the physical Galactic mid-plane, while the solid black lines indicate $\pm$ 20 pc from the Galactic mid-plane at the 3.9 kpc distance to filament 4, assuming the candidate is associated with the \citet{Dame_2011} Scutum-Centaurus model.}
\end{center}
\end{figure}

\begin{figure}[h!]
\textbf{Filament 4 (``BC\_021.25-0.15"): Grade ``C"}
\begin{center}
\plotone{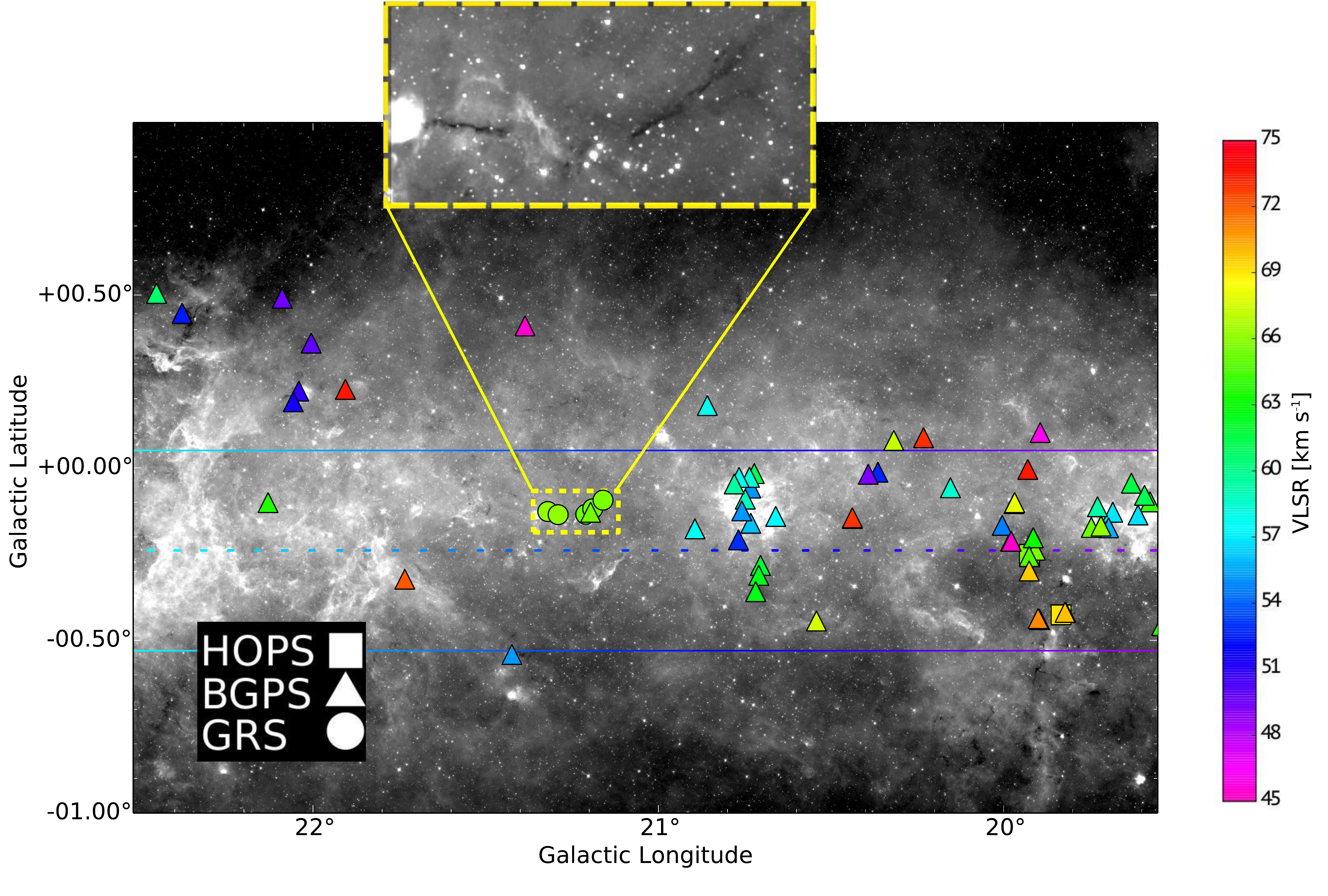}
\caption{Filament 4 lies within $\approx$ 8 pc of the physical Galactic mid-plane. The background is a GLIMPSE-\textit{Spitzer} 8 $\mu\textrm{m}$ image. The dashed line is color-coded by \citet{Dame_2011} LSR velocity and indicates the location of the physical Galactic mid-plane. The solid colored lines indicate $\pm$ 20 pc from the Galactic mid-plane at the 3.9 kpc distance to filament 4, assuming the candidate is associated with the \citet{Dame_2011} Scutum-Centaurus model. The squares, triangles, and circles correspond to HOPS, BGPS, and GRS sources, respectively.  A closer look at filament 4 can be seen in the inset.%
}
\end{center}
\end{figure}

\clearpage

\subsection{\large Filament 6 (``BC\_011.13-0.12"): Grade ``B"}
Filament 6 is not a confirmed bone candidate, receiving a quality grade of ``B." At 25:1, it has the smallest aspect ratio of all ten candidates, failing to satisfy criterion 6. Despite this, it lies within 5 km s$^{-1}$ of both the Scutum-Centaurus and Norma-4kpc fits in \textit{p-v} space, as well as within 15 pc of the physical Galactic mid-plane. Designated the ``snake", we note that filament 6 has been well-studied for its star formation properties, hosting over a dozen pre-stellar cores likely to produce regions of high mass star formation \citep{Wang_2014,Henning_2010}. 

\begin{figure}[h!]
\textbf{Filament 6 (``BC\_011.13-0.12"): Grade ``B"}
\begin{center}
\epsscale{.6}
\plotone{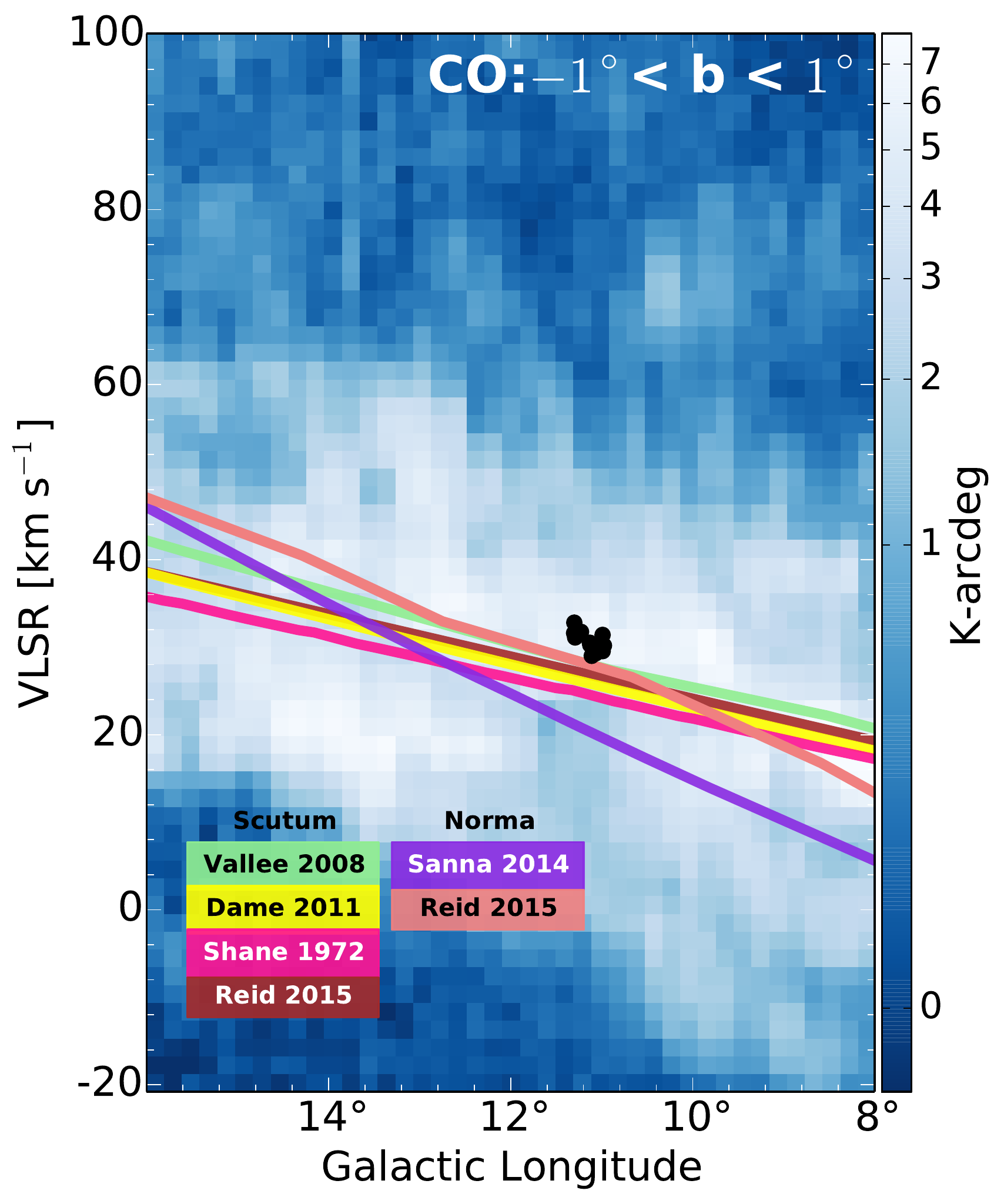}
\caption{Position-velocity diagram of CO, NH$_{3}$, N$_2$H$^+$ and HCO$^+$ emission for filament 6. Blue background shows $^{12}$CO (1-0) emission integrated between $-1^\circ < \textrm{b} < 1^\circ$ \citep{Dame_2001}. Black dots show HOPS, BGPS, and MALT90 sources associated with filament 6. Colored lines show fits for the Scutum-Centaurus and Norma-4kpc arms (see text for references).%
}
\label{fig:snake_pv}
\end{center}
\end{figure}

\begin{figure}[h!]
\textbf{Filament 6 (``BC\_011.13-0.12"): Grade ``B"}
\begin{center}
\plotone{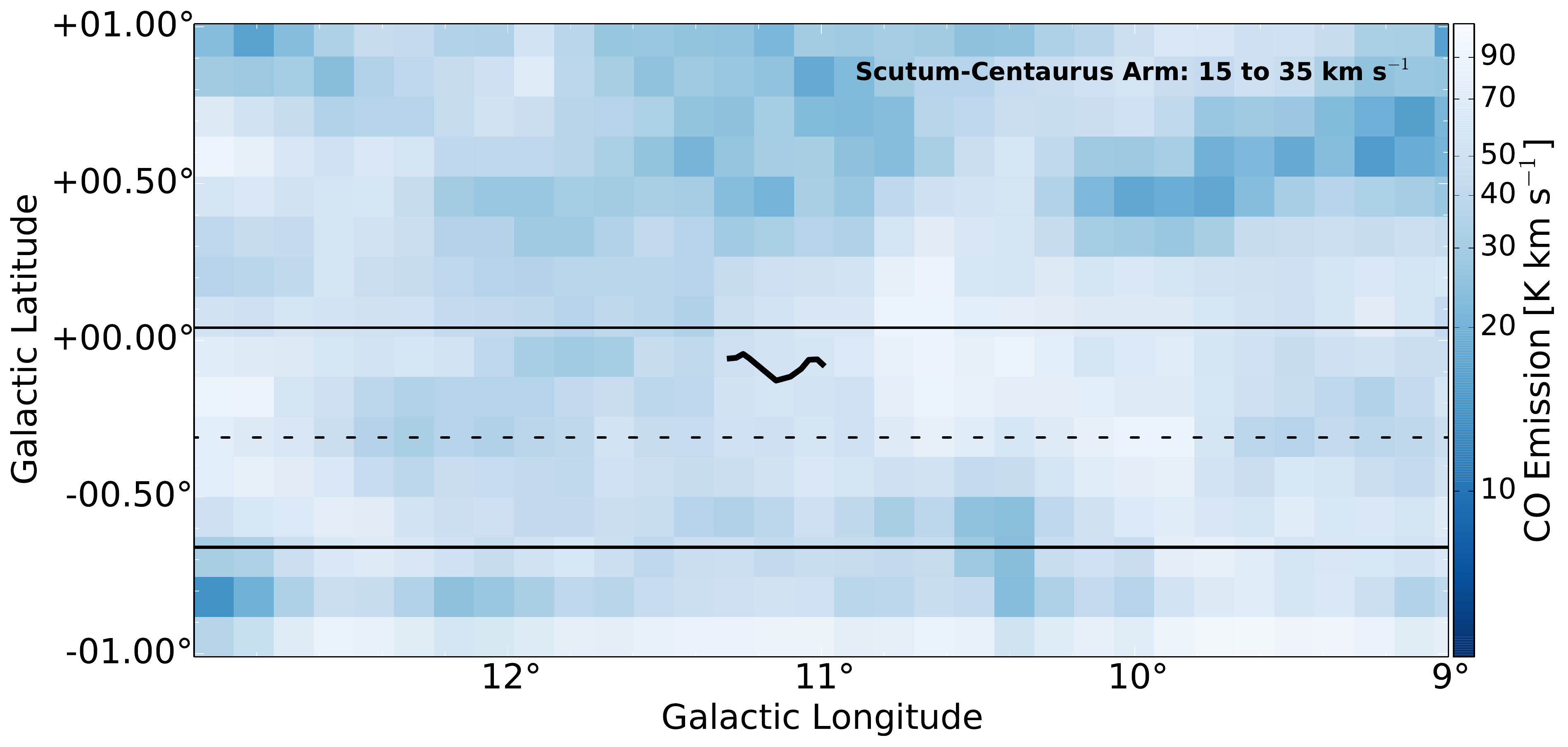}
\caption{Plane of the sky map integrated between 15 and 35 km s$^{-1}$, the approximate velocity range of the Scutum-Centaurus arm in the region around filament 6. A trace of filament 6, as it would appear as a mid-IR extinction feature, is superimposed on the $^{12}\rm{CO}$ emission map \citep{Dame_2001}. The black dashed line indicates the location of the physical Galactic mid-plane, while the solid black lines indicate $\pm$ 20 pc from the Galactic mid-plane at the 3.3 kpc distance to filament 6, assuming the candidate is associated with the \citet{Dame_2011} Scutum-Centaurus model.}
\end{center}
\end{figure}

\begin{figure}[h!]
\textbf{Filament 6 (``BC\_011.13-0.12"): Grade ``B"}
\begin{center}
\plotone{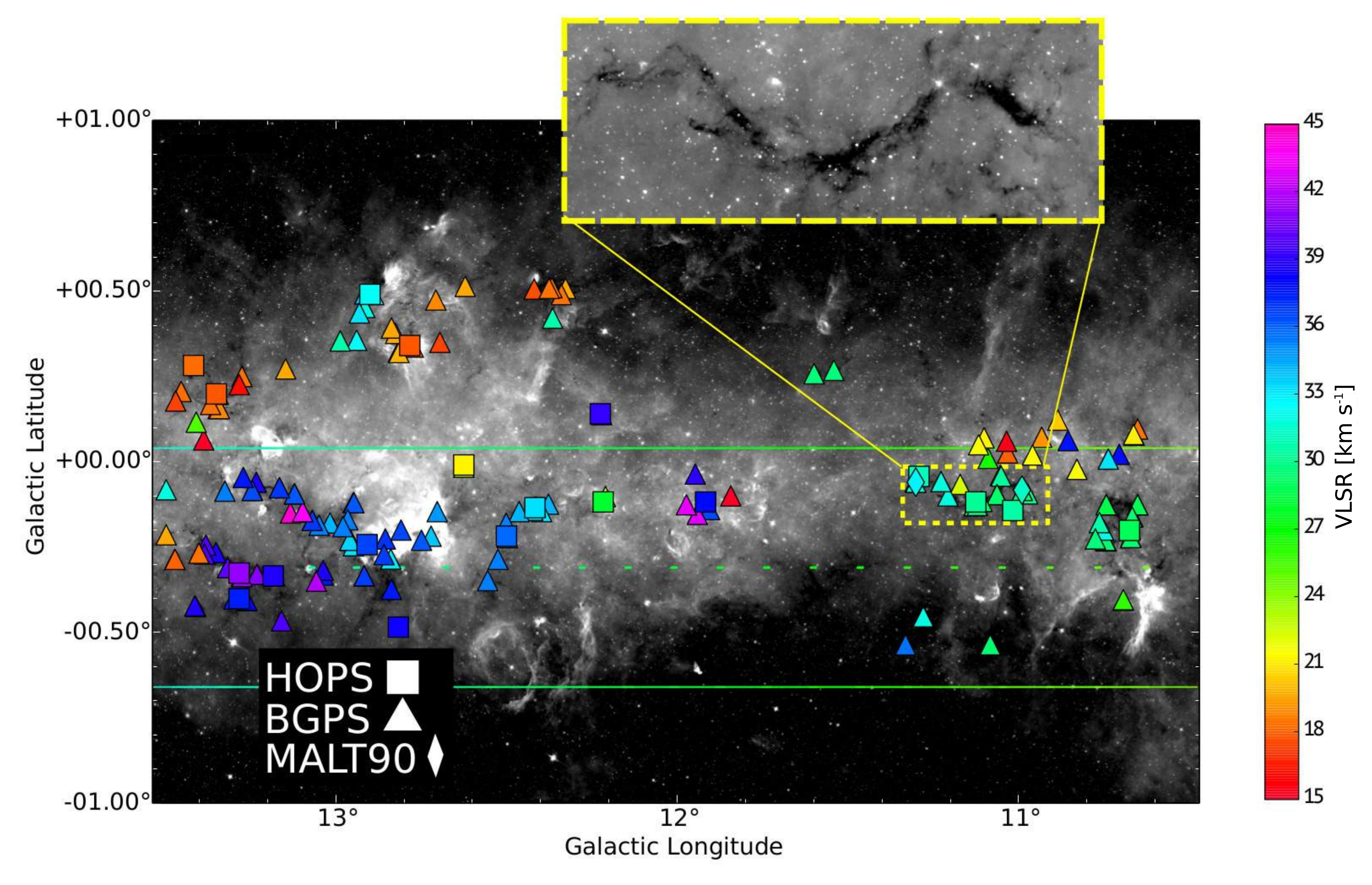}
\caption{Filament 6 lies within $\approx$ 15 pc of the physical Galactic mid-plane. The background is a GLIMPSE-\textit{Spitzer} 8 $\mu\textrm{m}$ image. The dashed line is color-coded by \citet{Dame_2011} LSR velocity and indicates the location of the physical Galactic mid-plane. The solid colored lines indicate $\pm$ 20 pc from the Galactic mid-plane at the 3.3 kpc distance to filament 6, assuming the candidate is associated with the \citet{Dame_2011} Scutum-Centaurus model. The squares, triangles, and diamonds correspond to HOPS, BGPS, and MALT90 sources, respectively.  A closer look at filament 6 can be seen in the inset.%
}
\end{center}
\end{figure}

\clearpage

\subsection{\large Filament 7 (``BC\_4.14-0.02"): Grade ``B"}
Filament 7 is a confirmed bone candidate, with a quality grade of ``B." There is a slight break in the extinction feature around $l=4^\circ$, so it weakly satisfies criterion 1. It moderately satisfies the other five criteria, though we note that we were unable to confirm contiguity in velocity space using lower density gas tracers, as it was outside the coverage range of both GRS and ThrUMMS. It does, however, express contiguity as traced by higher density gas from the HOPS and MALT90 surveys. It lies on top of the Scutum-Centaurus fits to CO and HI , but its velocity gradient is slightly angled with respect to these fits, suggesting it could be a potential spur of this arm. 

\begin{figure}[h!]
\textbf{Filament 7 (``BC\_4.14-0.02"): Grade ``B"}
\begin{center}
\epsscale{.6}
\plotone{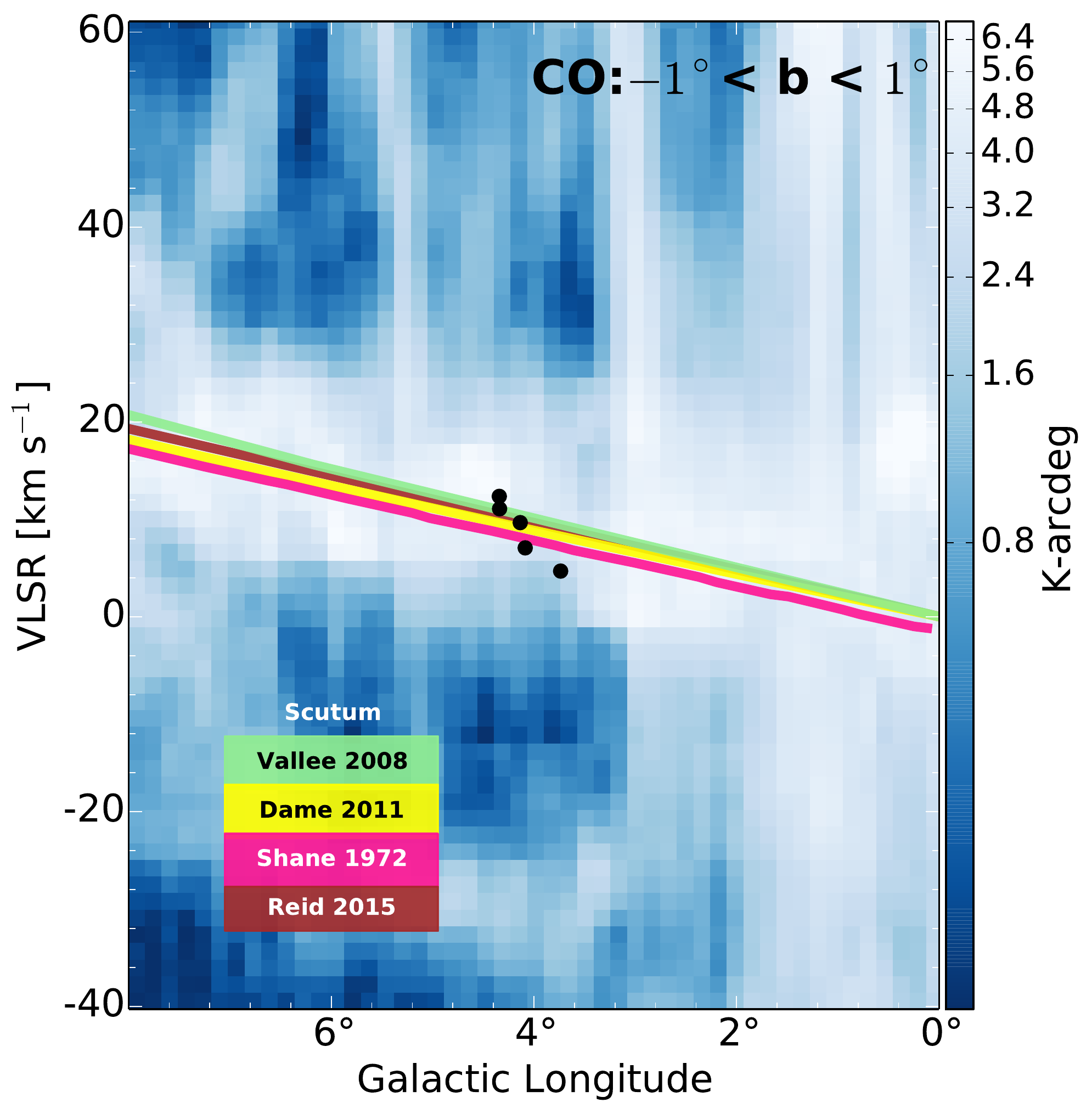}
\caption{Position-velocity diagram of CO, $\rm{NH}_3$, and $\rm{N}_2\rm{H}^+$ emission for filament 7. Blue background shows $^{12}\rm{CO} (1-0)$ emission integrated between ${-}1^\circ<\rm{b}<1^\circ$ (Dame et al., 2001). Black dots show HOPS and MALT90 sources associated with filament 7. Colored lines show fits for the Scutum-Centaurus arm (see text for references). %
}
\end{center}
\end{figure}

\begin{figure}[h!]
\textbf{Filament 7 (``BC\_4.14-0.02"): Grade ``B"}
\begin{center}
\plotone{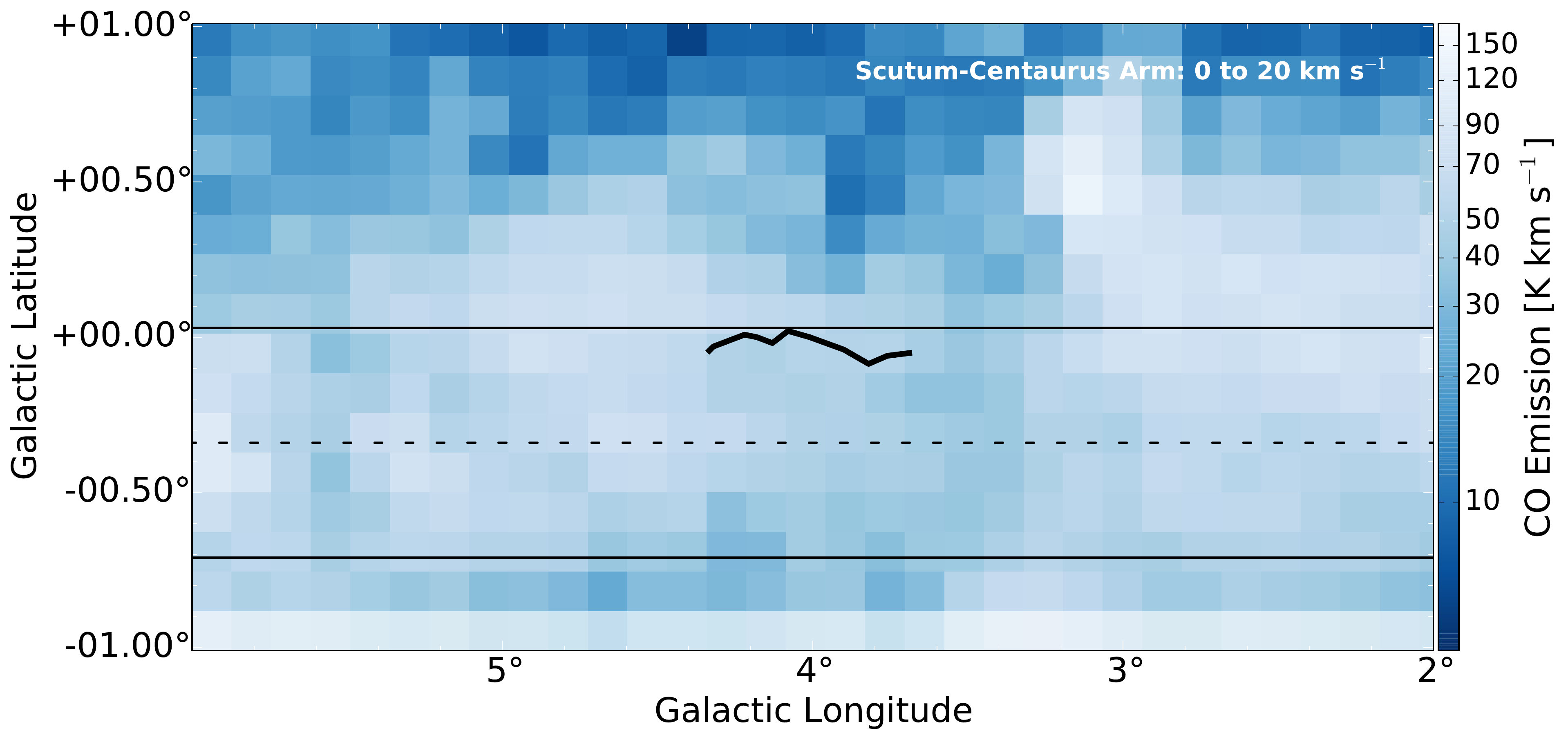}
\caption{Plane of the sky map integrated between 0 and 20 km s$^{-1}$, the approximate velocity range of the Scutum-Centaurus arm in the region around filament 7. A trace of filament 7, as it would appear as a mid-IR extinction feature, is superimposed on the $^{12}\rm{CO}$ emission map \citep{Dame_2001}. The black dashed line indicates the location of the physical Galactic mid-plane, while the solid black lines indicate $\pm$ 20 pc from the Galactic mid-plane at the 3.1 kpc distance to filament 7, assuming the candidate is associated with the \citet{Dame_2011} Scutum-Centaurus model.}
\end{center}
\end{figure}

\begin{figure}[h!]
\textbf{Filament 7 (``BC\_4.14-0.02"): Grade ``B"}
\begin{center}
\epsscale{.8}
\plotone{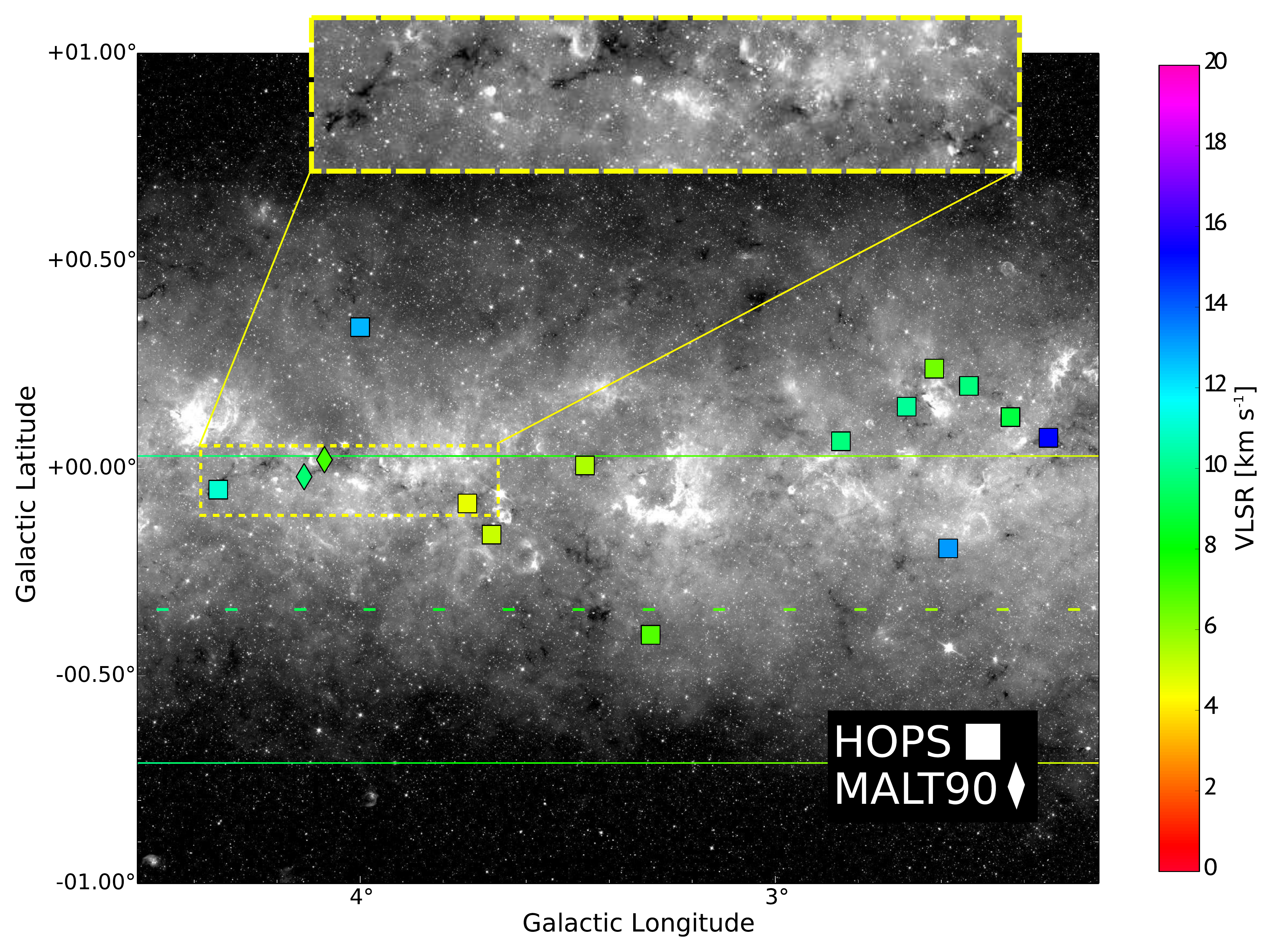}
\caption{Filament 7 lies within $\approx$ 20 pc of the physical Galactic mid-plane. The background is a GLIMPSE-\textit{Spitzer} 8 $\mu\textrm{m}$ image. The dashed line is color-coded by \citet{Dame_2011} LSR velocity and indicates the location of the physical Galactic mid-plane. The solid colored lines indicate $\pm$ 20 pc from the Galactic mid-plane at the 3.1 kpc distance to filament 7, assuming the candidate is associated with the \citet{Dame_2011} Scutum-Centaurus model. The squares and diamonds correspond to HOPS and MALT90 sources, respectively.  A closer look at filament 7 can be seen in the inset.%
}
\end{center}
\end{figure}

\clearpage

\subsection{\large Filament 8 (``BC\_357.62-0.33"): Grade ``B"}
Filament 8 is not a confirmed bone candidate, receiving a quality grade of ``B"; it fails to satisfy the 50:1 minimum aspect ratio criterion. Otherwise, it moderately or strongly satisfies the other five criteria. It lies almost \textit{exactly} on and parallel to the physical Galactic mid-plane. Though it lies about 6-8 km s$^{-1}$ from the Scutum-Centaurus arm, its exhibits a similar velocity gradient and traces a prominent peak of CO emisson in \textit{p-v} space. 

\begin{figure}[h!]
\textbf{Filament 8 (``BC\_357.62-0.33"): Grade ``B"}
\begin{center}
\plotone{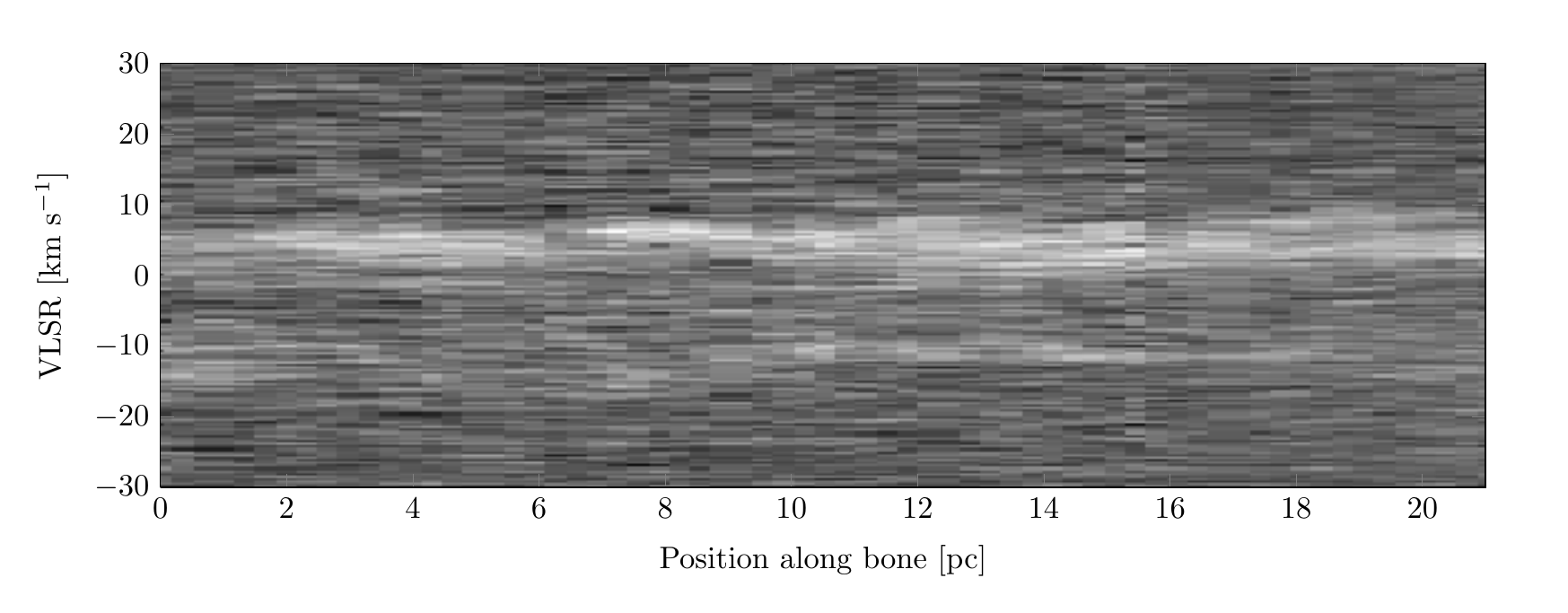}
\caption{Results of performing a slice extraction along the filamentary extinction feature of filament 8, using $^{13}\rm{CO}$ data from the ThruMMS survey.}
\end{center}
\end{figure}

\begin{figure}
\textbf{Filament 8 (``BC\_357.62-0.33"): Grade ``B"}
\begin{center}
\epsscale{.5}
\plotone{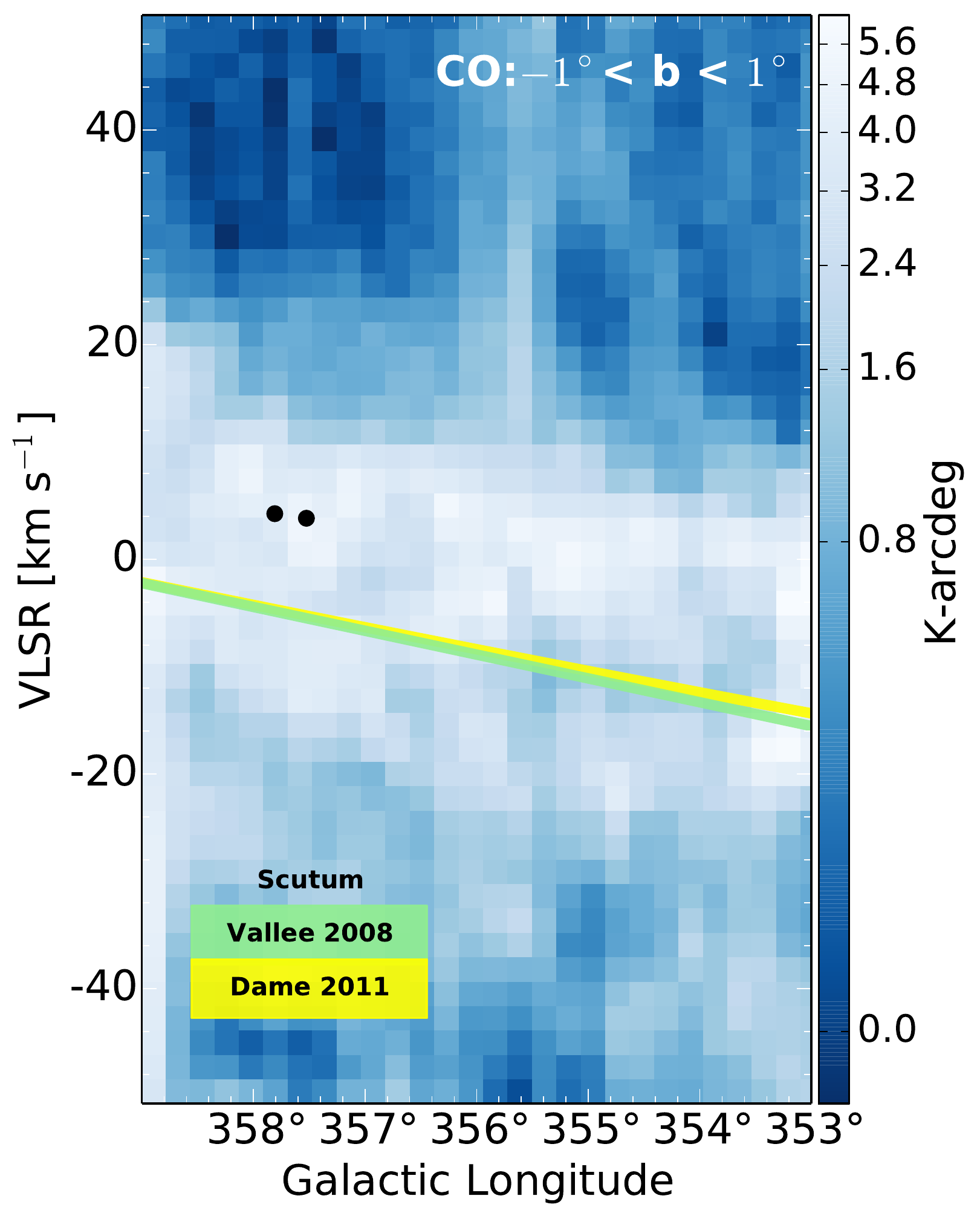}
\caption{Top: Position-velocity diagram of CO and $\rm{NH}_3$ emission for filament 8. Blue background shows $^{12}\rm{CO} (1-0)$ emission integrated between ${-}1^\circ<\rm{b}<1^\circ$ \citep{Dame_2011}. Black dots show HOPS sources associated with filament 8. The colored lines are fits for the Scutum-Centaurus arm (see text for references).%
}
\label{fig:candid8pv}
\end{center}
\end{figure}

\begin{figure}[h!]
\textbf{Filament 8 (``BC\_357.62-0.33"): Grade ``B"}
\begin{center}
\plotone{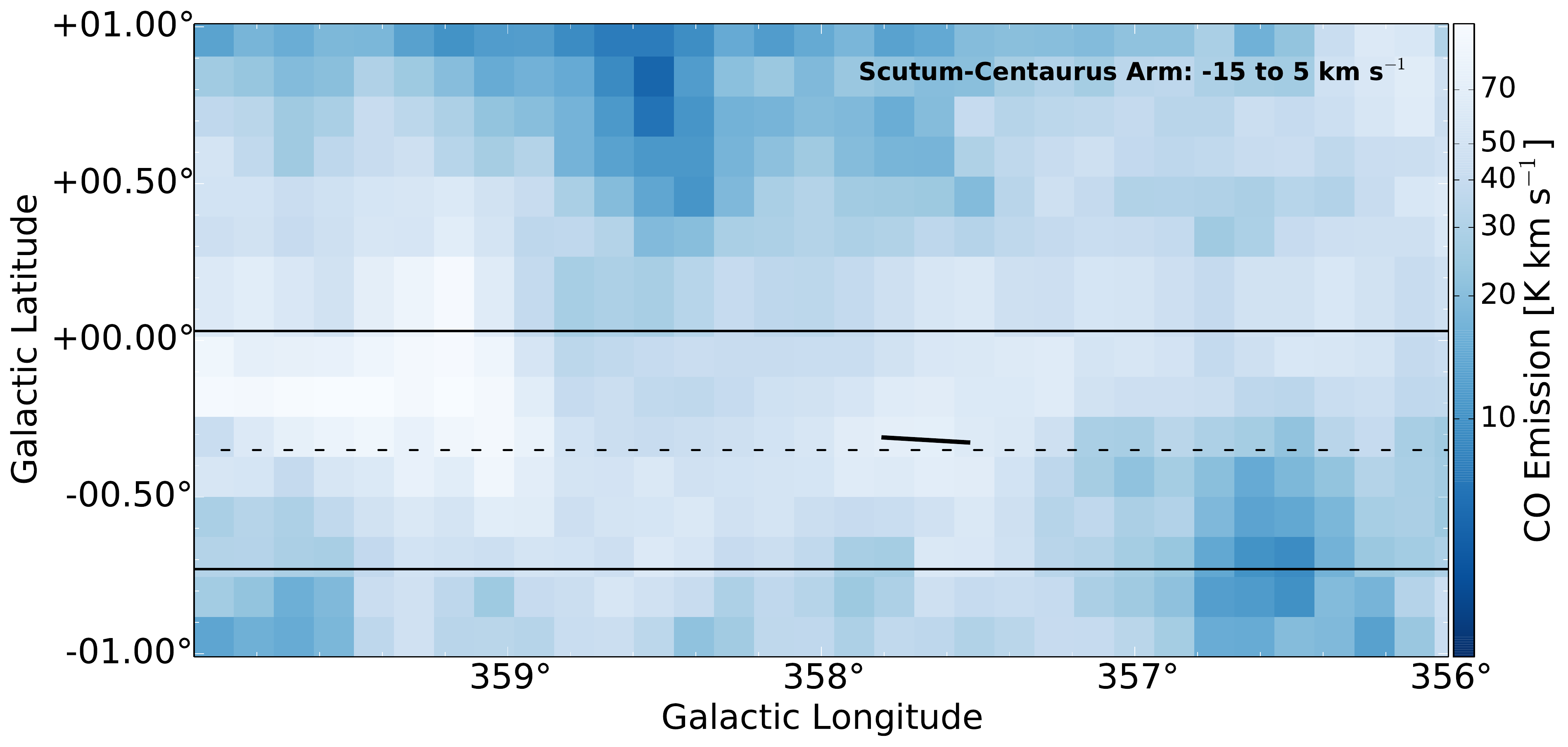}
\caption{Plane of the sky map integrated between $-15$ and 5 km s$^{-1}$, the approximate velocity range of the Scutum-Centaurus arm in the region around filament 8. A trace of filament 8, as it would appear as a mid-IR extinction feature, is superimposed on the $^{12}\rm{CO}$ emission map \citep{Dame_2001}. The black dashed line indicates the location of the physical Galactic mid-plane, while the solid black lines indicate $\pm$ 20 pc from the Galactic mid-plane at the 3.0 kpc distance to filament 8, assuming the candidate is associated with the \citet{Dame_2011} Scutum-Centaurus model.
}
\label{fig:candid8pp}
\end{center}
\end{figure}

\begin{figure}[h!]
\textbf{Filament 8 (``BC\_357.62-0.33"): Grade ``B"}
\begin{center}
\plotone{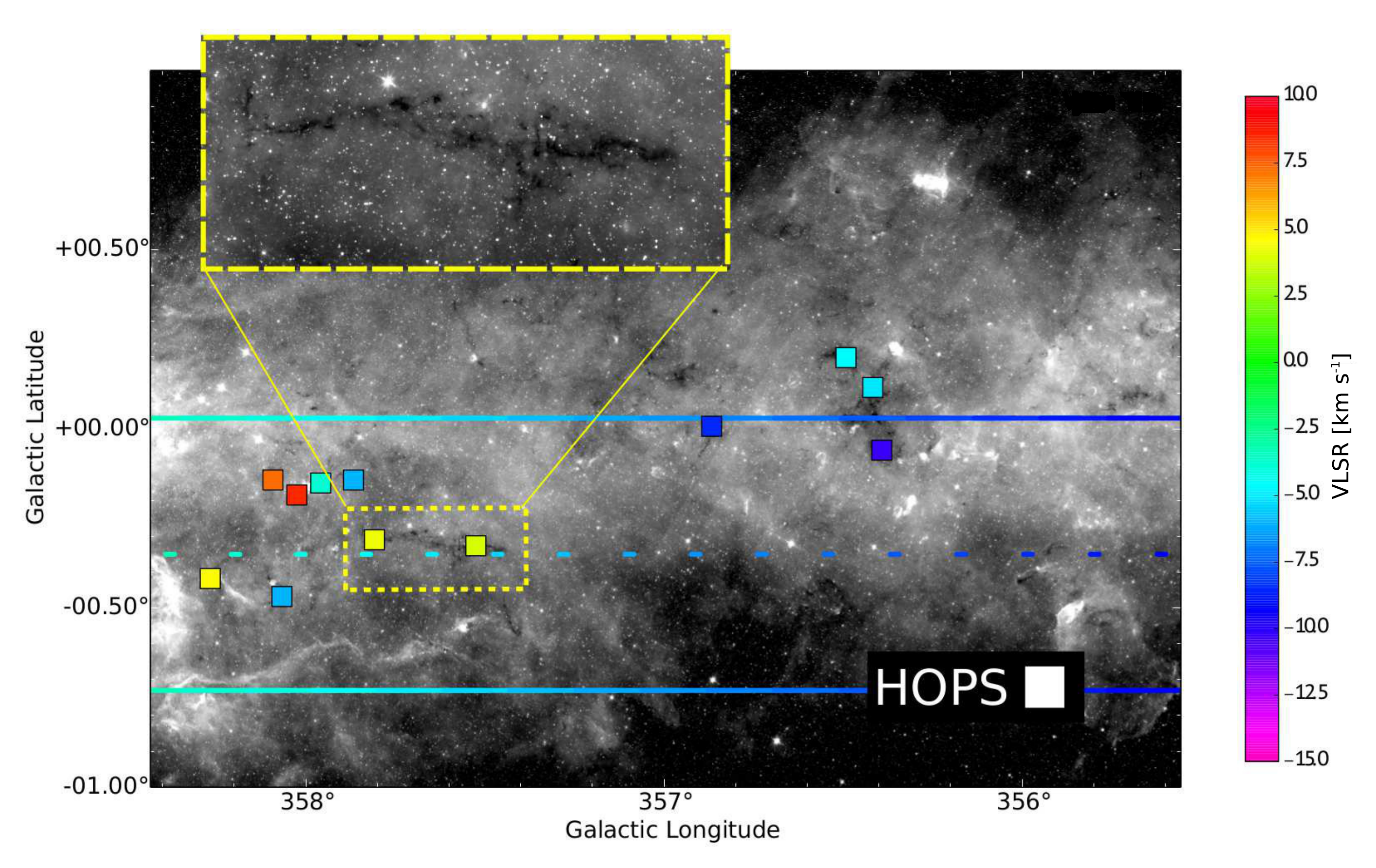}
\caption{Filament 8 lies right on the physical Galactic mid-plane. The background is a GLIMPSE-\textit{Spitzer} 8 $\mu$m image. The dashed line is color-coded by \citep{Dame_2011} LSR velocity and indicates the location of the physical Galactic mid-plane. The solid colored lines indicate $\pm$ 20 pc from the Galactic mid-plane at the 3.0 kpc distance to filament 8, assuming the candidate is associated with the \citet{Dame_2011} Scutum-Centaurus model. The squares correspond to HOPS sources. A closer look at filament 8 can be seen in the inset.%
}
\end{center}
\end{figure}

\clearpage

\subsection{\large Filament 9 (``BC\_335.31-0.29"): Grade ``B"}
Filament 9 is a confirmed bone candidate, receiving a quality grade of ``B." It strongly satisfies criterion 3 (lying exactly on the physical Galactic mid-plane) and moderately satisfies criterion 2 (lying at a slight $15^\circ$ angle with respect to the plane). Filament 9 is notable for lying perpendicular to the \citet{Dame_2011} Scutum-Centaurus fit in \textit{p-v} space, suggesting it could be a potential interarm filament. 

\begin{figure}[h!]
\textbf{Filament 9 (``BC\_335.31-0.29"): Grade ``B"}
\begin{center}
\plotone{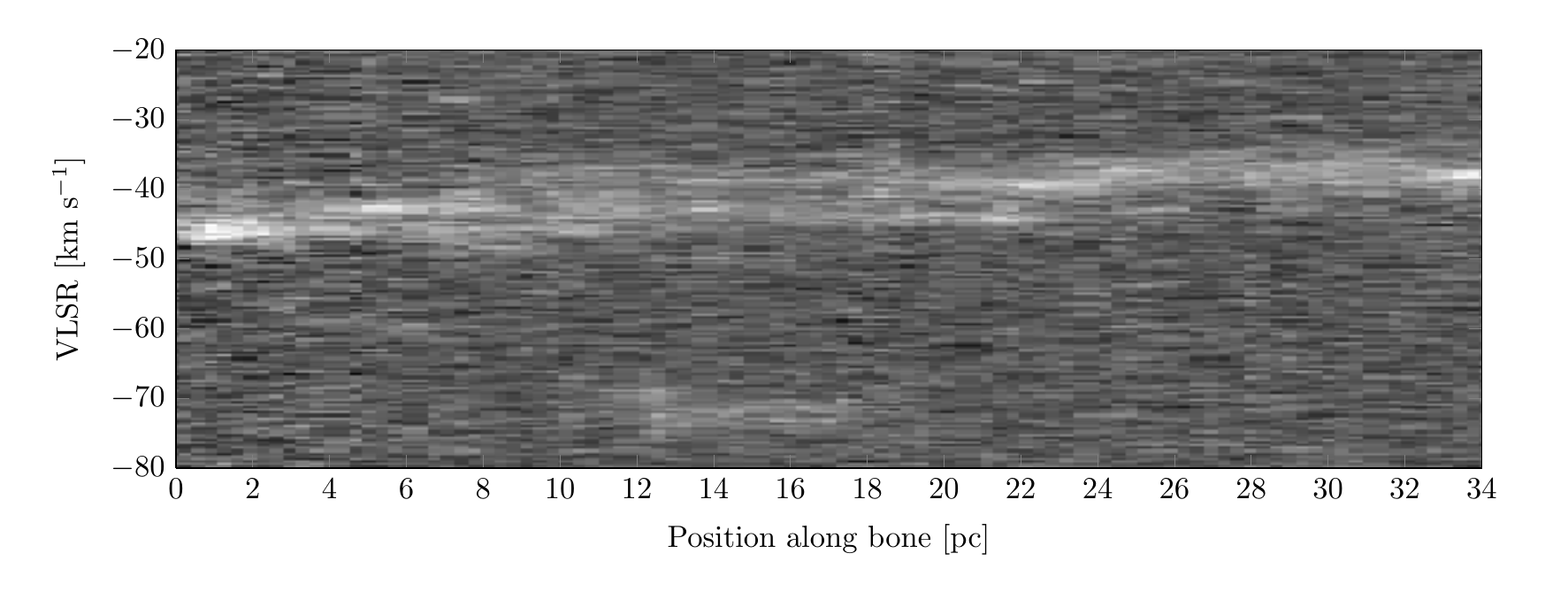}
\caption{Results of performing a slice extraction along the filamentary extinction feature of filament 9, using $^{13}\rm{CO}$ data from the ThrUMMS survey.}
\end{center}
\end{figure}

\begin{figure}[h!]
\textbf{Filament 9 (``BC\_335.31-0.29"): Grade ``B"}
\begin{center}
\epsscale{.6}
\plotone{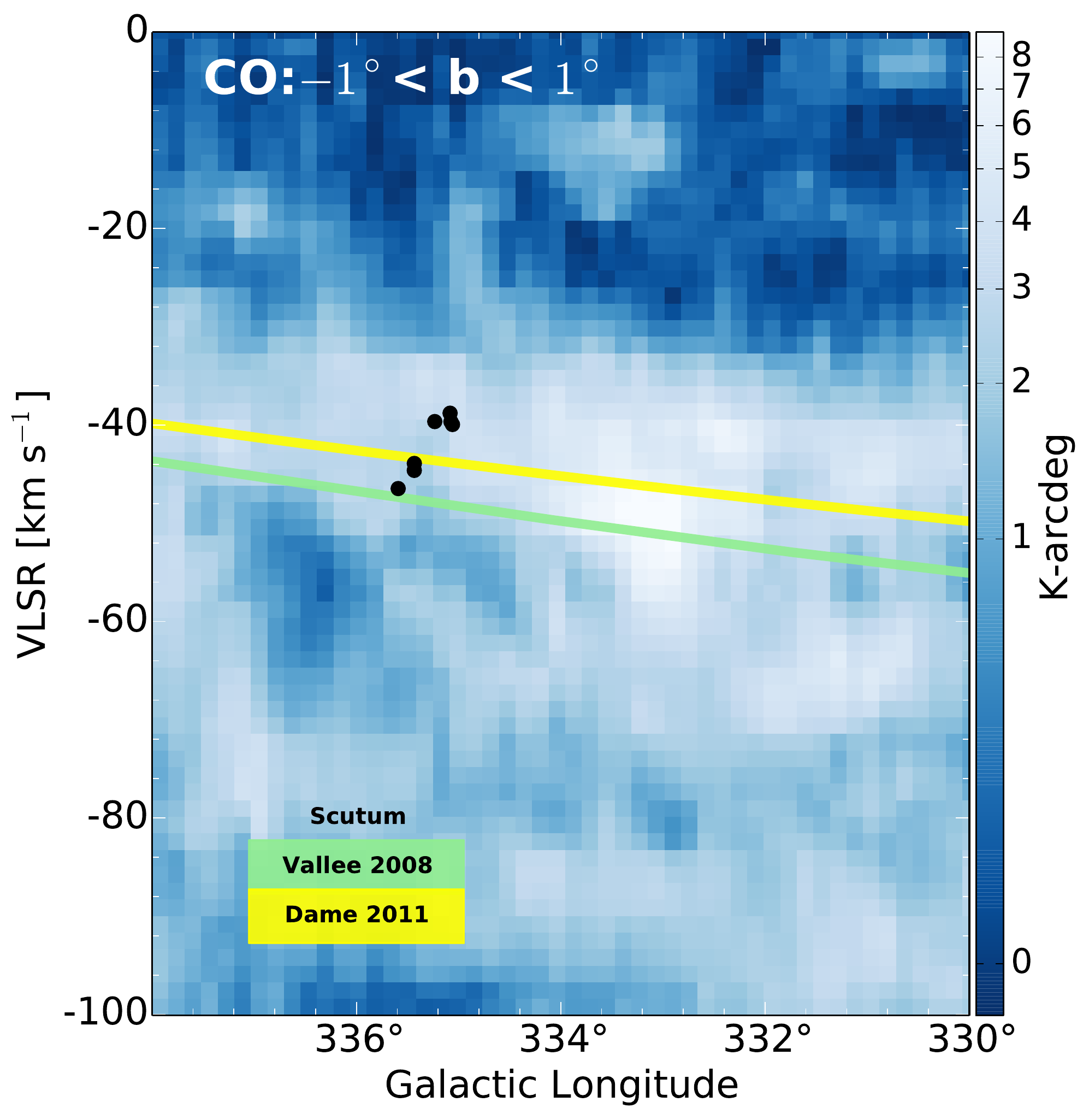}
\caption{Position-velocity diagram of CO, NH$_{3}$, and $\rm{N}_{2}\rm{H}^+$ emission for filament 9. Blue background shows $^{12}$CO (1-0) emission integrated between $-1^\circ < \textrm{b} < 1^\circ$ \citep{Dame_2001}. Black dots show HOPS and MALT90 sources associated with filament 9. Colored lines are fits to the Scutum-Centaurus arm (see text for references). %
}
\end{center}
\end{figure}

\begin{figure}[h!]
\textbf{Filament 9 (``BC\_335.31-0.29"): Grade ``B"}
\begin{center}
\plotone{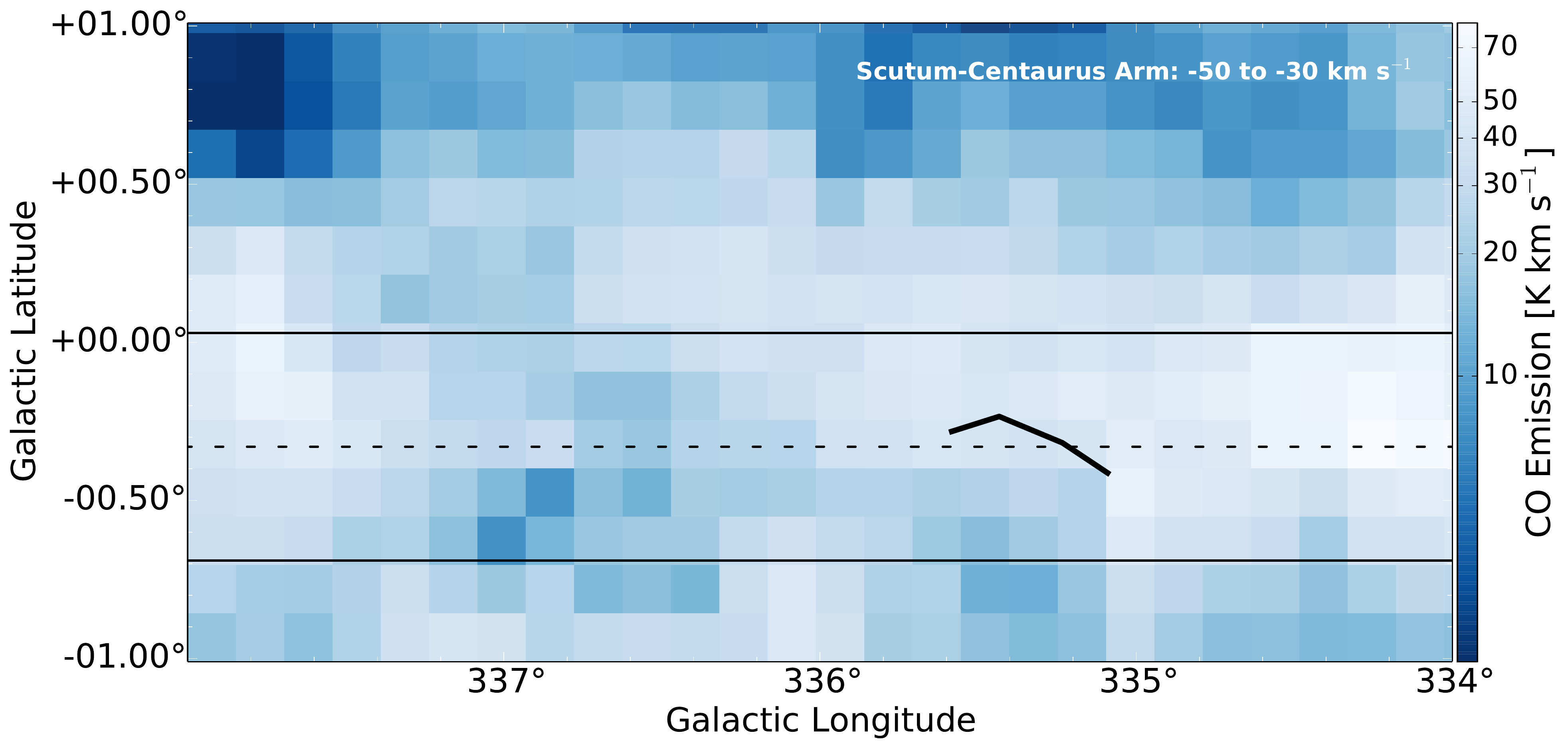}
\caption{Plane of the sky map integrated between $-50$ and $-30$ km s$^{-1}$, the approximate velocity range of the Scutum-Centaurus arm in the region around filament 9. A trace of filament 9, as it would appear as a mid-IR extinction feature, is superimposed on the $^{12}\rm{CO}$ emission map \citep{Dame_2001}. The black dashed line indicates the location of the physical Galactic mid-plane, while the solid black lines indicate $\pm$ 20 pc from the Galactic mid-plane at the 3.2 kpc distance to filament 9, assuming the candidate is associated with the \citet{Dame_2011} Scutum-Centaurus model.
}
\end{center}
\end{figure}

\begin{figure}[h!]
\textbf{Filament 9 (``BC\_335.31-0.29"): Grade ``B"}
\begin{center}
\plotone{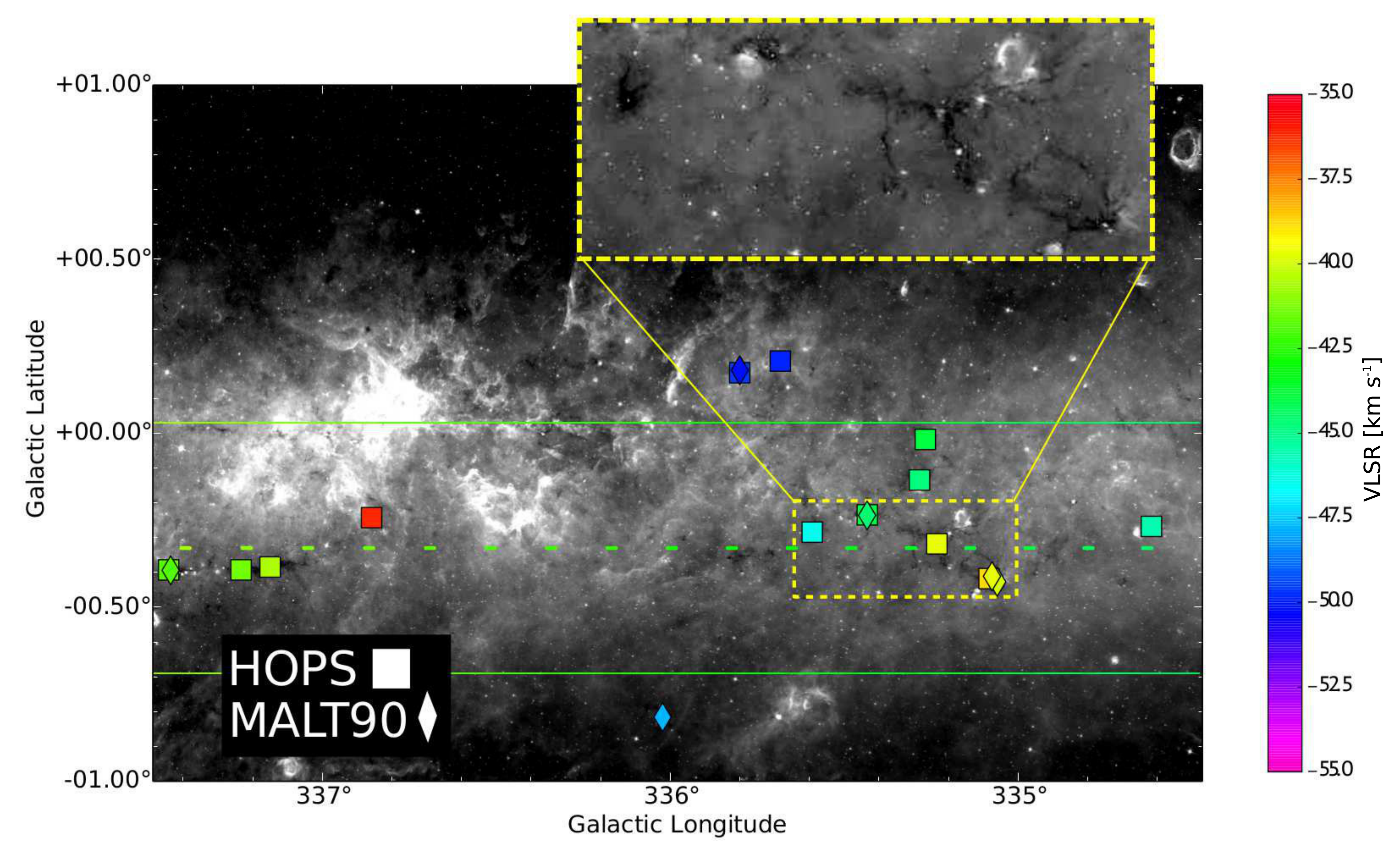}
\caption{Filament 9 lies right on the physical Galactic mid-plane. The background is a GLIMPSE-\textit{Spitzer} 8 $\mu\textrm{m}$ image. The dashed line is color-coded by \citet{Dame_2011} LSR velocity and indicates the location of the physical Galactic mid-plane. The solid colored lines indicate $\pm$ 20 pc from the Galactic mid-plane at the 3.2 kpc distance to filament 9, assuming the candidate is associated with the \citet{Dame_2011} Scutum-Centaurus model. The squares and diamonds correspond to HOPS and MALT90 sources, respectively.  A closer look at filament 9 can be seen in the inset.%
}
\end{center}
\end{figure}

\clearpage

\subsection{\large Filament 10 (``BC\_332.21-0.04"): Grade ``B"}
Filament 10 is a confirmed bone candidate, receiving a quality grade of ``B." It weakly satisfies criterion 1 (largely continuous mid-infrared extinction feature). We speculate that it is likely being broken apart by stellar feedback, making it more difficult to detect continuity in the extinction feature. Otherwise, it moderately or strongly satisfies the other five criterion, lying within 5 km s$^{-1}$ of the \citet{Dame_2011} Scutum-Centaurus fit in \textit{p-v} space, and within 10-15 pc of the physical Galactic mid-plane. 

\begin{figure}[h!]
\textbf{Filament 10 (``BC\_332.21-0.04"): Grade ``B"}
\begin{center}
\plotone{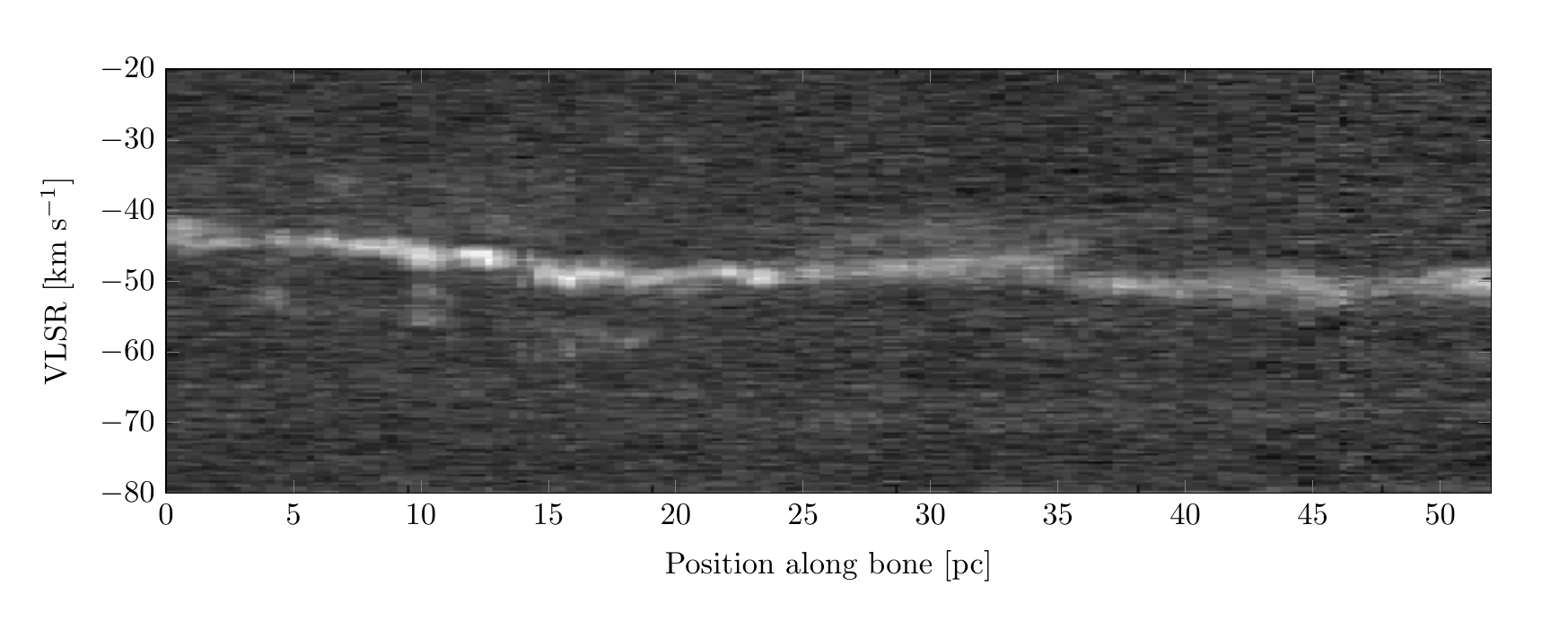}
\caption{Results of performing a slice extraction along the filamentary extinction feature of filament 10, using $^{13}\rm{CO}$ data from the ThrUMMS survey.}
\end{center}
\end{figure}

\begin{figure}[h!]
\textbf{Filament 10 (``BC\_332.21-0.04"): Grade ``B"}
\begin{center}
\epsscale{.6}
\plotone{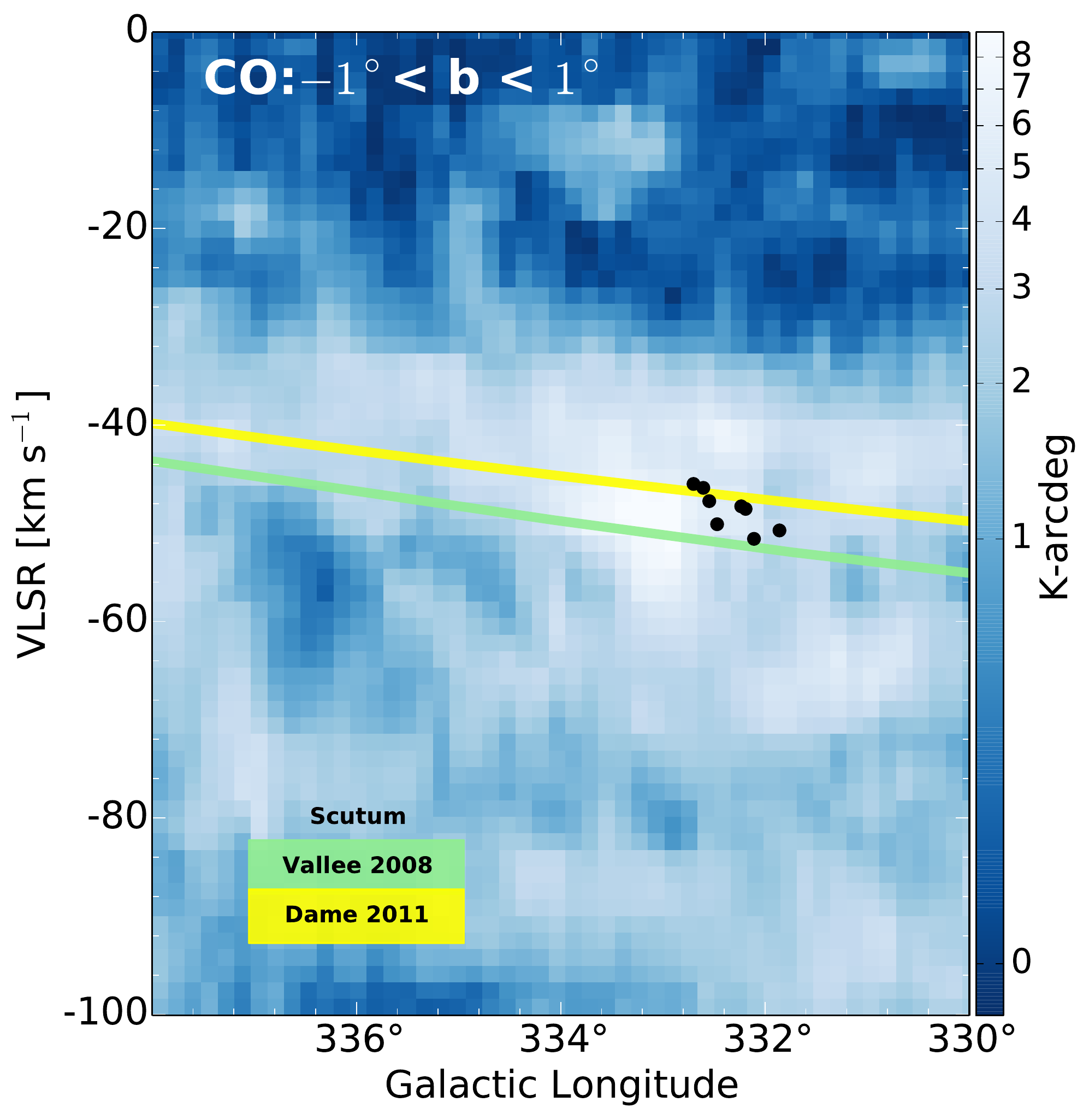}
\caption{ \textit{Top}: Position-velocity diagram of CO, NH$_{3}$, and N$_{2}$H$^+$ emission for filament 10. Blue background shows $^{12}$CO (1-0) emission integrated between $-1^\circ < \textrm{b} < 1^\circ$ \citep{Dame_2001}. Black dots show HOPS and MALT90 sources associated with filament 10. The colored lines are fits to the Scutum-Centaurus arm (see text for references). %
}
\end{center}
\end{figure}

\begin{figure}[h!]
\textbf{Filament 10 (``BC\_332.21-0.04"): Grade ``B"}
\begin{center}
\plotone{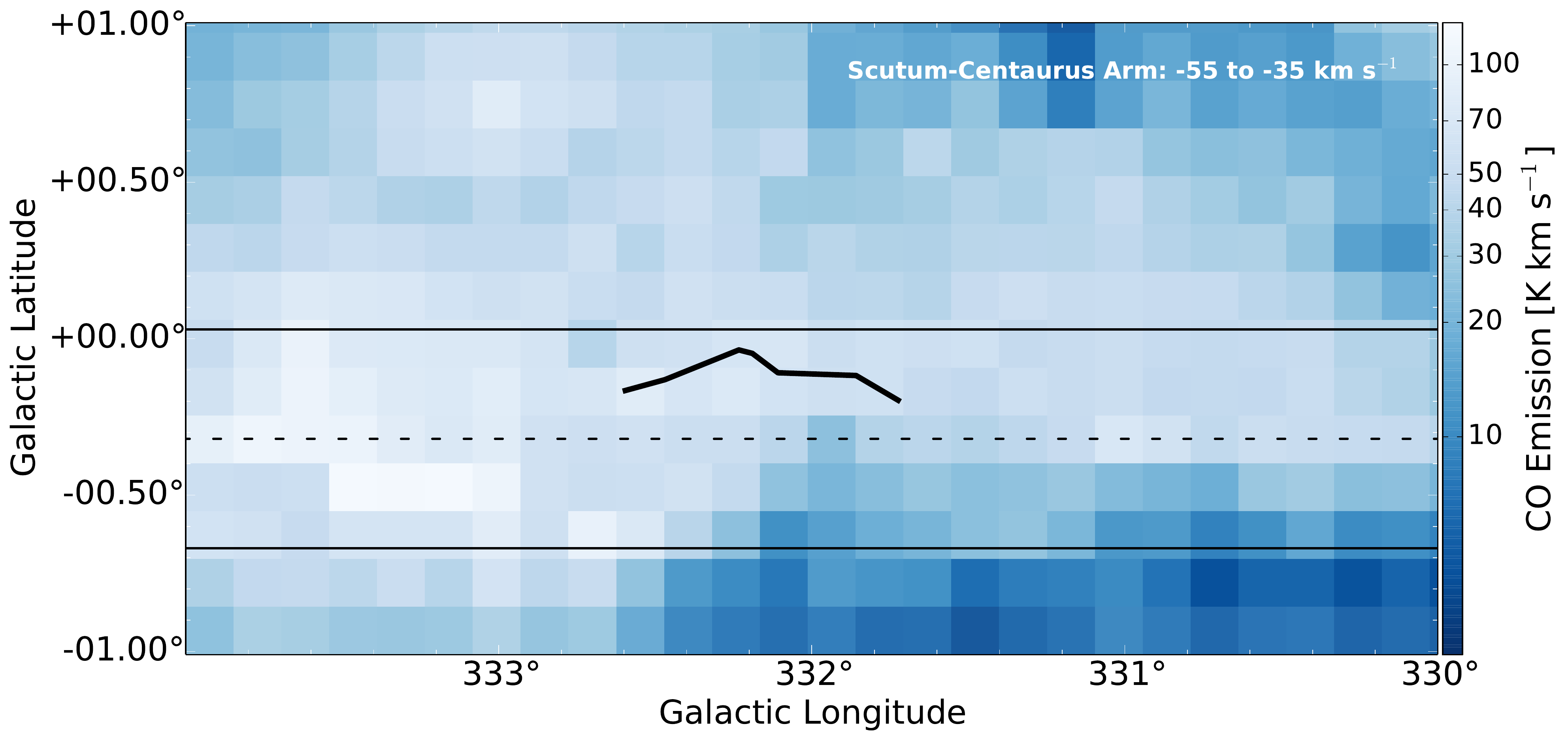}
\caption{Plane of the sky map integrated between $-55$ and $-35$ km s$^{-1}$, the approximate velocity range of the Scutum-Centaurus arm in the region around filament 10. A trace of filament 10, as it would appear as a mid-IR extinction feature, is superimposed on the $^{12}\rm{CO}$ emission map \citep{Dame_2001}. The black dashed line indicates the location of the physical Galactic mid-plane, while the solid black lines indicate $\pm$ 20 pc from the Galactic mid-plane at the 3.3 kpc distance to filament 10, assuming the candidate is associated with the \citet{Dame_2011} Scutum-Centaurus model.
}
\end{center}
\end{figure}

\begin{figure}[h!]
\textbf{Filament 10 (``BC\_332.21-0.04"): Grade ``B"}
\begin{center}
\plotone{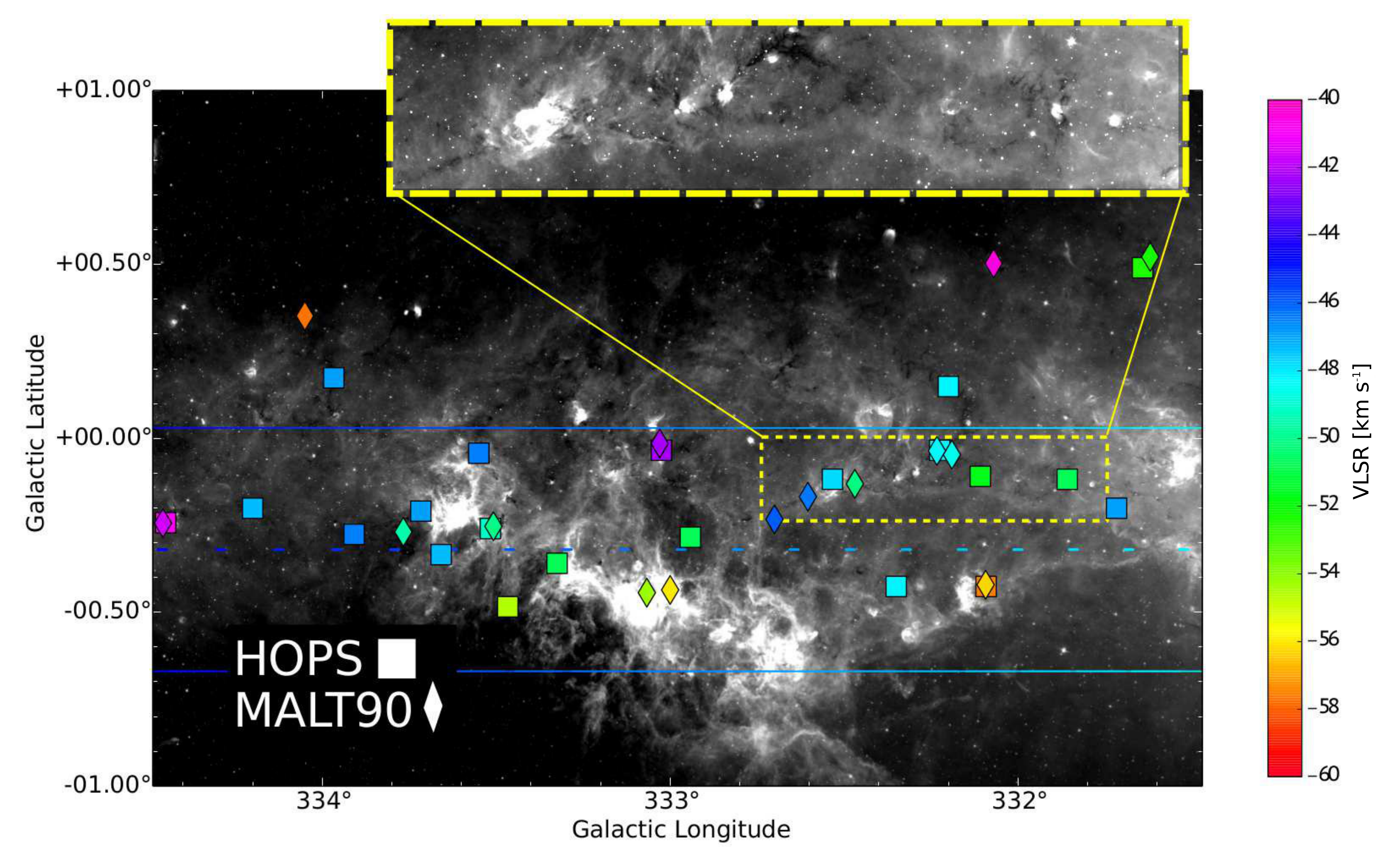}
\caption{Filament 10 lies within $\approx$ 15 pc of the physical Galactic mid-plane. The background is a GLIMPSE-\textit{Spitzer} 8 $\mu\textrm{m}$ image. The dashed line is color-coded by \citet{Dame_2011} LSR velocity and indicates the location of the physical Galactic mid-plane. The solid colored lines indicate $\pm$ 20 pc from the Galactic mid-plane at the 3.3 kpc distance to filament 10, assuming the candidate is associated with the \citet{Dame_2011} Scutum-Centaurus model. The squares and diamonds correspond to HOPS and MALT90 sources, respectively.  A closer look at filament 10 can be seen in the inset.%
}
\end{center}
\end{figure}

\end{document}